\documentclass[prd,showpacs,superscriptaddress,nofootinbib]{revtex4}
\tighten
\usepackage{epsf}
\def\la{\mathrel{\hbox{\rlap{\hbox{\lower4pt\hbox{$\sim$}}}\hbox{$<$}}}}
\def\ga{\mathrel{\hbox{\rlap{\hbox{\lower4pt\hbox{$\sim$}}}\hbox{$>$}}}}
\begin{document}

\title{\Large From the Big Bang Theory
to the Theory  of a Stationary Universe}

\author{Andrei Linde,$^{1}$ Dmitri Linde,$^{2}$ and Arthur Mezhlumian}
\affiliation{Department of Physics, Stanford University, Stanford, CA 94305,
USA\\
$^{2}$Gunn High School, 780 Arastradero Road, Palo Alto CA 94306,
 USA}

\begin{abstract}
We consider chaotic inflation in the theories with the effective
potentials
which at large $\phi$ behave either as $\phi^n$ or as
$e^{\alpha\phi}$.  In
such theories inflationary domains  containing  sufficiently large
and
homogeneous scalar field $\phi$   permanently produce new
inflationary domains of a similar type. This process may occur at
densities
considerably smaller than the Planck density.  Self-reproduction of
inflationary domains is responsible for the fundamental stationarity
which is
present in many   inflationary models: properties of the parts of the
Universe
formed in the process of self-reproduction  do not depend on the time
when this
process occurs. We call this property of the inflationary Universe
{\it
local  stationarity}. 

In addition to it, there may exist either a stationary distribution
of
probability $P_c$ to find a given field $\phi$ at a given time at a
given
point, or  a stationary distribution of probability $P_p$ to find a
given field
$\phi$ at a given time in a given physical volume. If any of these
distributions is stationary, we will be speaking of  a {\it global
stationarity} of the inflationary Universe.  

In all  realistic inflationary models which are known to us the
probability
distribution $P_c$ is not stationary. On the other hand,
investigation of the
probability distribution $P_p$ describing a self-reproducing
inflationary
Universe shows that the center of this distribution  moves towards
greater and
greater $\phi$ with increasing time. It
is argued, however,  that the probability of inflation  (and of the
self-reproduction of inflationary domains) becomes strongly
suppressed when
the energy density of the scalar field approaches the Planck density.
As a
result,  the  probability distribution $P_p$ rapidly approaches a
stationary regime, which we have found explicitly for the theories
${\lambda\over 4}\phi^4$ and $e^{\alpha\phi}$.  In this regime the
relative
fraction of the physical volume of the Universe
in a state  with given properties (with  given values of  fields,
with a given
density of matter, etc.)  does not depend on time,  both at the stage
of
inflation and after it. 

Each of the two types of stationarity mentioned above constitutes a
significant deviation of inflationary cosmology from the standard Big
Bang
paradigm. We compare our approach with other approaches to quantum
cosmology,
and illustrate some of the general  conclusions mentioned
above with the results  of a computer simulation of stochastic
processes in the
inflationary Universe.
\end{abstract}
\pacs{98.80.Cq \hskip 1.6 cm SU-ITP-93-13 \hskip 1.6 cm
gr-qc/9306035 \hskip 1.6 cm
June 28, 1993}
 \maketitle

\tableofcontents

%\voffset = -1.3truein
%\hoffset = -1.1truein

%\textwidth 470pt

\section{Introduction} \label{Intro}
The standard Big Bang theory  asserts that the Universe was born
at some moment $t = 0$ about 15 billion years ago, in a state of
infinitely
large density and temperature. With the rapid expansion  of the
Universe the
average energy
of particles, given by the temperature, decreased rapidly, and the
Universe
became cold.  This theory became especially popular after the
discovery of
the  microwave background radiation. However, by the end of the 70's
it was
understood that this theory is  hardly compatible with the  present
theory of
elementary particles  (primordial monopole problem, Polonyi fields
problem,
gravitino problem,  domain wall problem) and it has many internal
difficulties
(flatness problem,  horizon problem, homogeneity and isotropy
problems, etc.).

Fortunately,  all these problems can be solved simultaneously in the
context of
a relatively simple scenario of the Universe evolution --- the
inflationary
Universe  scenario \cite{b13}--\cite{b90}, \cite{MyBook}.  The main
idea of
this scenario is that the Universe at the very early stages of its
evolution
expanded quasi-exponentially (the stage of inflation) in a state with
energy
density dominated by the potential energy density $V(\phi)$ of some
scalar
field $\phi$. This rapid expansion made the Universe flat,
homogeneous and
isotropic and decreased exponentially the density of monopoles,
gravitinos and
domain walls. Later, the potential energy density of the scalar field
transformed into thermal energy, and  still later, the Universe was
correctly
described by the standard hot Universe theory predicting the
existence of the
microwave background radiation.

The first models of inflation were formulated in the context of the
Big Bang
theory. Their success  in solving internal problems of this theory
apparently
removed the last doubts concerning the  Big Bang cosmology.  It
remained almost
unnoticed that during the last ten years   inflationary theory
changed
considerably.  It has broken an umbilical cord connecting it with the
old Big
Bang theory, and acquired an independent life of its own. For the
practical
purposes of describing   the observable part of our Universe one may
still
speak about the Big Bang, just as one can still use Newtonian gravity
theory
to describe the Solar system with very high precision. However, if
one tries
to understand the beginning of the Universe, or its end, or its
global
structure, then some of the notions of the Big Bang theory become
inadequate.
For example, one of the main principles of the Big Bang theory is the
homogeneity of the Universe. The assertion of homogeneity seemed to
be so
important that it was called  ``the cosmological principle''
\cite{Peebles}. Indeed, without using this principle one could not
prove that
the whole Universe appeared  {\it at a single moment of time}, which
was
associated with the Big Bang. So far, inflation remains the only
theory which
explains why the observable part of the Universe is almost
homogeneous.
However, almost all versions of inflationary cosmology predict that
on a much
larger scale the Universe should be extremely inhomogeneous, with
energy
density varying from the Planck density to almost zero. Instead of
one single
Big Bang producing a single-bubble Universe, we are speaking now
about
inflationary bubbles producing new bubbles, producing new bubbles,
{\it ad
infinitum}   \cite{b19,b20}. Thus, recent development of inflationary
theory
considerably modified our cosmological paradigm \cite{MyBook}.  In
order to
understand better this modification, we should remember the main
turning points
in the evolution of the inflationary theory.

The first semi-realistic version of inflationary cosmology was
suggested
by Starobinsky \cite{b14}. However, originally it was not quite clear
what
should be the initial state of the Universe in this scenario.
Inflation in
this model could not occur if the Universe was hot from the very
beginning.
To solve this problem, Zeldovich in 1981 suggested that the
inflationary
Starobinsky Universe was created ``from nothing''  \cite{Zeld}. This
idea,
which is very popular now \cite{NothVil}--\cite{b59},  at that time
seemed too
extravagant, and most  cosmologists preferred to study inflation in
more
traditional context of the hot Universe theory.

One of the most important stages of the development of the
inflationary
cosmology was related to the old inflationary Universe scenario by
Guth
\cite{b15}. This scenario was based on three fundamental
propositions:

\begin{enumerate}
\item The Universe initially expands in a state with a very high
temperature,
which leads to the symmetry restoration in the early Universe, $\phi
(T) =
0$, where $\phi$ is some scalar field driving inflation (the inflaton
field).

\item The effective potential $V(\phi, T)$ of the scalar field $\phi$
has a
deep local minimum at $\phi = 0$ even at a very low temperature T. As
a result,
the Universe remains in a supercooled vacuum state $\phi = 0$ (false
vacuum)
for a long time. The energy-momentum tensor of such a state rapidly
becomes
equal to $T_{\mu \nu} = g_{\mu \nu} V(0)$, and the Universe expands
exponentially (inflates) until the false vacuum decays.

\item The decay of the false vacuum proceeds by forming  bubbles
containing the
field $\phi_0$ corresponding to the minimum of the effective
potential
$V(\phi)$. Reheating of the Universe occurs due to the bubble-wall
collisions.
\end{enumerate}

The main idea of the old inflationary Universe scenario was very
simple and
attractive, and the role of the old inflationary scenario in the
development of
modern cosmology was extremely important.  Unfortunately, as it was
pointed out
by Guth in \cite{b15}, this scenario had a major problem. If the rate
of the
bubble formation is bigger than the speed of the Universe expansion,
then the
phase transition occurs very rapidly and inflation does not take
place. On the
other hand, if the vacuum decay rate is small, then the Universe
after the
phase transition becomes unacceptably inhomogeneous.

All attempts to suggest a successful inflationary Universe scenario
failed
until
cosmologists managed to surmount a certain psychological barrier and
renounce
 the aforementioned assumptions, while retaining the main idea
of ref.\ \cite{b15} that the Universe has undergone inflation during
the early
stages of its evolution. The invention of the new inflationary
Universe
scenario \cite{b16} marked the departure from the assumptions (2),
(3). Later
it was shown that  the assumption (1)  also does not hold in all
realistic
models known so far, for two main reasons. First of all, the time
which is
necessary for the field $\phi$ to roll down to the minimum of
$V(\phi, T)$
is typically  too large, so that  either inflation occurs before the
field rolls
to the minimum of $V(\phi, T)$, or it does not occur at all. On the
other hand,
even if the field $\phi$ occasionally was near the minimum of
$V(\phi, T)$ from
the very beginning, inflation typically starts very late, when
thermal energy
drops from $M^4_p$ down to $V(0, T)$. In all realistic models of
inflation
$V(0, T) < 10^{-10} M^4_p$, hence inflation may start in a state with
$\phi =
0$ not earlier than at $t \sim 10^4 M^{-1}_p$. During such a time a
typical
closed Universe would collapse before the conditions necessary for
inflation
could be realized \cite{MyBook}.

The assumption (1)  was finally given up with the invention of the
chaotic
inflation scenario \cite{b17}. The main idea of this scenario was to
abandon
the  assumption that the Universe from the very beginning was hot,
and that the
initial state of the scalar field should  correspond to a minimum of
its
effective potential. Instead of that,  one should study various
initial
distributions of the scalar field $\phi$, including those which
describe the
scalar field outside of its equilibrium state, and investigate in
which case
the inflationary regime may occur.

In other words, the main idea of chaotic inflation was to remove all
unnecessary restrictions on inflationary models inherited from the
old theory
of the hot Big Bang. In fact, the first step towards this liberation
was
already made when it was suggested to consider quantum creation of
inflationary
Universe from nothing \cite{Zeld}--\cite{b59}. Chaotic inflation
naturally
incorporates this idea \cite{b56}, but it is more general: inflation
in this
scenario can also appear  under a more traditional assumption of
initial
cosmological singularity.

Thus, the main  idea of chaotic inflation is very simple and
general. One
should study all possible initial conditions without insisting that
the
Universe was in a state of thermal equilibrium, and that the field
$\phi$ was
in the minimum of its effective potential from the very beginning.
However,
this idea   strongly deviated from the standard lore and was
psychologically
difficult to accept. Therefore, for many years almost all
inflationary model
builders continued the investigation of the new inflationary scenario
and
calculated high-temperature effective potentials in all  theories
available. It
was argued that every inflationary model should satisfy the so-called
`thermal
constraints' \cite{TurStein}, that chaotic inflation requires
unnatural initial
conditions, etc.

At present the situation is quite opposite. If anybody ever discusses
the
possibility that inflation is initiated by high-temperature effects,
then
 typically the purpose of such a discussion is to show over and over
again that
this idea  does not work (see
e.g.  \cite{Thermal}), even though some exceptions from this rule
might still
be possible. On the other hand, there are many theories where one can
have
chaotic inflation  (for a review see \cite{MyBook}).

A particularly simple realization of this scenario can be achieved in
the
theory of a massive scalar field with the effective potential
${m^2\over 2}
\phi^2$, or in the theory ${\lambda\over 4}\phi^4$, or in any other
theory with
an effective potential which grows as $\phi^n$ at large $\phi$
(whether or
not there is a spontaneous symmetry breaking at small $\phi$).
Therefore a lot
of work illustrating the basic principles of chaotic inflation was
carried out
in the context of these simple models. However, it would be
absolutely
incorrect to identify the chaotic inflation scenario with these
models, just
as it would be  incorrect to identify new inflation with  the
Coleman-Weinberg
theory.  The dividing line between the new inflation and the chaotic
inflation
was not in the choice of a specific class of potentials, but in the
abandoning
of the idea that the high-temperature phase transitions should be a
necessary
pre-requisite of inflation.

Indeed, already in the first paper where the chaotic inflation was
proposed
\cite{b17}, it was emphasized that this scenario can be implemented
not only in
the theories $\sim \phi^n$, but in {\it any} model where the
effective
potential  is sufficiently flat at some $\phi$. In the second paper
on chaotic
inflation \cite{Chaot2} this scenario was implemented in a  model
with an
effective potential of the same type as those used in the new
inflationary
scenario. It was explained   that the standard scenario based on
high-temperature phase transitions cannot be realized in this theory,
whereas
the chaotic inflation can occur there, either due to the  rolling of
the field
$\phi$ from $\phi > 1$, or due to the rolling down from the local
maximum of
the effective potential at $\phi = 0$. In \cite{ExpChaot} it was
pointed out
that chaotic inflation can be implemented in many theories including
the
theories with exponential potentials $\sim e^{\alpha\phi}$ with
$\alpha <
\sqrt{16\pi}$. This is precisely the same class of theories which two
years
later was used in \cite{Exp} to obtain the  power  law inflation
\cite{Power}.
The class of models where chaotic inflation can be realized includes
also the
models based on the $SU(5)$ grand unified theory \cite{MyBook}, the
$R^2$
inflation  (modified Starobinsky model) \cite{StChaot},
supergravity-inspired
models with polynomial and non-polynomial potentials \cite{Supergr},
`natural
inflation' \cite{Natural}, `extended inflation'  \cite{b90},
\cite{EtExInf},
`hybrid inflation' \cite{Axions}, etc.\footnote{We discussed here
this
question at some length because recently there appeared  many papers
comparing observational consequences of different versions of
inflationary
cosmology. Some of these papers use their own definitions of new and
chaotic
inflation, which differ considerably from the original definitions
given by
one of the present authors at the time when these scenarios were
invented.
This
sometimes  leads to such misleading statements as the claim that  if
observational data will show that the inflaton potential is
exponential, or
that it is a pseudo-Goldstone potential used in the `natural
inflation' model,
this will disprove   chaotic inflation.   We should emphasize that
the
generality of chaotic inflation does not diminish in any way the
originality
and  ingenuity of any of its
particular realizations mentioned above. Observational data  should
make it
possible to chose between models  with different effective
potentials, such as
$\phi^n$, $e^{\alpha\phi}$,    the pseudo-Goldstone potential, etc.
However,
we are unaware of any possibility to
obtain inflation in the theories with exponential or pseudo-Goldstone
potentials outside of the scope of the chaotic inflation scenario.}

Several years ago it was discovered that chaotic inflation in many
models
including the theories $\phi^n$ has a very interesting property,
which  will be
discussed in the present paper \cite{b19,b20}. If the Universe
contains at
least one inflationary domain of a size of horizon ($h$-region) with
a
sufficiently large and homogeneous scalar field $\phi$, then this
domain will
permanently produce new $h$-regions of a similar type. During this
process the
total physical volume of the inflationary Universe (which is
proportional to
the total number of $h$-regions) will grow indefinitely. The process
of
self-reproduction of inflationary domains occurs  not only in the
theories with
the effective potentials growing at large $\phi$ \cite{b19,b20}, but
in some
theories with the effective potentials used in old, new and extended
inflation
scenarios as well \cite{b51,b52,b62,EtExInf}.  However, in the models
with the
potentials growing at large $\phi$ the existence of this effect was
most
unexpected, and it leads to especially interesting cosmological
consequences
\cite{MyBook}.

The process of the self-reproduction of inflationary Universe
represents a
major deviation of inflationary cosmology from the standard Big Bang
theory. In
this paper we will make an attempt to relate the existence of this
process to
the
long-standing problem of finding  a true  gravitational  vacuum, some
kind of a
stationary ground state of a system, similar  to the vacuum state in
Minkowski
space or to the ground state in quantum statistics.

{\it A priori} it is not quite clear that a stationary ground  state
in quantum
gravity may exist at all. This question becomes especially pronounced
being
applied to quantum cosmology. Indeed, how can one expect that at the
quantum
level there exists any kind of stationary state if all nontrivial
classical
cosmological solutions are non-stationary?

A possible (and rather paradoxical) answer to this question is
suggested by
the  observation made in \cite{DeWitt} that the wave function of the
whole
Universe  does not depend on time, since the total Hamiltonian,
including the
contribution from gravitational interactions, identically vanishes.
This
observation has led several authors to the idea of using scale factor
of the
Universe instead of time, which implied existence of many strange
phenomena
like time reversal and resurrection of the dead at the moment of
maximal
expansion of a closed Universe. The resolution of the paradox was
suggested by
DeWitt in \cite{DeWitt} (see also discussion of this question in
\cite{Page},
\cite{MyBook})\@. We do not ask why the Universe evolves in time
measured by
some nonexistent observer outside our Universe; we just ask why {\it
we see} it
evolving in time. At the moment when we make our observations, the
Universe is
divided into two pieces: an observer with its measuring devices and
the rest of
the Universe. The wave function of the rest of the Universe does
depend on the
time shown by the clock of the observer. Thus, at the moment we start
observing
the Universe it ceases being static and appears to us time-dependent.

{}From this discussion it follows that the condition of
time-independence  is not strong enough to pick up a unique wave
function of
the
Universe corresponding to its ground state: each wave function
satisfying the
Wheeler-DeWitt equation (being interpreted as a Schr\"odinger
equation
for  the wave function of the Universe) is time-independent. One way
to deal
with this problem is to assume that the standard Euclidean methods,
which help
to find the
wave function of the vacuum state in ordinary quantum theory of
matter fields,
will work for the wave function of the Universe as well. This
assumption was
made  by Hartle and Hawking \cite{HH}. Another way is to look for a
possibility
that with an account taken of quantum effects  our Universe in some
cases may
approach a stationary state, which could be called the ground state.

In some cases these two approaches lead to the same results. For
example, the
square  of the Hartle-Hawking wave function correctly describes
$P_c(\phi,t)$,
the stationary probability distribution for finding the scalar field
$\phi$ at
a
given time in a given point of de Sitter space with the Hubble
constant $H \gg
m$, where $m$ is the mass of the scalar field
\cite{b60}--\cite{Star}. Here
the subscript $c$ in $P_c(\phi,t)$ means that the distribution is
considered in
comoving coordinates, which do not take into account the exponential
growth of
physical volume of the Universe. However, later it was realized that
no
stationary solutions for $P_c(\phi,t)$ can exist in realistic
versions of
inflationary cosmology \cite{b19,b20}. The reason is that the
condition $H \gg
m$ is strongly violated near the minimum of the effective potential
$V(\phi)$
corresponding to the present state of the Universe. Indeed, this
condition
could be satisfied at present only if the Compton wavelength of the
inflaton
field were larger than the size of the observable part of the
Universe $\sim
H^{-1}$ and its mass were smaller than $10^{-22}$ eV. \@ In all
realistic models
of inflation this condition is violated by more than thirty orders of
magnitude.

Fortunately, another kind of stationarity may exist in many models of
the
inflationary Universe due to the process of self-reproduction of the
Universe
\cite{MyBook,b19,b20}.  The properties of inflationary domains formed
during the self-reproduction of the Universe do not depend on
the moment of time at which each such domain is formed; they depend
only on the
value of the scalar fields inside each domain, on the average density
of
matter  in this domain and on the physical length scale. For example,
all
domains of our Universe with energy density $\rho_0 \sim 10^{-29} g
\cdot
cm^{-3}$ filled with the same scalar fields look alike on the same
length
scale, independently of the time when they were formed. This kind
of stationarity, as opposed to the stationarity of the distribution
$P_c(\phi,t)$ corresponding to the Hartle-Hawking wave function,
cannot be
described in the minisuperspace approach. Hopefully, the existence of
this
stationarity  (which implies that our Universe has a fractal
structure) will
eventually help us to find the most adequate wave function describing
 the
self-reproducing inflationary Universe.

One way to describe this kind of stationarity is to use the methods
developed
in  the  theory of fractals. This question was studied in
\cite{ArVil} in
application to the theories where inflation occurs near a local
maximum of the
effective potential $V(\phi)$. In this case the expansion of the
whole
Universe,
though eternal, is almost uniform --- the scale factor of the
Universe $a(t)$
grows as $e^{Ht}$, where $H$ almost does not depend on the value of
the
inflaton field $\phi$. This makes it possible to factor out the
trivial overall
expansion factor, and come  to a time-independent fractal structure
without
taking into account the difference between $P_c(\phi, t)$ and
$P_p(\phi,t)$.
Unfortunately, it is rather difficult to apply
these methods to the most interesting and general case where
the effective potential $V(\phi)$ considerably changes during
inflation. This
is the case, for example, in the theories with potentials $\phi^n$
and
$e^{\alpha\phi}$ \cite{b17}. In such theories one has to treat the
boundary
conditions much more carefully in order to find the correct  rate of
the volume
growth.

Another approach is to investigate the probability distribution
$P_p(\phi,t)$,
which  takes into account the exponential growth of the volume of
domains
filled
by the inflaton field $\phi$ \cite{b19}.  Solutions for this
probability
distribution
were first examined in \cite{b19,b20} for the case of chaotic
inflation in the theories with potentials $\phi^n$. It
was shown  that if the initial value of the scalar field $\phi$ is
greater than
some critical value $\phi^*$, then the probability distribution
$P_p(\phi,t)$
permanently moves to greater and greater fields $\phi$. This process
continues
until the maximum of the distribution $P_p(\phi,t)$ approaches the
field
$\phi_p$, at which the effective potential of the field becomes of
the order of
Planck density $M_p^4$, where the standard methods of quantum field
theory
in a curved classical space are no longer valid. By the methods used
in
\cite{b19,b20} it was impossible to check whether $P_p(\phi,t)$
asymptotically approaches any stationary regime in the classical
domain $\phi
< \phi_p$.

More generally, one  may consider the probability distribution
$P_p(\phi,t|\chi)$, which shows the fraction of volume of the
Universe filled
by the field $\phi$ at the time $t$, if originally, at the time $t =
0$, the
part  of the Universe which we study was filled by the field $\phi_0
=
\chi$. Several important steps towards the investigation of this
probability
distribution in the theories $\phi^n$ were made  by Nambu and Sasaki
\cite{Nambu} (see also \cite{SNN,NewStationary}) and by Miji\'c
\cite{Mijic}.
Their papers
contain many beautiful insights, and we will use many results
obtained by these authors. However,  Miji\'c \cite{Mijic}  did not
have a
purpose in obtaining a complete expression for the stationary
distribution
$P_p(\phi,t)$\@. The corresponding expressions were obtained for
various
 potentials $V(\phi)$ in \cite{Nambu}. Unfortunately, the stationary
distribution $P_p(\phi,t)$ obtained in \cite{Nambu} was almost
entirely
concentrated at $\phi \gg \phi_p$, i.e. at $V(\phi) \gg M_p^4$, where
the
methods used in \cite{Nambu} were inapplicable.

In the present paper we will argue that the self-reproduction of the
inflationary Universe effectively kills itself at densities
approaching the
Planck density. This leads to the existence of a stationary
probability
distribution $P_p(\phi,t)$ concentrated entirely at sub-Planckian
densities
$V(\phi) < M_p^4$  in a wide class of theories leading to chaotic
inflation
\cite{MezhLinSmall,MezhLin}.

The new cosmological paradigm based on inflationary cosmology is very
unusual. Instead of the Universe looking like an expanding ball of
fire, we are
considering now a huge fractal consisting of many inflating balls
producing
new balls, producing new balls, etc. In order to make the new
cosmological concepts more visually clear, we include in this paper
a series of figures which show the results of computer simulations of
stochastic processes in the inflationary Universe \cite{LinLin}.

The plan of the paper is the following. In Section
\ref{SelfRepChaot} we give a short review of chaotic
inflation.  We  discuss the generation of density perturbations and
explain
how they lead to the process of self-reproduction of the Universe.
In Section
\ref{Wave+Stoch} we discuss the interrelations between stochastic and
Euclidean approaches to quantum cosmology and explain the main
advantages of
the  probability distribution $P_p(\phi,t) $  for the description of
the global
structure of  inflationary Universe.  Section \ref{Comput} briefly
describes
the results of the computer simulation of  stochastic  processes in
inflationary
Universe. Section \ref{Stationary}  contains the basics of our
approach to
finding a stationary probability distribution $P_p(\phi,t|\chi)$. In
Section
\ref{OtherTimes} we describe the analogous investigation for a
different
parametrization of time. Section \ref{Jumps} contains a general
discussion of
the consequences of our results, which are summarized in a short
concluding
Section \ref{Summary}.

To simplify our notation, throughout this paper we will use the
system of units
where  $M_p = 1$.

\section{Self-Reproducing Chaotic Inflationary Universe
\label{SelfRepChaot}}

\subsection{Chaotic Inflation \label{Chaotic}}

We will begin our discussion of chaotic inflation with the simplest
model based
on the  theory of a  scalar field $\phi$ minimally coupled to
gravity, with the
Lagrangian
\begin{equation}\label{E01}
L =  \frac{1}{16\pi}R + \frac{1}{2} \partial_{\mu} \phi
\partial^{\mu} \phi - V(\phi)  \ .
\end{equation}
Here $G = M^{-2}_p = 1$ is the gravitational constant,  $R$ is the
curvature
scalar, and $V(\phi)$ is the effective potential of the scalar field.
If the
classical field $\phi$ is sufficiently homogeneous in some domain of
the
Universe (see below), then its behavior inside this domain is
governed by the
equations
\begin{equation}\label{E02}
\ddot\phi + 3H\dot\phi = -dV/d\phi \ ,
\end{equation}
\begin{equation}\label{E03}
H^2 + \frac{k}{a^2} = \frac{8\pi}{3}\, \left(\frac{1}{2}
\dot\phi^2 + V(\phi)\right) \ .
\end{equation}
Here $H={\dot a}/a, a(t)$ is the scale factor of the
 Universe, $k=+1, -1,$ or $0$ for a closed, open or
flat Universe, respectively.

The simplest version of the  theory (\ref{E01}) is the theory of
a massive
noninteracting scalar field with the effective potential
$V(\phi)=\frac{1}{2}
m^2\phi^2$, where $m$ is the mass of the scalar field $\phi$, $m
\ll1$\@. If the
field $\phi$ initially is sufficiently large, $\phi \gg 1$, then one
can show that
the functions $\phi(t)$ and $a(t)$ rapidly approach the asymptotic
regime
\begin{equation}\label{E04} \phi(t)=\phi_0 - \frac{m}{2(3\pi)^{1/2}}
t \ ,
\end{equation} \begin{equation}\label{E05}
a(t) = a_0 \exp\left(2 \pi
\left(\phi^2_0 - \phi^2(t)\right) \right) \ .
\end{equation}
Note that in this regime the second derivative of the scalar field in
eq. (\ref{E02}) and  the term $k/a^2$ in  (\ref{E03})
can be neglected. The last statement  means
that the Universe becomes locally flat.

 According to (\ref{E04}) and
(\ref{E05}), during the time $\tau \sim \phi/m$ the relative change
of
the field
$\phi$ remains small, the effective potential $V(\phi)$ changes very
slowly and
the Universe  expands quasi-exponentially:
\begin{equation}\label{E06} a(t+\Delta t)\sim a(t) \exp
(H\Delta t) \end{equation}
for $\Delta t \leq \tau = \phi/m$\@. Here
\begin{equation}\label{E07}
H = 2 \sqrt \frac{\pi}{3} \, m \phi \ .
\end{equation}
Note that
$\tau \gg H^{-1}$ for $\phi \gg 1$.

The regime of the slow rolling of the field $\phi$ and the
quasi-exponential
expansion (inflation) of the Universe ends at $\phi \la \phi_e$. In
the theory
under consideration, $\phi_e \sim
0.2$\@.   At $\phi \leq \phi_e$ the field  $\phi$ oscillates rapidly,
and if this
field interacts with other matter fields (which are not written
explicitly in eq.
(\ref{E01})), its potential energy $V(\phi) \sim      \frac{m^2
\phi_e^2}{2}
 \sim {m^2\over 50} $ is transformed into heat. The
reheating temperature $T_R$ may be of the order
$m^{1/2}$ or somewhat smaller, depending on the
strength of the interaction of the field $\phi$ with other
fields. It is important that $T_R$ does not
depend on the initial value $\phi_0$ of the field $\phi$.
  The only parameter which depends on $\phi_0$ is the scale
factor $a(t)$, which grows $e^{2\pi\phi^2_0}$
times during inflation.

All results obtained above can be easily generalized for the theories
with more
complicated effective potentials. For example, during inflation in
the theories
with $V(\phi)=\frac{\lambda}{n}\, \phi^n$ one has
\begin{equation}\label{E04a}
\phi^{4-n\over 2}(t)=\phi^{4-n\over 2}_0 - \frac{4-n}{2}\,
\sqrt{\frac{n \lambda}{24 \pi}} \ t \ \ \ \ \mbox{for} \ n\not = 4\ ,
\end{equation} \begin{equation}\label{E04a'}
\phi(t)=\phi_0 \ \exp\left(-\sqrt{\lambda\over 6\pi} \, t\right) \ \
\ \ \
 \mbox{for} \ n = 4 \ .
 \end{equation}
For all $n$,
\begin{equation}\label{E05a}
a(t) = a_0 \exp \left({4 \pi\over n}
(\phi^2_0 - \phi^2(t))\right) \ .
\end{equation}
Inflation ends at
\begin{equation}\label{E05a'}
\phi_e \sim {n\over 4\sqrt{3\pi}} \ .
\end{equation}

Note, that  in all realistic models of elementary particles
spontaneous
symmetry breaking occurs on a scale which is many orders of magnitude
smaller
than 1 (i.e. smaller than $M_p$).
Therefore all results which we obtained remain valid in all theories
which have
potentials $V(\phi) \sim \frac{\lambda}{n}\, \phi^n$ at $\phi \ga 1$,
independently of the issue of spontaneous symmetry breaking which may
occur in
such theories at small $\phi$.

As was first pointed out in \cite{ExpChaot}, chaotic inflation occurs
as well
in the theories with exponential effective potentials $V(\phi)={V_0}\
{e^{\alpha\phi}}$ for sufficiently small $\alpha$.  It was shown
later
 \cite{Exp} that in the theories with \begin{equation}\label{E04au''}
V(\phi)={V_0}\ {e^{\alpha\phi}}
 \end{equation}
the Einstein equations and equations for the scalar field have an
exact
solution:
\begin{equation}\label{E04a''}
\phi(t)=\phi_0 - {2\over\alpha}\  \ln {t\over t_0}  \ ,
 \end{equation}
\begin{equation}\label{E05a''}
a(t) = a_0 \, t^p,  \ \  p =   \frac{16 \, \pi}{\alpha^2} \ .
\end{equation}
Note that here we are dealing with the power law expansion of the
Universe. It
can be called inflation  if $p \gg 1$, which
implies that
$\alpha \ll \sqrt{16\pi} \sim 7$ \cite{ExpChaot}. In this case the
Hubble
constant $H$ almost does not
change
within the Hubble time $\sim H^{-1}$, and expansion looks
quasiexponential,
like in eq. (\ref{E06})\@. Inflation in the theory with the
exponential
potential
never ends; it occurs at all $\phi$\@. In order to make this theory
realistic, one
should assume that at small $\phi$ the effective potential becomes
steep and
inflation ends at  $\phi < \phi_e$\@.  Without any loss of generality
one can
assume that $\phi_e = 0$\@. In this case the parameter $V_0$ gives
the
value of the
effective potential at the end of inflation.

\subsection{Initial conditions for inflation \label{IniCond}}
Let us consider first a closed Universe of initial  size $\l \sim 1$
(in Planck
units), which emerges  from the
space-time foam, or from singularity, or from `nothing'  in a state
with  the
Planck
density $\rho \sim 1$. Only starting from this moment, i.e. at $\rho
\la 1$,
can we describe this domain as  a {\it classical} Universe.  Thus,
at this
initial moment the sum of the kinetic energy density, gradient energy
density,
and the potential energy density  is of the order unity:\, ${1\over
2}
\dot\phi^2 + {1\over 2} (\partial_i\phi)^2 +V(\phi) \sim 1$.

We wish to emphasize, that there are no {\it a priori} constraints on
the
initial value of the scalar field in this domain, except for the
constraint
${1\over 2} \dot\phi^2 + {1\over 2} (\partial_i\phi)^2 +V(\phi) \sim
1$.  Let
us consider for a moment a theory with $V(\phi) = const$. This theory
is
invariant under the shift  $\phi\to \phi + a$. Therefore, in such a
theory {\it
all} initial values of the homogeneous component of the scalar field
$\phi$ are
equally probable. Note, that this expectation would be incorrect if
the scalar
field should vanish at the boundaries of the original domain. Then
the
constraint ${1\over 2} (\partial_i\phi)^2 \la 1$ would tell us that
the scalar
field cannot be greater than $1$ inside a domain of initial size $1$.
 However,
if the original domain is a closed Universe, then it has no
boundaries.  (We
will discuss a more general case shortly.)

The only constraint on the average amplitude of the field appears if
the
effective potential is not constant, but grows and becomes greater
than the
Planck density at $\phi > \phi_p$, where  $V(\phi_p) = 1$. This
constraint
implies that $\phi \la \phi_p$, but it does not give any reason to
expect that
$\phi \ll \phi_p$. This suggests that the  typical initial value
$\phi_0$ of
the field $\phi$ in such a  theory is  \begin{equation}\label{E08'}
\phi_0
\sim  \phi_p \ . \end{equation}

Thus, we expect that typical initial conditions correspond to
${1\over 2}
\dot\phi^2 \sim {1\over 2} (\partial_i\phi)^2\sim V(\phi) = O(1)$.
Note that if
by any chance ${1\over 2} \dot\phi^2 + {1\over 2} (\partial_i\phi)^2
\la
V(\phi)$ in the domain under consideration, then inflation begins,
and  within
the Planck time the terms  ${1\over 2} \dot\phi^2$ and ${1\over 2}
(\partial_i\phi)^2$ become much smaller than $V(\phi)$, which ensures
continuation of inflation.  It seems therefore that chaotic inflation
occurs
under rather natural initial conditions, if it can begin at $V(\phi)
\sim 1$
\cite{MyBook,b17,ExpChaot}.

The assumption that inflation may begin at a very large $\phi$ has
important
implications. For example, in the theory (\ref{E01}) one has
\begin{equation}\label{E08} \phi_0 \sim  \phi_p \sim m^{-1/2}\ .
\end{equation}
Let us consider for definiteness a closed Universe of a typical
initial size
$O(1)$. Then, according to (\ref{E05}), the total size of the
Universe after
inflation becomes equal to
\begin{equation}\label{E09}
l \sim  \exp \left(2\pi \phi^2_0\right)
\sim  \exp \left(\frac{2\pi}{m^2}\right) \ .
\end{equation}
For $m\sim 10^{-6}$ (which is necessary to produce
density perturbations $\frac{\delta{\rho}}{\rho} \sim 10^{-5}$, see
below)
\begin{equation}\label{E10}
l \sim  \exp \left(2\pi 10^{12}\right) > 10^{10^{12}} cm \ .
\end{equation}
Thus, according to this estimate, the smallest possible domain of the
Universe of
initial size \mbox{$O(M_p^{-1}) \sim 10^{-33} cm$}
 after inflation becomes
much larger
than the size of the observable part of the Universe $\sim 10^{28}
cm$\@. This is
the reason why our part of the Universe looks flat, homogeneous and
isotropic.
The same mechanism solves also the horizon problem. Indeed, any
domain of
the Planck size, which becomes causally connected within the Planck
time, gives
rise to the part of the Universe which is much larger than the part
which we
can see now.

In what follows we will return many times to our conclusion that the
most
probable initial value of the scalar field corresponds to  $\phi \sim
\phi_p$.
There were many objections to it. Even though all these objections
were
answered many years ago
\cite{MyBook,ExpChaot},  we need to discuss here one of these
objections again.
This is important for a proper understanding of the new picture of
the evolution of the Universe in inflationary cosmology.

Assume that the Universe is not closed but infinite, or at least
extremely
large from the very beginning. (This objection does not apply to the
closed Universe scenario discussed above.) In this case one could
argue that
our expectation that $\phi_0 \sim \phi_p \gg 1$ is not very natural
\cite{KolbTurn}.  Indeed, the conditions  $(\partial_i\phi)^2 \la 1$
and
$\phi_0 \sim \phi_p \gg 1$ imply that the field $\phi$ should be of
the same
order of magnitude $\phi \sim \phi_p \gg 1$ on a length scale at
least as large
as  $\phi_p$, which is much larger than the scale of horizon $l\sim
1$ at the
Planck time.  But this is highly improbable, since initially (i.e.,
at the
Planck time ) there should be no correlation between values of the
field
$\phi$ in different regions of the Universe separated from one
another by
distances greater than $1$.  The existence of such correlation would
violate
causality. As it is written in \cite{KolbTurn},  the scalar field
$\phi$ must
be smooth on a scale much greater than the scale of the horizon,
which does not
sound very chaotic.

The answer to this objection is very simple \cite{MyBook,ExpChaot}.
We have
absolutely no reason to expect that the overall energy density $\rho$
simultaneously becomes smaller than the Planck energy density in all
causally
disconnected regions of an infinite Universe, since that would imply
the
existence of an acausal correlation between values of $\rho$ in
different
domains of  Planckian size $l_p
\sim 1$.  Thus, each such domain at the Planck time after its
creation looks like an isolated island of
classical space-time, which emerges from the space-time foam
independently of
other such islands.  During inflation, each of these islands
independently
acquires a size many orders of magnitude larger than the size of the
observable part of the Universe.  A typical initial size of a domain
of
classical space-time with $\rho \la 1$ is of the order of the Planck
length.
Outside each of these domains the condition $\rho \la 1$ no longer
holds, and
there is no correlation between values of the field $\phi$ in
different
disconnected regions of classical space-time of size $1$.  But such
correlation
is not  necessary at all for the realization of the inflationary
Universe
scenario: according to the `no hair' theorem for de Sitter space, a
sufficient
condition for the existence of an inflationary region of the Universe
is that
inflation takes place inside a region whose size is of order
$H^{-1}$. In our
case this condition is satisfied.

We wish to emphasize once again that the confusion  discussed above,
involving
the correlation between values of the field $\phi$ in different
causally
disconnected regions of the Universe, is rooted in the familiar
notion of a
very large Universe that is instantaneously created from a singular
state with
$\rho= \infty$,  and instantaneously passes through a state with the
Planck
density $\rho = 1$.   The lack of justification for such a notion is
the very
essence of the horizon problem.  Now, having disposed of the horizon
problem
with the aid of the inflationary Universe scenario, we can possibly
manage to
familiarize ourselves with a different picture of the Universe. In
this picture
the simultaneous creation of the whole Universe is possible only if
its initial
size is of the order $1$, in which case no long-range correlations
appear.
Initial conditions should be formulated at the Planck time and on the
Planck
scale. Within each Planck-size island of the classical space-time,
the initial
spatial  distribution of the scalar field cannot be very irregular
due to the
constraint  $(\partial_i\phi)^2 \la 1$. But this does not impose any
constraints on the average values of the scalar field $\phi$ in each
of such
domains. One should examine all possible values of the field $\phi$
and check
whether they could lead to inflation.

Note, that the arguments given above \cite{b17,ExpChaot} suggest that
initial
conditions for inflation are quite natural only if inflation  begins
as close
as possible to the Planck density. These arguments do not give any
support to
the models where inflation is possible only at densities much smaller
than $1$.
And indeed, an investigation of this question shows, for example,
that a
typical closed Universe where inflation is possible only at $V(\phi)
\ll 1$
collapses before inflation begins. Thus, inflationary models of that
type
require fine-tuned initial conditions \cite{Piran}, and apparently
cannot
solve the flatness problem.

Does this mean that we should forget all models where inflation may
occur only
at  $V(\phi) \ll 1$? Does this mean, in particular, that the `natural
inflation'
model is, in fact, absolutely unnatural?

As we will argue in the last section of this paper, it may be
possible to
rescue such models either by including them into a more general
cosmological
scenario (`hybrid inflation' \cite{Axions}), or by using some ideas
of
quantum cosmology, which we are going to elaborate.

\subsection{Quantum Fluctuations and Density Perturbations
\label{Perturb}}

According to quantum field theory, empty space is not
entirely empty. It is filled with quantum fluctuations of
all types of physical fields. These fluctuations can be
considered as waves of physical fields with all possible
wavelengths, moving in all possible directions. If the
values of these fields, averaged over some macroscopically
large time, vanish, then the space filled with these fields
seems to us empty and can be called the vacuum.

In the exponentially expanding Universe the vacuum structure
is much more complicated. The wavelengths of all vacuum
fluctuations of the scalar field $\phi$ grow exponentially
in the expanding Universe. When the wavelength of any
particular fluctuation becomes greater than $H^{-1}$, this
fluctuation stops oscillating, and its amplitude freezes at
some nonzero value $\delta\phi (x)$ because of the large
friction term $3H\dot{\phi}$ in the equation of motion of the field
$\phi$\@. The amplitude of this fluctuation then remains
almost unchanged for a very long time, whereas its
wavelength grows exponentially. Therefore, the appearance of
such a frozen fluctuation is equivalent to the appearance of
a classical field $\delta\phi (x)$ that does not vanish
after averaging over macroscopic intervals of space and
time.

Because the vacuum contains fluctuations of all
wavelengths, inflation leads to the creation of more and
more new perturbations of the classical field with
wavelengths greater than $H^{-1}$\@. The average amplitude of
such perturbations generated during a time interval $H^{-1}$
(in which the Universe expands by a factor of e) is given
by\footnote{
To be more precise, the amplitude of each wave is greater by $\sqrt
2$, but this factor disappears after taking an average over all
phases.}
\begin{equation}\label{E23}
|\delta\phi(x)| \approx \frac{H}{2\pi}\ .
\end{equation}
 If the field is massless, the amplitude of each frozen wave does not
change in time at all. On the other hand, phases of each waves are
random.
Therefore,  the sum of all waves at a given point fluctuates and
experiences
Brownian jumps in all directions. As a result, the values of the
scalar
field in different points  become different from each other, and the
corresponding variance grows as
\begin{equation}\label{E23a0}
 {\left<\phi^2\right>} = {H^3\over 4\pi^2}\  t\ ,
\end{equation}
i.e.
\begin{equation}\label{E23a}
 \sqrt{\left<\phi^2\right>} = {H\over 2\pi}\ \sqrt{H\, t}\ .
\end{equation}
This result, which was first obtained in  \cite{b99}, is an obvious
consequence
of eq. (\ref{E23}), if one considers the Brownian motion of the field
$\phi$.
One should just remember that the field makes a step ${H\over 2\pi}$
each time
$\Delta t = H^{-1}$, and the total number of such steps during the
time $t$ is
given by $N = Ht$.

If the field $\phi$ is massive, with $m \ll H$, then the
amplitudes of the fluctuations frozen at each time interval $\Delta t
\sim H^{-1}$
are given by the same equation as for the massless field, but later
the
amplitudes of long-wavelength fluctuations begin to decrease slowly
and
the
variance of the long-wavelength fluctuations, instead of indefinite
growth
(\ref{E23a}), approaches the limit \cite{b99}
 \begin{equation}\label{E23b}
 \left<\phi^2\right> = {3H^4\over 8\pi^2\, m^2}\ .
\end{equation}
If perturbations grow while the mean value of the field decreases,
the
behavior of the variance becomes a little bit more complicated, see
Section \ref{SelfRepr}.

Fluctuations of the field $\phi$ lead to adiabatic density
perturbations
\begin{equation} \label{E23c}
\delta\rho \sim V^\prime(\phi) \delta \phi\ ,
\end{equation}
which grow after inflation, and at the stage of the cold matter
dominance acquire the  amplitude \cite{b42},  \cite{MyBook}
\begin{equation}\label{E24}
\frac{\delta\rho}{\rho} = \frac{48}{5}
   \sqrt{\frac{2 \pi}{3}} \frac{V^{3/2}(\phi)}{V^\prime(\phi)}\ .
\end{equation}
Here $\phi$ is the value of the classical field $\phi (t)$
(4), at which the fluctuation we consider has the
wavelength $l \sim k^{-1} \sim H^{-1}(\phi)$ and becomes
frozen in amplitude.
In the theory of the massive scalar field with $V(\phi) =
\frac{m^2}{2} \phi^2$

\begin{equation}\label{E25}
\frac{\delta \rho}{\rho} = \frac{24}{5}
     \sqrt{\frac{\pi}{3}}m
\phi^2\ .
\end{equation}

Taking into account (\ref{E04}), (\ref{E05}) and also the
expansion of the Universe by about $10^{30}$ times after the
end of inflation, one can obtain the following result for
the density perturbations with the wavelength $l(cm)$ at the
moment when these perturbations begin growing and the
process of the galaxy formation starts:
\begin{equation}\label{E26}
\frac{\delta \rho}{\rho} \sim 0.8 \ m \ln l(cm)\ .
\end{equation}
This implies that the density perturbations acquire the necessary
amplitude
$\frac{\delta \rho}{\rho} \sim  10^{-5}$ on the galaxy scale, $l_g
\sim 10^{22}$
cm, if $m \sim 10^{-6}$, in Planck units, which is equivalent to $
10^{13}$ GeV.

Similar constraints on the parameters of inflationary models can be
obtained in
all other theories discussed in the previous subsection. We will
mention here the
constraints on the theory with exponential effective potential
$V(\phi)={V_0}\
{e^{\alpha\phi}}$\@. Density perturbations in this theory have  the
following
dependence on $\alpha$, $V_0$ and the length scale $l$ measured in
cm:
\begin{equation} \label{E180.1}
\frac{\delta \rho(l)}{\rho} \sim
 15\,{\sqrt{V_0}\over \alpha}\ l^{\phantom{}^{\frac{1}{p-1}}} \; \; ;
\; \ p
= \frac{16 \, \pi}{\alpha^2} \ .
\end{equation}
This spectrum in
realistic models of inflation should not significantly (much more
than by a
factor of 2) change between the galaxy scale $l_g \sim 10^{22}$
cm and the
scale of horizon $l_h \sim 10^{28}$ cm, and should be of the order of
$10^{-5}$
on the horizon scale. This leads to the constraints $\alpha \la 1$
and $V_0 \sim
10^{-11} \alpha^2$.

In our previous investigation we only considered local
properties of the inflationary Universe, which was quite
sufficient for description of the observable part of the
Universe of the present size $ l\sim 10^{28}$ cm. For example,
in the model of a massive field $\phi$ with $m\sim 10^{-6}$ our
Universe, in
accordance with (\ref{E26}), remains relatively homogeneous up to the
scale
\begin{equation}\label{E28}
 l \leq l^* \sim e^{1/m} cm
      \sim 10^{5\cdot 10^5} cm\ .
\end{equation}
 The  density perturbations on the scale $l^*$ were
formed at the time when the scalar field $\phi(t)$
was equal to $\phi^*$, where (see (\ref{E25}))
\begin{equation}\label{E29}
\phi^*\sim {1\over 2\sqrt m}\sim  5\times10^2 \ ,
\end{equation}
Note that $V(\phi^*)\sim {m^2\over 2}
(\phi^*)^2 \sim m  \ll 1$\@.
On  scales $l > l^*$ the Universe becomes extremely inhomogeneous
due to quantum fluctuations produced during inflation.

Similar conclusion proves to be correct for all other models
discussed above. In
the models with \mbox{$V(\phi) = {\lambda\over n} \phi^n$}\, the
critical
value of the
field $\phi$ can be estimated as
\begin{equation}\label{E29a}
\phi^*\sim \left({n^3\over 134\,
\lambda}\right)^{\phantom{}^{\frac{1}{n+2}}} \
,
\end{equation}
whereas in the models with the exponential potential
\begin{equation}\label{E29aa}
\phi^*\sim {1\over \alpha} \ln {\alpha^2\over 134\, V_0} \ .
\end{equation}
The corresponding scale $l^* = l(\phi^*)$ gives us the typical scale
at which the
Universe still remains relatively homogeneous and can be described as
a
Friedmann  Universe. In all realistic models this scale is much
larger than the
size of the observable part of the Universe, but much smaller than
the naive
classical estimate of the size of a homogeneous part of the Universe
(\ref{E09}).

We are coming to a paradoxical conclusion that the global
properties of the inflationary Universe are determined not
by classical but by quantum effects.
 Let us try to understand the origin of such a behavior of
the inflationary Universe.

\subsection{\label{SelfRepr} Self-Reproducing Universe}

A very unusual feature of the inflationary Universe is that
the processes separated by distances $l$ greater than $H^{-1}$
proceed independently of one another. This is so because
during exponential expansion the distance between any two objects
separated by
more than $H^{-1}$ is growing with a
speed $v$ exceeding the speed of light.
As a result, an observer in the inflationary
Universe can see only the processes occurring inside the
horizon of the radius  $H^{-1}$.

An important consequence of this general result is that the
process of inflation in any spatial domain of radius
$H^{-1}$ occurs independently of any events outside it.
In this sense any inflationary domain of
initial radius exceeding $H^{-1}$ can be considered as a
separate mini-Universe, expanding independently of what
occurs outside it. This is the essence of the ``no-hair'' theorem for
de Sitter
space, which we already mentioned in Section 2.2.

To investigate the behavior of such a mini-Universe, with an
account taken of quantum fluctuations, let us consider an
inflationary domain of initial radius $H^{-1}$
containing sufficiently homogeneous field with initial
value $\phi \gg 1$ (assume $V(\phi) = \frac{m^2 \phi^2}{2}$ for
simplicity)\@.
Equation (\ref{E04}) tells us
that during a typical time interval $\Delta t=H^{-1}$ the field
inside this domain will be reduced by
\begin{equation}\label{E30}
\Delta\phi = \frac{1}{4\pi\phi}\ .
\end{equation}
By comparison of (\ref{E23}) and (\ref{E30}) one can easily see
that if $\phi$ is much less than $\phi^* \sim {1\over 2\sqrt{m}} $,
 then the decrease of the field $\phi$
due to its classical motion is much greater than the average
amplitude of the quantum fluctuations $\delta\phi$ generated
during the same time. But for greater $\phi$ (up to the
classical limit of about $10^7$), $\delta\phi (x)$ will exceed
$\Delta\phi$, i.e. the Brownian motion of the field $\phi$
will become more rapid than its classical motion. Because the
typical wavelength of the fluctuations $\delta\phi (x)$
generated during this time is $H^{-1}$, the whole domain
after $\Delta t$ effectively becomes divided
into $e^3$ separate domains (mini-Universes) of radius
$H^{-1}$, each containing almost homogeneous field $\phi -
\Delta\phi+\delta\phi$\@. We will call these domains ``$h$-regions''
\cite{ArVil,Nambu},
to  indicate   that each of them has the radius coinciding with the
radius of the event
horizon $H^{-1}$\@. In almost half of these domains (i.e. in $e^3/2
\sim 10$
$h$-regions) the field $\phi$ grows by $|\delta\phi(x)|-\Delta\phi
\approx
|\delta\phi (x)| = H/2\pi$, rather than decreases. During the next
time interval
$\Delta t = H^{-1}$ the field grows again in the half of the new
$h$-regions. Thus,
the total number of $h$-regions containing growing field $\phi$
becomes equal
to $(e^3/2)^2 = e^{2\,(3 - \ln 2)}$\@.  This means that until the
fluctuations of field
$\phi$ grow sufficiently large, the total physical volume occupied by
permanently growing field $\phi$ (i.e. the total number of
$h$-regions containing the growing field $\phi$)  increases with time
like
$\exp[(3 - \ln 2)\,Ht]$\@. The growth of the volume of the Universe
at
later stages
becomes even faster.

For example, let us consider those   $h$-regions where the field
grows
permanently, i.e. where the jumps of the field $\phi$ are always
positive,
$\delta\phi \sim   H/2\pi$.
Since this process occurs each time $H^{-1}$, the average
speed of    growth of the scalar field in such domains is given by
\begin{equation}
\label{19}
{d\phi\over d t} = {H^2(\phi)\over 2\pi} = {4 V(\phi)\over 3 } \ .
\end{equation}
In the theory of a massive noninteracting field with $V(\phi) =
m^2\phi^2/2$
the solution of this equation is
\begin{equation}\label{20}
\phi^{-1}(t) = \phi^{-1}_0 - {2m^2 t\over 3} \ ,
\end{equation}
where $\phi_0$ is the initial value of the field. Thus, within the
time
\begin{equation}\label{21}
t = {3\over2m^2\phi_0}
\end{equation}
the field $\phi$ in those domains becomes infinitely large.

Of course, this solution cannot be trusted when the field approaches
the
value $\phi_p \sim m^{-1} \gg 1$, where the effective potential
becomes of the
order of one (i.e. when the energy density becomes comparable with
the Planck
density)\@. In fact, we are going to argue in Section \ref{1/N} that
the growth of
the field $\phi$ stops when it approaches $\phi_p$\@. What is
important
though
is that within a finite time (\ref{21}) a finite part of the volume
of the
Universe approaches a state of the maximally high energy density.
After that
moment, a finite portion of the total volume of the Universe will
stay in a state
with the Planck density and will expand with the maximal possible
speed, $a(t)
\sim e^{Ct}$, where $C = O(1)$ (i.e. $C \sim M_p$)\@. All other
parts of the Universe will expand much  slower. Consequently, the
parts of
the Universe with $\phi \sim \phi_p$ will give the main contribution
to the
growth of the total volume of the Universe, and this volume will grow
as
fast as if  the whole Universe were almost at the Planck density.

Thus, if these considerations are correct, the main part of the
physical
volume of the Universe is the result of the expansion of
domains with nearly the maximal possible field value,
$\phi  \sim \phi_p$, for which $V(\phi)$ is close to the Planck
boundary $V
\sim 1$\@. There are also exponentially many domains with smaller
values of $\phi$\@. Those domains, in which $\phi$ eventually
becomes sufficiently small, give rise to the
mini-Universes of {\em our} type. In such domains, $\phi$
eventually rolls down to the minimum of $V(\phi)$, and
these mini-Universes are subsequently describable by the
usual Big Bang theory.
However a considerable part of the physical volume of the
entire Universe remains forever in the inflationary phase
\cite{b19}.

Let us return to the discussion of the properties of our stochastic
processes. Eq. (\ref{21}) suggests that it takes time $t \sim
{3\over2m^2\phi_0}$ until   the Planck density domains will dominate
the speed of the growth of the volume of the Universe. To be sure
that
this conclusion is correct, one should take into account that at the
beginning of the Planckian expansion of the domains where the field
$\phi$ was permanently growing, their initial volume could be
somewhat
smaller than the total volume of all other domains, where the field
$\phi$
was not permanently growing. Indeed, only in a small portion of the
original
space the scalar field jumps up permanently; typically it will jump
in both
directions. However, one can show that for $m \ll 1$
the corresponding corrections are small. One should check also that
this result is not modified by the possibility of larger quantum
jumps
of the scalar field. Indeed, the typical amplitude of the quantum
jump
within the time $H^{-1}$ is given by $\delta\phi \sim {H\over 2\pi}$.
However, there exists a nonvanishing probability that the field
$\phi$
within the same time will jump  much higher. Even though this
probability is exponentially suppressed, such domains will start
their
expansion with the Planckian speed earlier, and their relative
contribution to the volume of the Universe could become significant.

Assuming that this probability of a jump $\phi \rightarrow \phi +
\delta\phi$ is given by the Gaussian distribution with variance
${H\over 2\pi}$, one obtains
\begin{equation}\label{22} P(\delta\phi) =
\exp\Bigl(-{2\pi^2 \delta\phi^2\over H^2}\Bigr)\ .
\end{equation}
For example, the probability of a jump from $\phi_0 \ll \phi$ to
$\phi$
can be estimated by  \begin{equation}\label{23}
P(\phi) = \exp\Bigl(-{2\pi^2 \phi^2\over H(\phi_0)^2}\Bigr) \ .
\end{equation}
For the theory of a massive scalar field this gives the following
probability of a jump to $\phi = \phi_p$:
 \begin{equation}\label{24}
P(\phi_p) \sim \exp\Bigl(-{3\pi \over \phi_0^2\,m^4}\Bigr) \ .
\end{equation}
Even though the volume of this part of the Universe grows with the
Planckian  speed, $a \sim e^{3ct}$, $c = O(1)$, it takes longer than
$t
\sim {3\pi \over \phi_0^2\,m^4}$ for this part of the Universe to
grow
to the same size as the  rest of the Universe.  This time is much
larger
than the time $t = {3\over2m^2\phi_0}$ which we obtained neglecting
the
possibility of large jumps.

In fact, our estimate of the probability of large jumps was even too
optimistic. Indeed, if the jump is very high, then  the gradient
energy
of the corresponding quantum fluctuation on a scale $H^{-1}$ is given
by
${\phi^2\over H(\phi_0)^2}$. This quantity may become greater than
the
potential energy density of the scalar field $V(\phi)$. In such a
case the
domain where the jump occurs does not inflate. To overcome this
difficulty
one should consider jumps on a larger length scale, which further
reduces their
probability to
 \begin{equation}\label{24aa}
P(\phi) \sim \exp \Bigl(- c {\pi^2 \phi^4\over V(\phi)}\Bigr) \ ,
\end{equation}
where $c = O(1)$ \cite{Creation}. One could conclude, therefore,
that large
quantum jumps are really unimportant, and the main features of the
process can
be understood neglecting the possibility  of jumps greater than
${H\over
2\pi}$.

However, one should keep in mind that eq. (\ref{19}) is valid only
for
$\phi \gg \phi^*$. For  $\phi < \phi^*$, the mean decrease of the
field $\phi$
during the typical time $H^{-1}$ is always more significant than the
typical amplitude of the jump ${H\over 2\pi}$. Therefore, any process
of self-reproduction of the Universe with a small initial value of
the
field $\phi$ is possible only with an account taken of large quantum
jumps. The probability of such jumps is exponentially suppressed.
Indeed, by comparing eqs. (\ref{23}) and (\ref{E09}) one can see that
the strong suppression of large jumps cannot be compensated even by
the
exponentially large inflation after such a jump. Therefore in a
typical
$h$-region with $\phi < \phi^*$ the process of the Universe
self-reproduction does not occur and inflation eventually ceases to
exist. However, importance of large jumps should not be overlooked:
{\it one}\, such jump may be enough to start an infinite process of
the
Universe self-reproduction. This process may lead to spontaneous
formation of an inflationary Universe from the ordinary Minkowski
vacuum and to a possibility of the Universe creation in a laboratory
\cite{Creation}.  We will return to this question in Section 7.2.

Note that the process of the self-reproduction of the Universe may
occur
not only in the chaotic inflation scenario, but in the old \cite{b51}
and the
new inflationary scenarios \cite{b52} as well. In the corresponding
models some
part of the  volume of the Universe can always remain in a state
corresponding to a local extremum of $V(\phi)$ at $\phi = 0$\@. In
the
chaotic inflation scenario in the theories $\phi^n$ or
$e^{\alpha\phi}$, the
results are  more surprising. Not only can
the Universe stay permanently on the top of a hill as in the old and
new
inflationary  scenario; it can also climb perpetually up the wall
towards the
largest possible values of its potential energy density \cite{b19}.

\section{Stochastic Approach to Inflation and the Wave Function
of the Universe \label{Wave+Stoch}}

\subsection{Euclidean Approach \label{Euclid}}

Now we would like to compare our methods with other approaches to
quantum cosmology. One of the most ambitious approaches to cosmology
is based on the investigation of the Wheeler-DeWitt equation for the
wave function $\Psi$ of the Universe \cite{DeWitt}. However, this
equation has many different solutions, and {\it a priori}\, it
is not quite clear which one of these solutions describes our
Universe.

A very interesting idea was suggested by Hartle and Hawking
\cite{HH}. According to their work, the wave function of the
ground state of the Universe with a scale factor $a$ filled
with a scalar field $\phi$ in the semi-classical approximation
is given by
\begin{equation}\label{E31}
\Psi_0(a,\phi)\sim \exp\left(-S_E(a,\phi)\right)\ .
\end{equation}
Here $S_E(a,\phi)$ is the Euclidean action corresponding to
the Euclidean solutions of the Lagrange equation for
$a(\tau)$ and $\phi(\tau)$ with the boundary conditions
$a(0)=a, \phi(0)=\phi$.
The reason for choosing this particular solution of the
Wheeler-DeWitt equation was explained as follows \cite{HH}. Let us
consider the Green's function of a particle which moves from
the point $(0,t')$ to the point ${\bf x},t$:
\begin{eqnarray}\label{E32}
<{\bf x},0|0, t'>
 = \sum_n \Psi_n ({\bf x})\Psi_n(0)
\exp\left(iE_{n}(t-t')\right)\nonumber\\
 = \int d{\bf x}(t) \exp\left(iS({\bf x}(t))\right)\ ,
\end{eqnarray}
where $\Psi_n$ is a complete set of energy eigenstates
corresponding to the energies $E_n\geq 0$.

To obtain an expression for the ground-state wave function
$\Psi_0({\bf x})$, one should make a rotation
           $t \rightarrow -i\tau$ and take the limit as
$\tau \rightarrow -\infty$\@. In the summation (\ref{E32})
only the term $n=0$ with the lowest eigenvalue $E_0 = 0$
survives, and the
integral transforms into $\int dx(\tau)\exp(-S_E(\tau))$.
Hartle and Hawking have argued that the generalization of
this result to the case of interest in the semiclassical
approximation would yield (\ref{E31}).

The gravitational action corresponding to the Euclidean
section $S_4$ of de Sitter space $dS_4$ with $a(\tau) =
H^{-1}(\phi)\cos H\tau$ is negative,
\begin{equation}\label{E33}
S_E(a, \phi) = - \frac{1}{2}
    \int d\eta\Bigl[\Bigl(\frac{da}{d\eta}\Bigr)^2 - a^2 +
\frac{\Lambda}{3}a^4\Bigr]
    \frac{3\pi}{2} = - \frac{3}{16 V(\phi)}\ .
\end{equation}

Here $\eta$ is the conformal time, $\eta = \int {dt\over a(t)}$,
$\Lambda = 8\pi V$\@. Therefore, according to \cite{HH},
\begin{equation}\label{E34}
\Psi_0(a,\phi)\sim \exp{\Bigl(-S_E(a,\phi)\Bigr)} \sim
\exp\left(\frac{3
}{16V(\phi)}\right) \ .
\end{equation}

This means that the probability $P$ of finding the Universe
in the state with $\phi = const$, $a=H^{-1}(\phi) =
\sqrt{3\over 8\pi V(\phi)}$ is   given by
\begin{equation}\label{E35}
P(\phi) \sim |\Psi_0|^2\sim \exp\left(\frac{3 }{8V(\phi)}\right) \ .
\end{equation}

This expression has a very sharp maximum as $V(\phi) \rightarrow
0$\@. Therefore the probability of finding the Universe in a
state with a large field $\phi$ and having a long stage of
inflation becomes strongly diminished. Some authors consider this as
a strong
argument against the Hartle-Hawking wave function. However, nothing
in the
`derivation' of this wave function tells that it describes initial
conditions
for inflation; in fact, in the only case where it was possible to
obtain eq.
(\ref{E34}) by an alternative method, the interpretation of this
result was
quite different, see Section 3.2.

There exists an alternative choice of the wave function of
the Universe. It can be argued that the analogy between the
standard theory (\ref{E32}) and the gravitational theory (\ref{E33})
is incomplete. Indeed, there is an overall minus sign in the
expression for $S_E(a, \phi)$ (\ref{E33}), which indicates that the
gravitational energy associated with the scale factor $a$ is
negative. (This is related to the well-known fact that the
total energy of a closed Universe is zero, being a sum of
the positive energy of matter and the negative energy of the
scale factor $a$.) In such a case, to suppress terms with $E_n < E_0$
and to
obtain $\Psi_0$ from
(\ref{E33}) one should rotate $t$ not to $-i\tau$, but to
$+i\tau$. This gives  \cite{b56}
\begin{equation}\label{E35a}
\Psi_0(a,\phi) \sim \exp\Bigl(-|S_E(a,\phi)|\Bigr) \sim\exp
\left(- \frac{3}{16V(\phi)}\right)\ ,
\end{equation}
and
\begin{equation}\label{E36}
P(\phi) \sim|\Psi_0(a,\phi)|^2\sim \exp\Bigl(-2|S_E(a,\phi)|\Bigr)
\sim\exp
\left(- \frac{3}{8V(\phi)}\right) \ .
\end{equation}

An obvious objection against this result is that it may be incorrect
 to use  different  ways of rotating $t$ for quantization of the
scale
factor and of the  scalar field. However, the idea that quantization
of
matter coupled to gravity can be accomplished just by a proper choice
of a complex contour of integration, though very appealing
\cite{b202}, may be too optimistic.  We know, for example,  that
despite
many attempts to suggest  Euclidean formulation (or just {\it any}
simple set of Feynman rules) for nonequilibrium quantum
statistics
or for the field theory in a nonstationary background,   such
formulation does not exist yet.
It is quite clear from (\ref{E32}) that the $t \rightarrow -i\tau$
trick
would not work if the spectrum $E_n$ were not bounded from below.
Absence of equilibrium, of any simple
stationary ground state seems to be  a typical situation in quantum
cosmology. In
some cases where a stationary or quasistationary ground state does
exist, eq. (\ref{E35}) may be correct, see Section 3.2.
In a more general situation
it may be very difficult to obtain any simple expression for the wave
function of the Universe. However, in certain limiting cases this
problem is relatively simple. For example, at present the scale
factor
$a$ is very big and it changes very slowly, so one can consider it
to be a C-number and quantize matter fields only by rotating $t
\rightarrow
-i\tau$\@. On the other hand, in the
inflationary Universe the evolution of the scalar field is  very
slow;
during the typical time intervals $O(H^{-1})$ it behaves essentially
as
a  classical field, so one can describe the process of the creation
of
an inflationary Universe  filled with a homogeneous scalar field by
quantizing
the
scale factor $a$ only and by rotating $t \rightarrow i\tau$. Eq.
(\ref{E36}),
which was first obtained  in \cite{b56} by the method described
above, later
was obtained also by another method  by Zeldovich and Starobinsky
\cite{b57},
 Rubakov \cite{b58}, and Vilenkin \cite{b59}. This result can be
interpreted as the probability of quantum tunneling of the
Universe from $a=0$ (from ``nothing'') to $a=H^{-1}(\phi)$\@.
Therefore the
wave function (\ref{E35a}) is called `tunneling wave function'. In
complete agreement with our previous argument, eq.  (\ref{E36})
predicts that a
typical initial value of the field $\phi$ is given by
$V(\phi)\sim 1$ (if one does not speculate about the
possibility that $V(\phi) \gg 1)$, which leads to a very
long stage of inflation.

Unfortunately, there is no rigorous derivation of either
 (\ref{E35}) or (\ref{E36}), and the physical meaning of creation
of everything from `nothing' is far from being clear.
However, one may give a simple physical argument explaining that
(\ref{E36})
has a better chance to describe the probability of  quantum creation
of the
Universe than  (\ref{E35}) \cite{b56}.

    Examine a closed de Sitter space with energy density
$V(\phi)$.  Its
size
behaves as $H^{-1}\, \cosh{Ht}$. Thus, its  minimal volume at the
epoch of
maximum contraction  is of order $H^{-3} \sim V^{-3/2}$, and the
total energy
of the scalar field contained in de Sitter space at that instant is
approximately  $V\times V^{-3/2}\sim V^{-1/2}$.  Thus, to create the
Universe
with the Planckian energy density $V(\phi) \sim 1$ one needs only a
Planckian
fluctuation of energy $\Delta E \sim 1$ on the Planck scale $H^{-1}
\sim 1$,
whereas to create the Universe with $V(\phi) \ll 1$ one needs a very
large
fluctuation of energy  $\Delta E \sim V^{-1/2}$ on a large scale
$H^{-1} \sim
V^{-1/2}$. This means that, in accordance with (\ref{E36}), the
probability of
quantum creation of the Universe with $V(\phi) \ll 1$ should be
strongly
suppressed, whereas  no such suppression is expected  for the
probability of
creation of an inflationary Universe with $V(\phi) \sim 1$.

This argument suggests that, in agreement with our discussion of
initial
conditions in Section 2.2,  inflation appears in a  much more natural
way in
the theories where it is possible for $V(\phi) \sim 1$ (e.g. in the
theories
with the effective potentials $\phi^n$ or $e^{\alpha\phi}$), rather
than in the
theories where it may occur only at $V(\phi) \ll1$ (theories of this
class
include `natural inflation' \cite{Natural}, `hyperextended inflation'
\cite{Hext}, etc.).

A deeper understanding of the physical processes in the
inflationary Universe is necessary in order to investigate
the wave function of the Universe $\Psi_0(a,\phi)$, to
suggest a correct interpretation of this wave function and to
understand a possible
relation of this wave function to our results concerning the process
of
self-reproduction of inflationary Universe.  With this purpose we
will try to
investigate the global structure of the inflationary Universe, and go
beyond the
minisuperspace approach used in the derivation of (\ref{E31}) and
(\ref{E36}).
This can be done with the help of a stochastic approach to inflation
\cite{b60,b20}, which is a more formal way to investigate the
Brownian motion of
the scalar field $\phi$. As we will see, this approach may somewhat
modify our
conclusions concerning the possibility to realize various models of
inflation.

\subsection{\label{Stochastic} Stochastic Approach to Inflation}

The evolution of the fluctuating field $\phi$ in any given domain
can be described with the help of its distribution function
$P(\phi)$, or in terms of its average value $\bar\phi$ in this
domain and its variance
$\Delta = \sqrt{\left\langle \delta \phi^2\right\rangle}$\@.
However, one will obtain different results
depending on the method of averaging. One can consider the
probability distribution $P_c(\phi,t)$ over the non-growing
coordinate
volume of the domain (i.e. over its physical volume at some
initial moment of inflation)\@. This is equivalent to the probability
to
find a given field at a given time at a given {\it
point}.\footnote{Note,
that by the field $\phi$ we understand here its
classical long-wavelength component, with the wavelength $l >
H^{-1}$.
Thus,
this field remains (almost) constant over each $h$-region.}
Alternatively, one may consider the distribution $P_p(\phi,t)$ over
its
{\em physical} (proper) volume, which grows exponentially at a
different rate in different parts of the domain. In many interesting
cases
the variance $\Delta_c$ of the field $\phi$ in the coordinate volume
remains much smaller than $\bar\phi_c$.
In such cases the evolution of the averaged field $\bar\phi_c$, can
be
described approximately by  (\ref{E02})-(\ref{E05})\@. However, if
one
wishes to know the resulting spacetime structure and the distribution
of the field $\phi$ after (or during) inflation, it is more
appropriate
to take an average $\bar\phi_p$ over the physical volume, and in some
cases the behavior of $\bar\phi_p$ and $\Delta_p$ differs
considerably
from the behavior of $\bar\phi_c$ and $\Delta_c$.

We will begin with the description of the distribution $P_c(\phi,t)$.
This is a relatively simple problem. As we learned already, the field
$\phi$ at a given point behaves like a Brownian particle. The
standard
way
to describe the Brownian motion is to use the diffusion equation,
where
instead of the position of the Brownian particle $x$ one should write
$\phi$: \begin{equation}\label{E37a}
\frac{\partial P_c}{\partial t} =
\frac{\partial}{\partial \phi}  \left({\cal
D}^{1-\beta}\frac{\partial\,({\cal
D}^\beta P_c)}{\partial \phi} + \kappa \frac{d V}{d \phi}
{P_c}\right) \ .
\end{equation}
Here ${\cal D}$ is the coefficient of diffusion, $\kappa$ is the
mobility
coefficient, $- \frac{dV}{d\phi}$ is the analog of an
external
force $F$\@. The parameter $\beta$ reflects some ambiguity which
appears
when one
derives this equation in the theories with ${\cal D}$ depending on
$\phi$\@. The
value $\beta = 1$ corresponds to the Ito version of stochastic
approach, whereas
$\beta = 1/2$ corresponds to the version suggested by Stratonovich.

The standard way of deriving this equation is based on the  Langevin
equation,
which in our case can be written in the following way \cite{b60,b61}:
 \begin{equation} \label{m1}
\frac{d}{dt} \, \phi = - \frac{V'(\phi)}{3H(\phi)} +
\frac{H^{3/2}(\phi)}{2\pi}
\, \xi(t) \ .
\end{equation}
Here $\xi(t)$ is some function representing  the effective white
noise generated
by quantum  fluctuations, which leads to the Brownian motion of the
classical
field $\phi$\@. The derivation of this equation, finding the function
$\xi(t)$ and
the subsequent  derivation of the diffusion equation is rather
complicated
\cite{b60,b61}. Meanwhile, the main purpose of the derivation is to
establish the
functional form of ${\cal D}$ and $\kappa$\@. This can be done in a
very simple
and intuitive way \cite{MyBook}.

Indeed, let us study two limiting cases. The first case is a
classical motion without any quantum corrections. In this case the
field $\phi$ during inflation satisfies equation (\ref{m1}) without
the
last term; see eq.  (\ref{E02}) and discussion after it.
Comparing this equation and the standard definition of the
mobility
coefficient ($\dot x = \kappa F$) one easily establishes that in our
case $\kappa = {1\over 3 H(\phi)}$.

Determination of the diffusion coefficient is slightly more
complicated but still quite elementary. This coefficient describes
the
process of adding (and freezing) new long-wavelength perturbations.
As we
mentioned already, the speed of this process does not depend on $m$
at
$m^2 \ll H^2$\@. Therefore we may get a correct expression for $\cal
D$
by
investigating the theory with a flat effective potential,  where the
last term in eq. (\ref{E37a}) disappears. According to eq.
(\ref{E23a0}), in such a theory
\begin{equation} \label{m1a}
\frac{d}{dt} \, \left<\phi^2\right> = {H^3\over 4\pi^2}\  .
\end{equation}
On the other hand, eq. (\ref{E37a}) without the last term gives
\begin{eqnarray}\label{E37b}
\frac{d}{dt} \, \left<\phi^2\right> &=& \int {d\phi\ \phi^2\
\frac{\partial
P_c}{\partial t}} = \int {d\phi\ \phi^2\ \frac{\partial}{\partial
\phi }
\left(\frac{\partial\,({\cal D}P_c)}{\partial \phi }\right)}
 = 2\ \int {d\phi\   {\cal D}P_c}\\ \nonumber &=&  2\ {\cal D} \int
{d\phi\
P_c} =2\  {\cal D}\ .
\end{eqnarray}
When going from the first to the second line of this equation, we
used the fact that  ${\cal
D} = const$ in  the theory with the flat potential.
 From  equations  (\ref{m1a}) and (\ref{E37b})
it follows that  ${\cal D} = {H^3\over 8\pi^2} = {2\sqrt 2 V^{3/2}
\over 3\sqrt {3\pi}}$\@. This gives us the  diffusion
equation
\begin{equation}\label{E3711}
\frac{\partial P_c}{\partial t} =
\frac{\partial}{\partial \phi} \left({H^{3(1-\beta)}(\phi)\over
8\pi^2} \
\frac{\partial}{\partial \phi}\, \left(H^{3\beta}(\phi) P_c \right) +
{V'(\phi)\over 3H(\phi)} \, P_c\right) \ ,
\end{equation}
or, in an expanded form,
\begin{equation}\label{E37111}
\frac{\partial P_c(\phi,t)}{\partial
t} = {2 \sqrt 2\over 3\sqrt {3\pi}}\ \frac{\partial}{\partial \phi}
\left(V^{3(1-\beta)/2}(\phi) \
\frac{\partial}{\partial
\phi}\left({V^{3\beta/2}(\phi)}P_c(\phi,t)\right) +
{3V'(\phi)\over 8V^{1/2}(\phi)} \, {P_c(\phi,t)}\right)
\ .
 \end{equation}

 This equation for
the case $H(\phi) = {\rm const}$ was first derived by Starobinsky
\cite{b60}; for a more detailed derivation see \cite{b61},
\cite{Star}. For the
special case $\frac{d V}{d \phi} = 0$
this equation was obtained by Vilenkin \cite{b62}.
This equation can be represented in the following useful form, which
reveals
its physical meaning: \begin{equation}\label{E37111a}
\frac{\partial P_c}{\partial
t} + \frac{\partial J_c}{\partial
\phi} = 0\ ,
 \end{equation}
where the probability current is given by
\begin{equation}\label{E37111b}
J_c = - {2 \sqrt 2\over 3\sqrt {3\pi}}\
\left(V^{3(1-\beta)/2}(\phi) \
\frac{\partial}{\partial \phi}\left({V^{3\beta/2}(\phi)}P_c\right) +
{3V'(\phi)\over 8V^{1/2}(\phi)} \, {P_c}\right) \ .
 \end{equation}
Equation (\ref{E37111a}) can be interpreted as a continuity equation,
which
follows from the   conservation of  probability. We will use this
representation in our discussion of the boundary conditions for the
diffusion
equations.

We still did not specify the value of the parameter $\beta$ in this
equation. The
uncertainty in $\beta$ is related to the precise definition of the
white noise
term $\frac{H^{3/2}(\phi)}{2\pi} \, \xi(t) $ in the Langevin
equation. At $H =
const$ (de Sitter space) there is no ambiguity in definition of the
white noise
$\xi(t)$\@. However, if one wishes to take into account the
dependence of
$H$ on
$\phi(t)$, one may include some part of this dependence into the
definition of
white noise. Alternatively (and equivalently), one may use several
different
ways to obtain the diffusion equation from the Langevin equation,
which also
leads to the same uncertainty. (A similar
ambiguity is
known to appear at the level of operator ordering in the
Wheeler-DeWitt
equation.) In what follows we will use the diffusion equation in the
Stratonovich
form ($\beta = 1/2$),
\begin{equation}\label{E37}    \frac{\partial
P_c(\phi,t)}{\partial t} = {2 \sqrt 2\over 3\sqrt {3\pi}}\
\frac{\partial}{\partial \phi}
\left(V^{3/4}(\phi) \
\frac{\partial}{\partial \phi}\Bigl({V^{3/4}(\phi)}P_c(\phi,t)\Bigr)
+
{3V'(\phi)\over 8V^{1/2}(\phi)} \, {P_c(\phi,t)}\right) \ .
 \end{equation}
but one should keep in mind that most of the results which we will
obtain are
not very sensitive to a particular choice of $\beta$\@. For example,
a simplest
stationary solution ($\partial P_c/\partial t=0$) of equation
(\ref{E37111})
would be \cite{b60}
\begin{equation}\label{E38} P_c(\phi)\sim V^{-3\beta/2}(\phi)\cdot
\exp\left({3\over 8 V(\phi)}\right) \ ,
\end{equation}
The whole dependence on $\beta$ here is concentrated
in the subexponential factor $V^{-3\beta/2}(\phi)$.

In fact, one can make another step and derive an equation describing
the
probability distribution $P(\phi,t|\chi)$ that the value of the field
$\phi(t)$
initially (i.e. at $t=0$) was equal to  $\chi$. The derivation of a
similar
equation (which is called `backward Kolmogorov equation') will be
contained in
Section \ref{Stationary}. Now we are just presenting the result:
\begin{equation} \label{b60b} \frac{\partial
P_c(\phi,t|\chi)}{\partial t} =
{2 \sqrt 2\over 3\sqrt {3\pi}}\left( V^{3/4}(\chi) \frac{\partial
}{\partial\chi}
 \left( V^{3/4}(\chi) \frac{\partial  P_c(\phi,t|\chi)
}{\partial\chi}\right)
 - \frac{3V'(\chi)}{8V^{1/2}(\chi)} \frac{\partial
P_c(\phi,t|\chi)}{\partial\chi} \right)  \ .
\end{equation}
Note that in this equation one considers $\phi$ as a constant,
and finds the time dependence of the probability that this value of
the scalar
field $\phi$ was produced by the diffusion from its initial value
$\chi$ during
the time $t$.  The simplest stationary solution of both eq.
(\ref{E37}) and eq.
(\ref{b60b}) (subexponential factors being omitted) would be
\begin{equation}\label{E38a}
P_c(\phi,t|\chi) \sim  \exp\left({3\over 8
V(\phi)}\right)\cdot \exp\left(-{3\over 8 V(\chi)}\right) \ .
\end{equation}

This function is extremely interesting. Indeed, the first term  in
(\ref{E38a}) is equal to the square of the Hartle-Hawking wave
function of the
Universe (\ref{E35}), whereas the second one gives the square of the
tunneling
wave function  (\ref{E36})\hskip .1cm !

  At  first glance, this result gives a direct confirmation and a
simple
physical interpretation of  both the Hartle-Hawking
 wave function of the Universe {\it and} the tunneling wave function.
However, in all realistic cosmological theories, in which $V(\phi)=0$
at its minimum, the distributions (\ref{E38}),  (\ref{E38a}) are not
normalizable. The source of this difficulty can be easily
understood: any stationary distribution may exist only due
to a compensation of a classical flow of the field $\phi$
downwards to the minimum of $V(\phi)$ by the diffusion motion
upwards. However, diffusion of the field $\phi$ discussed
above exists only during inflation, i.e. only for $\phi \geq
1$, $V(\phi)\geq V(1)\sim m^2\sim 10^{-12}$ for $m \sim 10^{-6}$.
Therefore (\ref{E38a}) would correctly describe the stationary
distribution $P_c(\phi,t|\chi)$ in the inflationary Universe only if
$V(\phi)\geq 10^{-12} \sim 10^{80}$ g$\cdot $cm$^{-3}$ in the
absolute minimum of
$V(\phi)$, which is, of course, absolutely unrealistic \cite{b20}.

Of course, (\ref{E38a}) is not the most general stationary solution
for $P_c$.
For example, eq. (\ref{E37111}) in the case $\beta = 0$, $V(\phi) =
{m^2\phi^2\over 2}$ has a simple constant solution
\begin{equation}\label{E38b}
P_c(\phi,t|\chi) =  -J_c\, {2\sqrt{3\pi}\over m} \ .
\end{equation}
Similar solutions were described in \cite{Star}; for a very recent
discussion
see also \cite{Spok}. However, an interpretation of such solutions is
obscure,
see
e.g. \cite{b19,MyBook}. Indeed, eq. (\ref{E38b}) is a solution of eq.
(\ref{E37111}) even in the absence of any diffusion; one may simply
ignore the
first term in eq. (\ref{E37111}). It may exist only if from the very
beginning
there were {\it infinitely } many $h$-regions with {\it all} possible
values of
the field $\phi$ equally distributed everywhere from $0$ to $\infty$.
In this
case the motion of the whole distribution  (\ref{E38b}) towards small
$\phi$
due to the classical rolling of the field $\phi$ does not change the
distribution.
But this presumes that the Universe from the very beginning was
infinitely
large, that our probability distribution was fine-tuned, and that our
diffusion
equations are valid at infinitely large energy density. There was an
attempt to
interpret these solutions as representing creation of new domains of
the
Universe from nothing. Such an interpretation could be possible for
the
distribution $P_p$, which takes into account increase of the volume
of the
Universe. Meanwhile, the derivation of the diffusion equation for
$P_c$
presumes that there is no creation of  new space in comoving
coordinates.  For
all these reasons (see also below) we will
not discuss  `stationary solutions' of the type of (\ref{E38b}) in
our paper.

To get a better understanding of the situation, we will  study a
general
non-stationary solution of (\ref{E37}) describing the time-evolution
of the
initial distribution $P_c(\phi,t=0)=\delta(\phi-\phi_0)$\@. This is a
rather
general form of the initial distribution $P_c(\phi,t)$ in a domain of
initial
size
of the order of
$H^{-1}$\@. Indeed, the typical deviation from homogeneity in such
domains
is given by the amplitude of quantum perturbations on this scale,
$\delta
\phi \sim {H\over 2 \pi}$\@. One can easily check that for $m \ll 1$
this
amplitude is always much smaller than $\phi$ during inflation.
This explains why the delta-functional initial conditions are
relevant to our
problem.

 The  solution of the diffusion equation
for the theory $\frac{m^2}{2}\phi^2$ with these initial conditions is
\cite{b20}:
\begin{equation}\label{E39}
P_c(\phi, t) =  \exp\left(-\frac{3\, (\phi-\phi(t))^2}
{2m^2\, (\phi^4_0-\phi^4(t))}\right) \ ,
\end{equation}
where $\phi(t)$ is the slow-rollover solution of the classical
equations (\ref{E02}), (\ref{E03}).
This equation shows that at the
first stage of the process, during the time $\Delta t \leq
\frac{\phi_{0}2\sqrt{3\pi}}{m}$ (see eq. (\ref{E04})), the maximum of
the
distribution $P_c(\phi, t)$ almost does not move, whereas the
variance grows
linearly. Then, at $t \gg \Delta t$, the maximum of $P_c(\phi, t)$
moves to $\phi = 0$, just as the classical field $\phi(t)$
(\ref{E04}).
This shows that in the realistic situations with reasonable initial
conditions there are no nontrivial stationary solutions of the
diffusion equation for $P_c$: the field $\phi$ just moves towards the
absolute minimum of the effective potential and stays there
\cite{b19}.

For completeness we shall mention here another  solution of
 (\ref{E37})\@. If the initial value of the field $\phi$ is very
large, $\phi_0 \ga m^{-1}$, i.e. if one starts with the
spacetime foam with $V(\phi_0)\geq 1$, then the evolution
of the field $\phi$ in the first stage (rapid diffusion)
becomes more complicated (the naively estimated variance
$\Delta^2_c\sim H^3 t$ soon becomes greater than $\phi^2_0$)\@. In
this case the distribution of the field $\phi$ is not
Gaussian. The solution of eq. (\ref{E37}) at the stage of
diffusion from $\phi_0$ to some field $\phi$ with $V(\phi)
\ll 1$ is given by
\begin{equation}\label{E41}
P_c(\phi) \sim \exp\left(-\frac{3\sqrt{3}\pi }{m^3\phi \, t}\right) \
{}.
\end{equation}
This solution describes quantum creation of domains of a
size $l \geq H^{-1}(\phi)$, which occurs due to the diffusion
of the field $\phi$ from $\phi_0\ga m^{-1}$ to $\phi \ll
\phi_0$\@. Direct diffusion with formation of a domain filled
with the field $\phi$ is possible only during the time
$ 2c \sqrt{3}\pi \phi/m$, $c =O(1)$\@. At larger times a
more rapid process is diffusion to some field $\tilde{\phi}
> \phi$ and a subsequent classical rolling down from
$\tilde{\phi}$ to $\phi$\@. Therefore one may interpret a
distribution $P_c(\phi)$ formed after a time $\Delta t= (2c\sqrt
{3}\pi\phi/m $) as the probability of quantum creation of a
mini-Universe filled with a field $\phi$ \cite{b20},
\begin{equation}\label{E42}
P_c(\phi) \sim \exp\left(- \frac{3 c}{2m^2\phi^2}\right)
\sim \exp\left(- \frac{3c}{4V(\phi)}\right) \ ,
\end{equation}
which is in agreement with the previous estimate for the
probability of quantum creation of the Universe, eq. (\ref{E36})
\cite{b56}--\cite{b59}. However, this result is not a true
justification of our
expression for the probability of the quantum creation of the
Universe. First
of all, we did not actually determine the constant $c$\@. Moreover,
the
careful analysis shows that for the effective potentials steeper than
$\lambda \phi^4$ the simple expression (\ref{E42}) should be
modified.
For
example, for $V \sim \phi^n$, $n > 4$, an improved result is
\cite{MyBook}
\begin{equation}\label{E42impr}
P_c(\phi) \sim  \exp\left(- \frac{3c}{4V(\phi)}
\Bigl({\phi_0\over\phi}\Bigr)^{(n-4)/2}\right) \ .
 \end{equation}

Now let us try  to write  an
equation of the type of (\ref{E37}) for $P_p$\@. Let us first forget
about the
normalization of the distribution $P_p$\@. Then the distribution
$P_p(\phi,t)$
has a meaning of a total number of ``points'' with a given $\phi$,
provided
that the number of points not only changes due to diffusion, but also
grows proportionally to the increase of volume. This means that
during a small
time interval $dt$ the total number of   points with the field $\phi$
additionally increases by
 the factor $3H(\phi)\,dt$\@. (For  $\phi = const$ this would lead to
an
exponential
growth
of the number of points corresponding to the expansion of the volume
$\sim
e^{3Ht}$)\@. This leads to the following equation for the
(unnormalized)
distribution $P_p$:
\begin{equation}\label{E371}
\frac{\partial P_p}{\partial t} =
\frac{\partial}{\partial \phi} \left({\cal D}^{1/2}\
\frac{\partial\,({\cal D}^{1/2}P_p)}{\partial \phi}
+ \kappa V'(\phi)\,{P_p}\right)  + 3H(\phi)\,P_p \ ,
\end{equation}
or, in an expanded form,
\begin{eqnarray}\label{E372}
& &\frac{\partial P_p(\phi,t)}{\partial t} = \frac{\partial
}{\partial\phi}
 \left( {H^{3/2}(\phi)\over 8\pi^2} \frac{\partial }{\partial\phi}
\left(
 {H^{3/2}(\phi)}P_p(\phi,t) \right)
 +  \frac{V'(\phi)}{3H(\phi)} \, P_p(\phi,t) \right)
 +  3H(\phi)   P_p(\phi,t) ~~~~  ~~~~  ~~~~  ~ \nonumber \\
&=& {2 \sqrt 2\over 3\sqrt
{3\pi}}\left( \frac{\partial}{\partial \phi} \Bigl(V^{3/4}(\phi) \
\frac{\partial\,({V^{3/4}(\phi)}P_p(\phi,t))}{\partial \phi} +
\frac{3V'(\phi)}{8V^{1/2}(\phi)} \, P_p(\phi,t) \Bigr)\ + 9\pi
V^{1/2}(\phi)\ P_p(\phi,t) \right) .
\end{eqnarray}
This simple derivation was proposed by Zeldovich and one of the
present
authors
immediately after the discovery of the self-reproduction of the
Universe in the
chaotic inflation scenario \cite{ZelLin}. An alternative (and more
rigorous)
derivation was given in \cite{Nambu,SNN}\@. In  Section
\ref{Branching}
of this paper we will give another derivation of this equation and
its generalizations, using methods of the theory of  branching
diffusion processes.

In the first papers on the self-reproduction of the Universe this
equation was
not used. Instead of that, it was found useful to study
solutions of equation (\ref{E37}) and then make some simple estimates
which gave a qualitatively correct description of the behavior of
$P_p$
\cite{b19,b20}. For example, one can use eq. (\ref{E39}) and take
into account
that  during the first period of time  $\Delta t \leq
\frac{\phi_{0}2\sqrt{3\pi}}{m}$, when the field almost does not move,
the volume
of domains filled by the field $\phi$ increases approximately by
$\exp\bigl(3H(\phi)\Delta t\bigr) \sim
   \exp \bigl( c_1\phi_0 \phi \bigr)$ where $c_1 =
O(1)$\@. After this time the distribution $P_p(\phi,\Delta t)$
is (approximately)  given by
\begin{equation}\label{E40}
P_p (\phi,\Delta t)\sim P_c(\phi,\Delta t)\cdot
\exp\bigl(3H(\phi)\Delta
t\bigr)
\sim \exp\left(-\frac{\phi^2}{c_2m^2\phi^4_0} +
c_1 \phi_0\phi\right) \ ,
\end{equation}
where $c_2 = O(1)$\@. One can easily verify that, if
$\phi_0 \ga \phi^* \sim {1\over 2\sqrt m}$, then the maximum of
$P_p(\phi, t)$ during the time $\tau$ becomes shifted to some field
$\phi$ which is bigger than $\phi_0$\@. This just corresponds to
the process of eternal self-reproduction of the inflationary
Universe studied in Section \ref{SelfRepr}.

Eq. (\ref{E40}) suggests that the total volume of the part
of the Universe
which experiences one jump to $\phi$ from $\phi_0$, grows more than \
$\exp\left(-\frac{\phi^2}{c_2m^2\phi^4_0} + c_1 \phi_0\phi\right)$
times during the subsequent classical rolling of the field $\phi$
back to
$\phi_0$. But this factor is much larger than $1$ for $\phi >
\phi^*$\@.
Even if such domains
later do not jump up but move down according to the classical
equations of
motion, their total volume at the moment when the energy density
inside them
becomes equal to $\rho_0 \sim 10^{-29} g\cdot
cm^{-3}$ is \ $\exp\left(-\frac{\phi^2}{c_2m^2\phi^4_0} +  c_1
\phi_0\phi\right)$  times larger than the volume of those (typical)
domains
which did not experience any jumps up at all. And the volume of those
domains which experienced two, three or more jumps up will be even
larger!

This means that almost all  physical volume of the
Universe in a state with given density (for example, on the
hypersurface
$\rho_0 \sim 10^{-29} g\cdot cm^{-3}$) should result from the
evolution of those
relatively rare but additionally inflated regions in which the field
$\phi$, over
the longest possible time, has been fluctuating about its maximum
possible values, such that $V(\phi) \sim 1$.

\subsection{\label{tau} Expansion of the Universe as a measure of time}
In the previous Sections we used  the time measured by synchronized
clock of
comoving  observers as a time coordinate $t$. However, in general
relativity
one may use different ways of measuring time intervals. For example,
instead of
synchronizing clocks of different observers, we may ask them to
measure a local
growth of the scale factor of the Universe near each of them. Namely,
the
local  value of the scale factor in the inflationary Universe grows
as follows:
\begin{equation}\label{c0} a\left(x, t \right) = a(x,0)\, \exp\left(
{\int_0^t{H(\phi(x,t_1))\ dt_1}} \right)\ .
\end{equation}
Then one can define a new time
coordinate $\tau$ as a logarithm of the growth of the scale factor,
\begin{equation}\label{c1} \tau  =  \ln{{a\left(x, t \right) \over
a(x,0)}} =  \int_0^t{H(\phi(x,t_1))\ dt_1}\ .
\end{equation}
If one neglects the space and time dependence
of $H$, then $\tau = Ht$.  However, in  a more  general case the
difference between $t$ and $\tau$ may be significant, and, as we will
see in
the next Section, the time $\tau$ often is more convenient.

Now let us try to derive equations describing diffusion as a function
of  the
new  time  $\tau$.

First of all, instead of eq. (\ref{m1a}) we have now
\begin{equation} \label{m1aa}
\frac{d}{d\tau} \, \left<\phi^2\right> =  {H^2(\phi(x,t))  \,
\over 4\pi^2}\ ,
\end{equation}
and the ordinary classical equation of motion of a homogeneous scalar
field
during  inflation acquires the following form:
 \begin{equation} \label{m1ab}
\frac{d}{d\tau} \, \phi = - \frac{V'(\phi) }{3H^2(\phi)}  \ .
\end{equation}
This gives $\kappa = {1\over 3 H^2(\phi)}$ and  ${\cal D} =
{H^2\over 8\pi^2} = { V(\phi)  \over 3\pi}$. The final form of the
diffusion
equation is \cite{Star}
\begin{equation}\label{eq37aa}
\frac{\partial P_c(\phi,\tau)}{\partial \tau} =  {1\over 3\pi}\
\frac{\partial}{\partial
\phi} \left(\sqrt {V(\phi)}\ \frac{\partial\,({\sqrt
{V(\phi)}}\,P_c(\phi,\tau))}{\partial \phi} +
\frac{3V'(\phi)}{8V(\phi)} \, P_c(\phi,\tau) \right) \ .
\end{equation}

There are two important advantages of this choice of time over the
standard  one \cite{Star,SB}.  First of all, if we consider evolution
of a domain with initial value of the field $\phi = \phi_0$, then the
physical wavelength $\lambda_p$ of perturbations which are frozen at
each given
moment of time $t$ is  proportional to $H^{-1}(\phi)$. This
corresponds to
the wavelength  $\lambda_c = H^{-1}(\phi)
\exp\left(-{\int_0^t{H(\phi(x,t_1))\ dt_1}} \right)$ in the comoving
coordinates. The dependence of  $H^{-1}$ on $\phi$ is not very
important.
Indeed,
the deviation of $\phi$ from $\phi_0$ becomes significant only on an
exponentially large length scale. However, the second term
significantly changes
even for small deviations of $\phi$ from $\phi_0$. Therefore, when we
will perform our computer simulations of stochastic processes in the
inflationary Universe, we will face a complicated problem of adding
to each
other waves of the field $\phi$ with wavelengths which rapidly change
from one
point to another (see next Section). The only way of doing it which
we found is
to use the recently developed concept of wavelets, waves with a
compact
support, see e.g. \cite{David}. There is no such problem in the time
$\tau$,
since with this time parametrization the integral
$\int_0^t{H(\phi(x,t_1))\ dt_1}$ is the same for all points with
given $\tau$.

Another important advantage of using the time $\tau$ is an extremely
simple
transition from the distribution $P_c(\phi,\tau)$ to
$P_p(\phi,\tau)$:
\begin{equation}\label{eq37aaa}
P_p(\phi,\tau) = P_c(\phi,\tau)\ e^{3 \tau} \ .
\end{equation}
Naively, one would expect that the normalized distribution $\tilde
{P}_p(\phi,\tau)$ is
trivially related to $P_c(\phi,\tau)$,
\begin{equation}\label{eq37aaaa}
\tilde{P}_p(\phi,\tau) = P_c(\phi,\tau) \ .
\end{equation}
However, this is  true  only  if the distribution $P_c(\phi,\tau)$ is
normalized. It does not make much sense to normalize $P_c$
since this probability distribution is not stationary, and decreases
in time in
all realistic models  of inflation.
Meanwhile, the normalized probability distribution $\tilde{P}_p$ may
become
stationary. Therefore the normalized probability distribution
$\tilde{P}_p$
will be proportional to $P_c$, but the coefficient of proportionality
will
depend on $\tau$.

The first step towards finding the normalized probability
distribution
$\tilde{P}_p$ is to write a
diffusion equation which describes the non-normalized distribution
$P_p$
(\ref{eq37aaa}):
\begin{equation}\label{eq37aau}
\frac{\partial P_p}{\partial \tau} =  {1\over 3\pi}\
\frac{\partial}{\partial
\phi} \left(\sqrt {V(\phi)}\ \frac{\partial\,({\sqrt
{V(\phi)}}\,P_p(\phi,\tau))}{\partial \phi} + {3V'(\phi)\over
8V(\phi)}\,P_p(\phi,\tau)
\right) + 3P_p(\phi,\tau) \ .
\end{equation}

Despite all advantages of using the time $\tau$, we should remember
that
this   is {\it not} the time which can be measured  by usual clock of
a local
observer. Rather it is a peculiar time which an observer measures
by his rulers. In what follows we will discuss both time $t$ and time
$\tau$, but we will mainly concentrate on the evolution in time $t$.

\section{Computer Simulations \label{Comput}}

All concepts discussed in this paper are rather unusual. We are used
to
think about the Universe as if it were an expanding spherically
symmetric
ball of fire. Now we are trying to explore the possibility that the
Universe
looks uniform only locally, but on a much larger scale it looks like
a fractal.
The properties of fractals are not very simple, but there is some
beauty and
universality in them, which transcends the beauty and universality of
simple
spherically symmetric objects. The simplest way to grasp the new
picture of
the Universe would be to make a computer simulation of its structure.

Unfortunately, it is very difficult to make any kind of graphical
illustration
of physical processes in the inflationary Universe, since the
Universe is
curved, and, moreover, it is curved differently in its different
parts. What
we made is just a small step in this direction. We performed a
computer
simulation of stochastic evolution of the scalar field $\phi$ in one-
and
two-dimensional inflationary Universes, both in time $t$ and in time
$\tau
\sim \log a(t)$\@.

\subsection{Fluctuations in  a  one-dimensional Universe}
As a first step, we considered  a part of   a one-dimensional
inflationary
Universe, of initial size $H_0 = H^{-1}(\phi_0)$ (i.e. the size of
the
horizon)\@.
Then we followed generation of perturbations of the scalar field
$\phi$ during
expansion of this part of the Universe. We did it by dividing the
process of
expansion into small steps of duration $\Delta t = u\,
H^{-1}(\phi_0)$,  where
$u$ is some small parameter.   At each step we solved the
equations of motion for   classical field $\phi$ at all points of the
grid,
and then added to the result a `quantum fluctuation' $\delta \phi$\@.
The
idea was to represent such quantum fluctuations by waves with random
phases
and directions, with an amplitude ${H(\phi)\over \sqrt 2 \pi}\,{\sqrt
{uH(\phi) \over  H(\phi_0)}}$, and with the (physical) wavelength
$\lambda_p$, which would be equal to $H^{-1}(\phi)$ at each point.
Note that
the amplitude of a fluctuation is equal to ${H(\phi)\over \sqrt 2
\pi}$ for $u
=
1$, $H = H_0$. This gives correct contribution to
$<(\delta\phi)^2>\sim
{H^3\over 4\pi^2}t $ after summation over the contributions with
different
phases.\footnote{In fact, the amplitude of fluctuations in the
one-dimensional
Universe would be slightly smaller \cite{KofStar}, but in our
computer
simulations we preserve its normal (three-dimensional) amplitude.}
The extra
factor $\sqrt {uH(\phi)}\over \sqrt {H(\phi_0)}$ appears due to the
fact that
we
add each wave not within the time $H^{-1}(\phi)$, but within the time
$uH^{-1}(\phi_0)$\@. Introduction of the small phenomenological
parameter $u$
provides a better model of  white noise, since it allows for a
possibility of
jumps higher than the average amplitude ${H(\phi)\over \sqrt 2 \pi}$
within
the time $H^{-1}$ (even though the probability of large jumps will be
exponentially suppressed). We took $u \sim 10^{-1}$ in most of the
calculations.

All calculations were performed in comoving coordinates, which did
not change
during the expansion of the Universe. In such coordinates,
expansion of the Universe results in an exponential shrinking of
wavelengths,
\begin{equation}\label{c11}
\lambda_c(x,t) = \lambda_p(x,t)~ e^{\phantom
{}^{-\int_0^t{H(\phi(x_1,t_1))\,
dt_1}}} ~ .
\end{equation}
Perturbations which have a wavelength $H^{-1}$ in physical
coordinates, have
the wavelength ${\lambda_c\over \lambda_p}\, H^{-1}$ in comoving
coordinates.
Thus, at each step of calculations we were adding the perturbation
\begin{eqnarray}\label{c2}
\delta \phi(x,t) &=& a \cdot
\sin\left(\int_0^x H(\phi(x_1,t)) {\lambda_p(x_1,t)\over
\lambda_c(x_1,t)}\,
dx_1 +\alpha_n\right)
\nonumber \\
&=& a \cdot \sin\left(\int_0^x H(\phi(x_1,t))\
e^{\phantom {}^{\int_0^t{H(\phi(x,t_1))\, dt_1}}}dx_1+\alpha_n\right)
\ .
\end{eqnarray}
Here $\alpha_n$ are random numbers, and the amplitude of the
perturbation is
given by
\begin{equation}\label{c}
a = {H(\phi(x,t))\over \sqrt 2 \pi}\,{\sqrt
{uH(\phi(x,t))\over H(\phi_0)}} \  ~.
\end{equation}

We have made our one-dimensional calculations using the grids
containing up
to $3\times 10^5$ points.
 At each step of calculations we   added sinusoidal waves
(\ref{c2}) corresponding to perturbations of the field $\phi$
generated during
the time $u H^{-1}_0$,  and then subtracted the decrease of the field
$\phi$ due to its classical motion,
\begin{equation}\label{c3}
\Delta \phi(x,t) = -\, {u V'(\phi)\over 3H_0\, H(\phi(x,t))} ~~.
\end{equation}

In our computer simulations we considered the simplest theory of a
massive
scalar field with $V(\phi) = {m^2\phi^2\over 2}$.  We took the mass
$m$ very
big ($m = 0.5$), since otherwise, according to eq. (\ref{E28}), we
would need
exponentially large pages to show our results. In most of the
calculations the
initial value of the scalar field was about $1$, which  corresponds
to the
energy density which is one order of magnitude smaller than the
Planck density.

The results of the computer simulation are shown in the set of
Figures {1}.
Some explanations are necessary here.  We present a sequence of 3 panels corresponding to different moments of time $t$. Each panel consists of three
figures. On
the first panel we show the spatial distribution of the scalar
field $\phi$
in comoving coordinates. The value of the scalar field at each point
is given
by the upper boundary of the shaded area.  The distribution of the
scalar field
in these coordinates averaged over the whole domain corresponds to
the
probability distribution $P_c$.

\begin{figure}[h!]
\centering\leavevmode{\epsfysize=6cm{\epsfbox{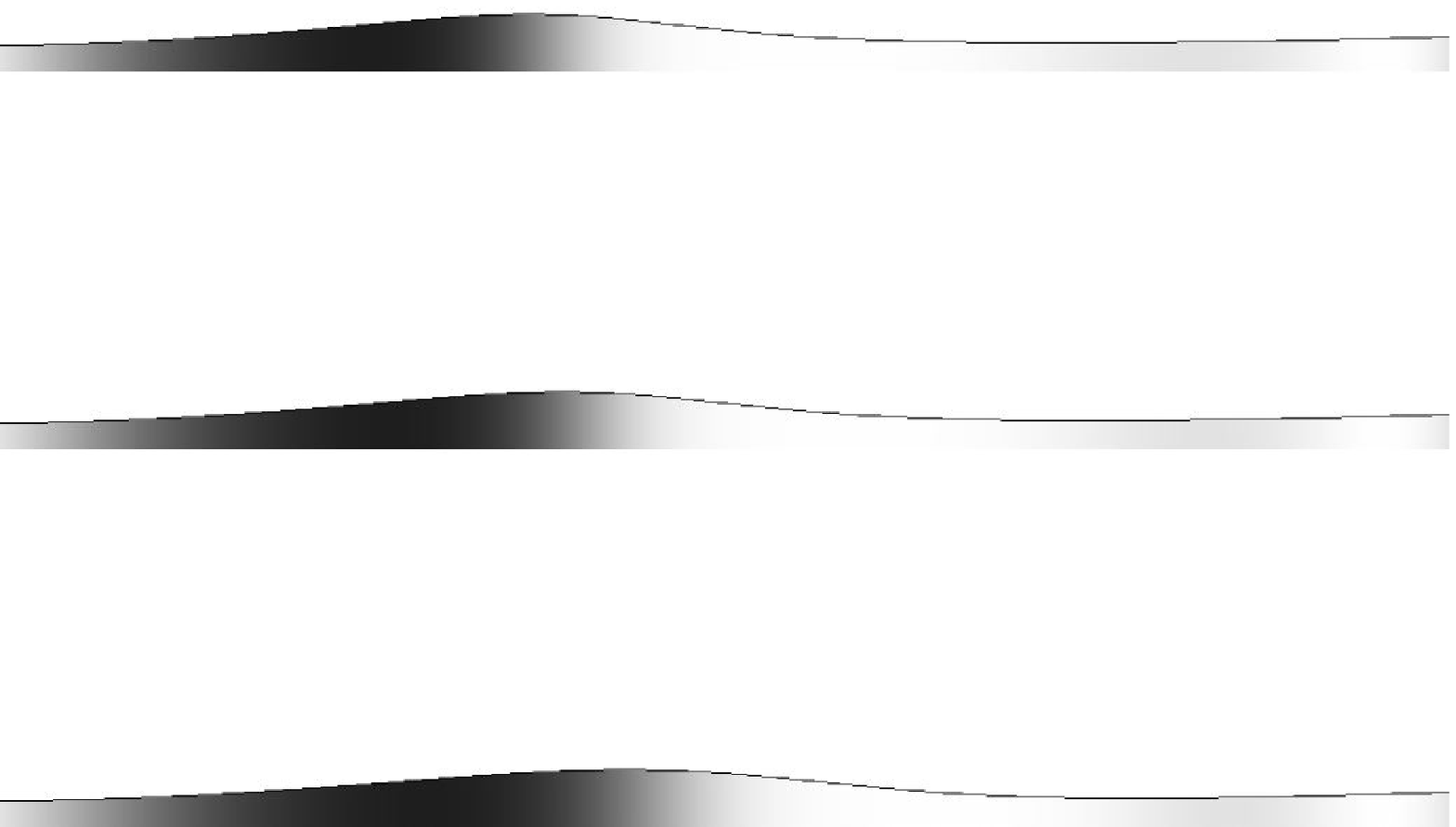} } \hskip 0.8cm \epsfysize=6cm{\epsfbox{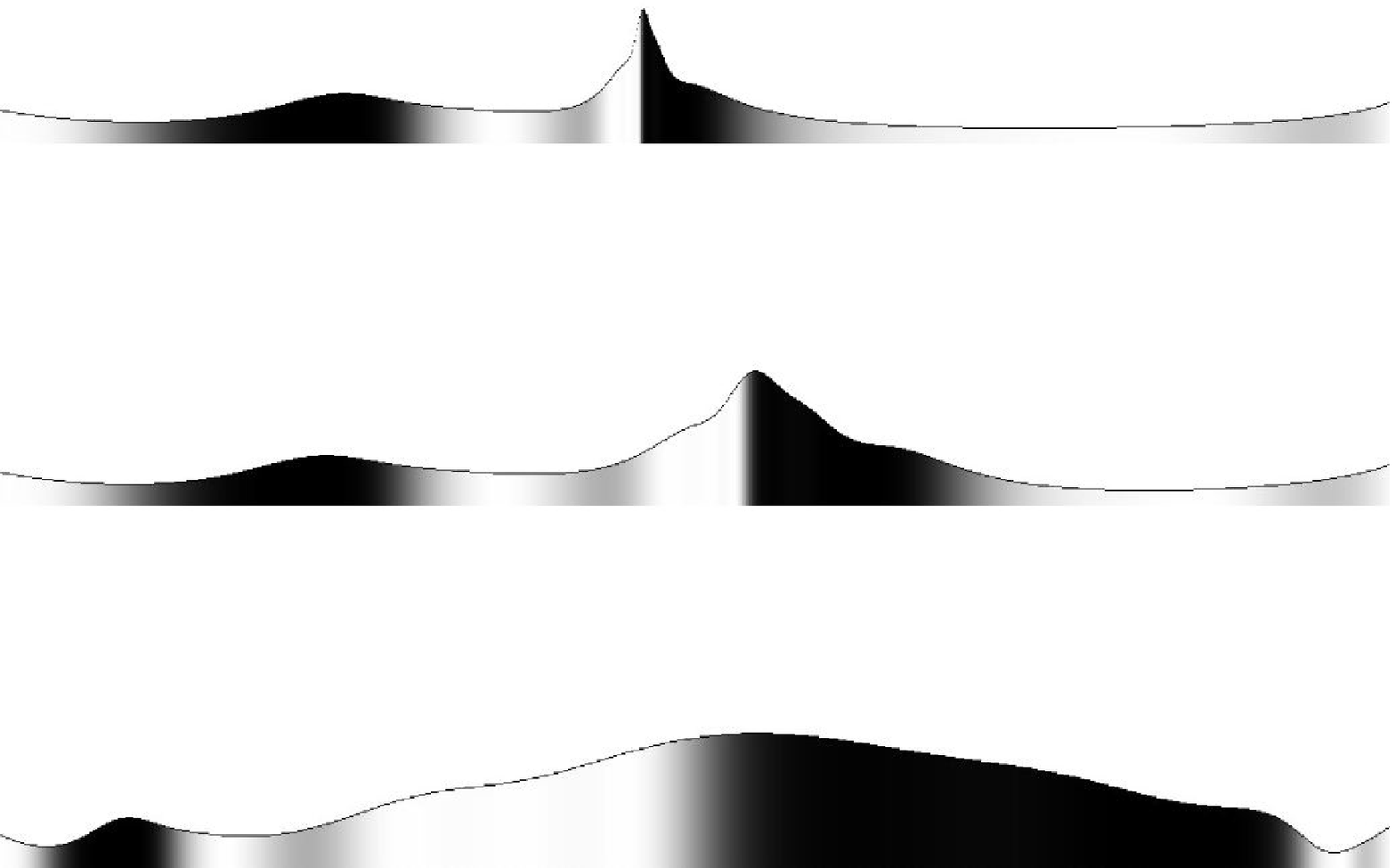}}}\\  \vskip  1cm \centering\leavevmode{\epsfysize=9cm{\epsfbox{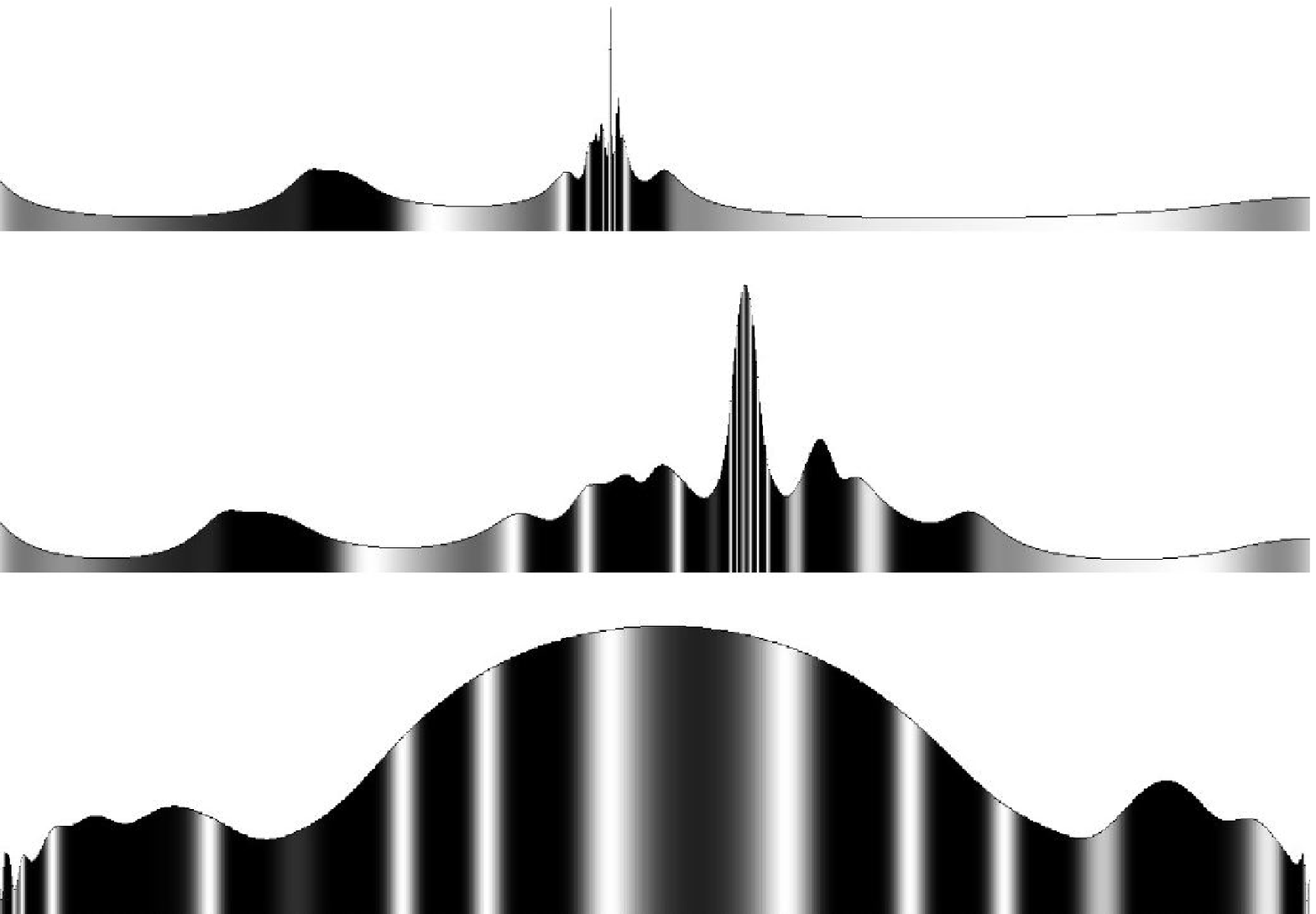}}}
\par
\ 
\caption{Time  evolution of the distribution of the inflaton scalar
field
$\phi$,  shown by the upper boundary of the shaded area, and the
scalar  field
$\Phi$, shown by the color (black -- white).  This distribution is
shown as a
function of one coordinate $x$, in a domain of initial size
$H^{-1}(\phi_0)$.
We present a sequence of three panels for each time $t$. The first
one shows
the distribution of the fields in the comoving coordinates, which is
related to
the probability distribution $P_c$. The second one exhibits the same
fields in
the coordinates which
show the physical distance from one point to another, divided by the
total
distance between the two sides of the domain. The third one shows the
distribution of the fields per unit  physical three-dimensional
volume, which
would correspond to the distribution  $P_p$ in a three-dimensional
Universe.
}
\label{fig:Figa}
\end{figure}

The second panel shows the same distribution of the
scalar field,
but in  different coordinates, which take into account different rate
of
expansion in different domains.
To achieve this goal, at each step $\Delta t$ of our calculations we
expand the distance
between the nearby points  by $e^{H(\phi(x,t))\Delta t}$. After that,
the computer displays the resulting
distribution at the interval of the original size, as if there were
no change
in the total length of the domain. The distribution of the field
$\phi$ in
these figures corresponds to the probability distribution $P_p$, and
the
squeezing of the interval back to its original size automatically
normalizes
this distribution. This means that the distribution of the scalar
field  shown
at the lower figure gives a good idea of the relative fraction  of
volume of
the Universe filled by the  field $\phi$.

Well, this is not {\it quite} correct. These figures give a good idea
of the
relative fraction  of {\it length} of a {\it one-dimensional}
Universe filled
by the field $\phi$. Indeed, our computer simulations here were
one-dimensional. To get a correct idea of how the probability
distribution
$P_p$ behaves in a three-dimensional Universe, each time $\Delta t$
one should expand the distance
between the points by $e^{3H(\phi(x,t))\Delta t}$. This gives a
distorted spatial
distribution of the scalar field, but this distribution shows in a
correct way
the   distribution  of the three-dimensional volume of the Universe
filled by
the field $\phi$.
This is shown at the third panel (the third series of figures in Fig. 1, the lowest ones).

Black and white regions in these figures correspond to the
distribution of
another scalar field, which we added to our model. This is a scalar
field
$\Phi$ with the effective potential $V(\Phi)$ which may have several
different
minima. It is assumed that during the stage of inflation driven by
the field
$\phi$, the effective potential $V(\Phi)$ is much smaller than
$V(\phi)$.
Therefore, at this stage one may neglect the contribution of the
field $\Phi$
to the rate of expansion of the Universe. However, this field may be
responsible for the symmetry breaking in the theory of elementary
particles.
The field $\Phi$ also experiences quantum fluctuations and Brownian
motion.

This Brownian motion can push the field $\Phi$ from one minimum of
$V(\Phi)$ to
another. Then, after
the end of inflation, the field $\Phi$ becomes trapped by the minimum
to which
it jumped, and
it cannot move from it anymore. However, in different  exponentially
large
domains of the Universe this field may be trapped in different
minima. {\it As a
result, the Universe becomes divided into exponentially large domains
with
different types of elementary particle physics inside each of them.}

To give a particular example of such theory, we will consider a model
with the
effective potential
\begin{equation}\label{c4}
V(\Phi) = V_0~\left (1 - \cos {N\Phi\over \Phi_0 }\right )~,
\end{equation}
where $N$ is some integer. The ratio  $\Phi\over \Phi_0$ is
considered as an
angular variable.  Such potentials appear, e.g., in the theory of the
axion
field, as well as in the model of `natural inflation' driven by the
field
$\Phi$. The amplitude of the quantum jumps of the field $\Phi$ is
described by
the same equation (\ref{c2}) as for the field $\phi$; to study the
classical
motion of the field $\Phi$ one should apply eq. (\ref{c3}) to this
field.

In our series of computer simulations of the one-dimensional
inflationary
Universe, we take a particular value $N=2$. In this case the
effective
potential (\ref{c4}) has two minima of equal depth, $V(\pi\Phi_0) =
V(-\pi\Phi_0) = 0$. We use black color to show the parts of the
Universe  with
$\Phi \sim \pi\Phi_0$, we use white color to show the parts with
$\Phi \sim
-\pi\Phi_0$, and we use various shades of grey to show intermediate
regions.

Our calculations begin with a domain of initial size $H_0^{-1}$
filled by some
homogeneous (or almost homogeneous) fields $\phi$ and $\Phi$. After
few steps
the distribution of the scalar field $\phi$ becomes slightly
inhomogeneous. At
the same time the inhomogeneities generated in the field $\Phi$ may
be
relatively large, if their amplitude $\delta\Phi \sim \sqrt u H$ is
large in
comparison with $\Phi_0$. As a result, the field $\Phi$ rapidly jumps
from one
minimum of its effective potential  to another, and the Universe
becomes
divided into approximately equal number of black and white regions.

After few steps of calculation the scalar field $\phi$ also becomes
very
inhomogeneous. However, this inhomogeneity looks different on the
three
different series of figures we produced. In comoving coordinates
(upper
figures) the field $\phi$ in the most part of space decreases; we
see only few
 hills, which in the course of time grow and produce many  thin
spikes.
However, in the physical coordinates, which take into account
exponential
expansion of the Universe, these thin spikes  look like large
mountains, see
figures in the middle. Gradually these mountains occupy more and more
space,
and their physical size grows exponentially. Also, these mountains
are never
sharp; they are build from the sinusoidal waves which always look
very smooth
on the scale $H^{-1}$. (The two last statements are only partially
illustrated
by the figures, since the  mountains look more narrow than they
really are
after the computer shrinks the expanded interval back to its original
size.)

Finally, the last series of figures shows a very rapid growth of the
relative
fraction  of volume of the Universe occupied by the  growing field
$\phi$. This
exactly corresponds to the process of self-reproduction of the
Universe. We see
also  that quantum fluctuations of the scalar field $\Phi$ lead to
formation of
exponentially large number of black and white domains, corresponding
to
different types of symmetry breaking in the theory (\ref{c4}).

\subsection{One-dimensional Universe in the $\tau$-parametrization of
time}
Similar calculations can be done for the $\tau$-parametrization of
time, where
$\tau = \ln{{a\left(x, t \right) \over a(x,0)}} $. This immediately
reveals an
important difference between this parametrization of time and the
standard one.
 First of all, now all three sets of figures look absolutely the
same, since,
by definition, at a given time $\tau$ the degree of exponential
expansion is
the same for all points $x$. An important difference between the
corresponding
distributions $P_c$ and $P_p$ is that the distribution $P_c$ is not
stationary:
the comoving volume of the regions filled by the large field $\phi$
decreases.
Meanwhile,   the total volume of the domains  filled by the large
field $\phi$,
which is given by $P_p$, increases exponentially. As we will show in
this
paper, after we divide $P_p$ by the overall growth of the volume of
the
Universe, we obtain a stationary normalized distribution $\tilde
P_p$.

Unfortunately, this difference between $P_c$ and $P_p$ cannot be seen
in our
figures. Therefore we will present only one figure (Fig. 2) instead
of the
three series of figures. This figure shows the distribution of the
scalar
fields $\phi$ and $\Phi$ after a few steps of our computer
simulations. It
illustrates a specific difference between the distributions at a
given $t$ and
at a given $\tau$. The typical wavelength of inhomogeneities  of the
field
$\phi$  in the $\tau$-parametrization of time is approximately
constant in
different parts of the Universe, whereas  the field $\phi$ in the
$t$-parametrization is much more inhomogeneous near the maxima of its
distribution in comoving coordinates.

\begin{figure}[h!]
\centering\leavevmode{\epsfysize=5 cm{\epsfbox{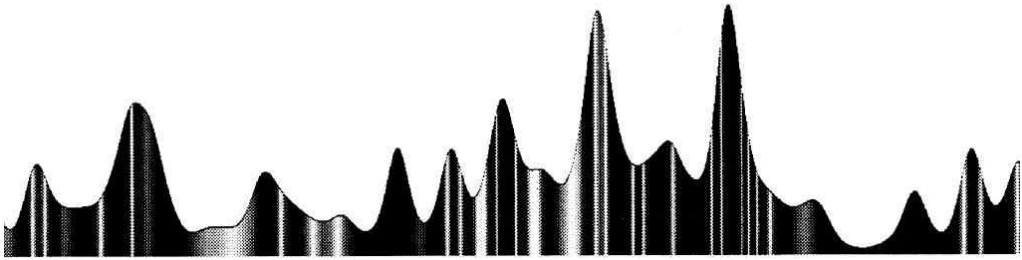}}}
\par
\ 
\caption{The distribution of the fields $\phi$ and $\Phi$ after
several steps
 in time $\tau$.}
\label{f2}
\end{figure}

From the technical point of view, the difference between the two
calculations
is reflected in the equations for the perturbations $\delta\phi$ and
$\delta\Phi$, and for their classical shift during the time interval
$\Delta t
= u$ corresponding to one step of our calculations. For example, for
the field
$\phi$ we have
\begin{equation}\label{c5}
\delta\phi(x,y)={H(\phi(x))\sqrt u \over   \sqrt{2}\pi}\cdot
\sin\Bigl(H(\phi(x))~ e^{\tau}\, x + \alpha_n\Bigr) ~,
\end{equation}
and
\begin{equation}\label{c6}
\Delta\phi(x,y)=-\, {u V'(\phi)\over 3 H^2(\phi(x,t))} \ \, .
\end{equation}

\subsection{Fluctuations in  a two-dimensional Universe}
 When generalizing our methods for two- or three-dimensional Universe
in time
$t$,  one meets several complications. First of all,  it is possible
to plot
the distribution of the field in  comoving coordinates, but it is
very
difficult to stretch these coordinates to show the distribution of
the fields
per unit physical volume, as we did for the one-dimensional Universe.
This
seems to be a rather general problem, since it is not always possible
to show
the curvature of space by representing it as a curved surface in a
flat space
of higher dimension. Therefore we represented all our results in the
comoving
coordinates only.

Another problem is related to the possibility to represent quantum
fluctuations
of the scalar field by plane waves. In one-dimensional Universe this
was an
easy problem, but in a two dimensional curved Universe  this is
impossible. In
some sense, the waves of the scalar field look as if they were
propagating in a
medium with an exponentially large refraction coefficient. This
refraction
coefficient  changes from point to point. If one starts with a plane
wave at
some point $x$, it
does not remain a plane wave at some distance from this point: the
front of the wave becomes strongly bent  because of gravitational
lensing.

This is not a real problem if one studies only relatively small
domains, but
even if one starts with a very small domain, eventually its size
becomes exponentially large, and the effects of bending become more
and more
significant. Therefore, if one wishes to investigate evolution of
the scalar field in the whole domain,  one should find some other
way to simulate quantum fluctuations of the scalar field, which would
lead to the same (or almost the same) correlation functions
(\ref{E23})-(\ref{E23b}) as the ordinary plane wave fluctuations,
while
being free from the problems mentioned above.  Such a method does
exist: one should use functions $\delta \phi(x)$ which look like
plane waves inside domains of a  sufficiently large radius $r_0$,
where
$r_0 \gg
H^{-1}$. However, these functions should rapidly decrease at $r >
r_0$.
Such functions are called wavelets \cite{David}.

A simplest example of a wavelet which we used in our calculations
looks as
follows:
\begin{equation}\label{c7}
\delta \phi(x,t) = a\,  \exp \left( - C^2 ({\bf r} -{\bf r}_n)^2
\cdot
H^2({\bf r}_n,t))  \right) \times \, \sin \left(H({\bf r}_n,t))
{\lambda_p({\bf r}_n,t)\over
\lambda_c({\bf r}_n,t)} (x \sin\theta_n + y \cos \theta_n +
\alpha_n)\right) ~.
\end{equation}
Here  $a = {H(\phi({\bf r}_n,t))\over \sqrt 2 \pi}\,{\sqrt
{uH(\phi({\bf r}_n,t))\over H(\phi_0)}}$,\,  $\theta_n$ and
$\alpha_n$ are random
numbers, $C $ is some constant, $({\bf r} -{\bf r}_n)^2$ is the
square of the
physical distance from the center of the wavelet    ${\bf r_n}$ to
the point
${\bf r}$. In the comoving coordinates $x$ and $y$
\begin{equation}\label{c8}
({\bf r} -{\bf r}_n)^2 = \Bigl( (x-x_n)^2 + (y-y_n)^2 \Bigr)\cdot
{\lambda^2_p({\bf r}_n,t)\over \lambda^2_c({\bf r}_n,t)}   \ .
\end{equation}

At  $C\ll 1$ this field configuration looks like a sinusoidal wave
with an
amplitude which is given by $a$ at  ${\bf r}= {\bf r_n}$, and which
becomes
exponentially small at a distance about $(C H)^{-1}$ from the center
of the
wavelet. In the limit
$C\to 0$ wavelets  would behave as ordinary plane waves. Thus, one
should take
$C$ small, but not too small, to avoid the effect of gravitational
lensing
inside the wavelets.

The procedure of simulation of quantum
perturbations of the scalar field consisted of generation of a proper
number
density of
wavelets   and of their distribution at random points ${\bf r_n}$ in
such a way
that the number of the
wavelets in a given domain is proportional to its physical volume
(i.e. to $e^{\int 3H({\bf r}_n,t)\,dt }$)\@.  By choosing a proper
number density of
wavelets we mean that the variance of perturbations produced by
wavelets in the
limit when the Universe is homogeneous should coincide with the
standard result
$<(\delta\phi)^2> = H^3t/4\pi^2$.

 Our choice of the form of the wavelets  is not perfect, but it is
good enough
for our purposes.  Ideally, one should consider wavelets which form a
complete
orthonormal set of functions, like it is done in the usual Fourier
analysis. It
is well known that the wavelet transform can give a much better
information
about very inhomogeneous structures than the standard Fourier
transform
\cite{David}. It is also known that the usual momentum
representation, which is
 related to the Fourier transform, is not suitable for formulation of
quantum
field theory in curved space.  We were forced to use wavelets instead
of plane
waves  to perform  a consistent computer simulation of quantum
fluctuations in
the usual $t$-parametrization of time.  This gives us a hint that the
  wavelet
transform, rather than the standard Fourier transform,  may become a
very
useful tool  for   field quantization   in   curved space.

On the other hand, in some simple but important cases one may escape
from the
difficulties associated with gravitational lensing by choosing a
different set
of coordinates. In our case every\-thing becomes very simple in the
$\tau$-parametrization of time, where
$\tau \sim \log a(t)$\@. Computer simulations of the evolution in
time $\tau$
are much easier \cite{SB}. In particular, one can  use
ordinary plane waves, but with the amplitude proportional to
${H(\phi({\bf r}_n))\sqrt u\over \pi \sqrt 2}$. The reason is that
different parts of
inflationary Universe expand by the same number of  $e$-foldings at a
given value of
the time $\tau$. From the point of view of computer simulations this
means
that expansion of the Universe does not introduce any distortions
into the
original grid of comoving coordinates $x$ and $y$.\footnote{There
still remains
a small distortion, since the physical wavelength $\sim H^{-1}(\phi)$
takes
different values in different parts of the Universe. However, this is
a
relatively small effect as compared with the exponential distortion
discussed
above; in  computer simulations one may simply keep the wavelength
proportional
to $H^{-1}(\phi_0)$.}
However, it would be incorrect to make computer simulations in time
$\tau$ {\it
instead} of simulations in time $t$. These two approaches are
complimentary to
each other, emphasizing different features of the same process. For
example, as
we have already seen in our study of one-dimensional Universe,
sometimes it is
 easier to illustrate the process of self-reproduction of
inflationary domains
using ordinary $t$-parametrization of time. Figures {3}--{6}
contained in this
section were obtained by computer simulations of the evolution of
scalar fields
in time $t$, whereas Fig. {7} corresponds to the time $\tau$.

\begin{figure}[h!]
\centering\leavevmode{\epsfysize=6.8 cm{\epsfbox{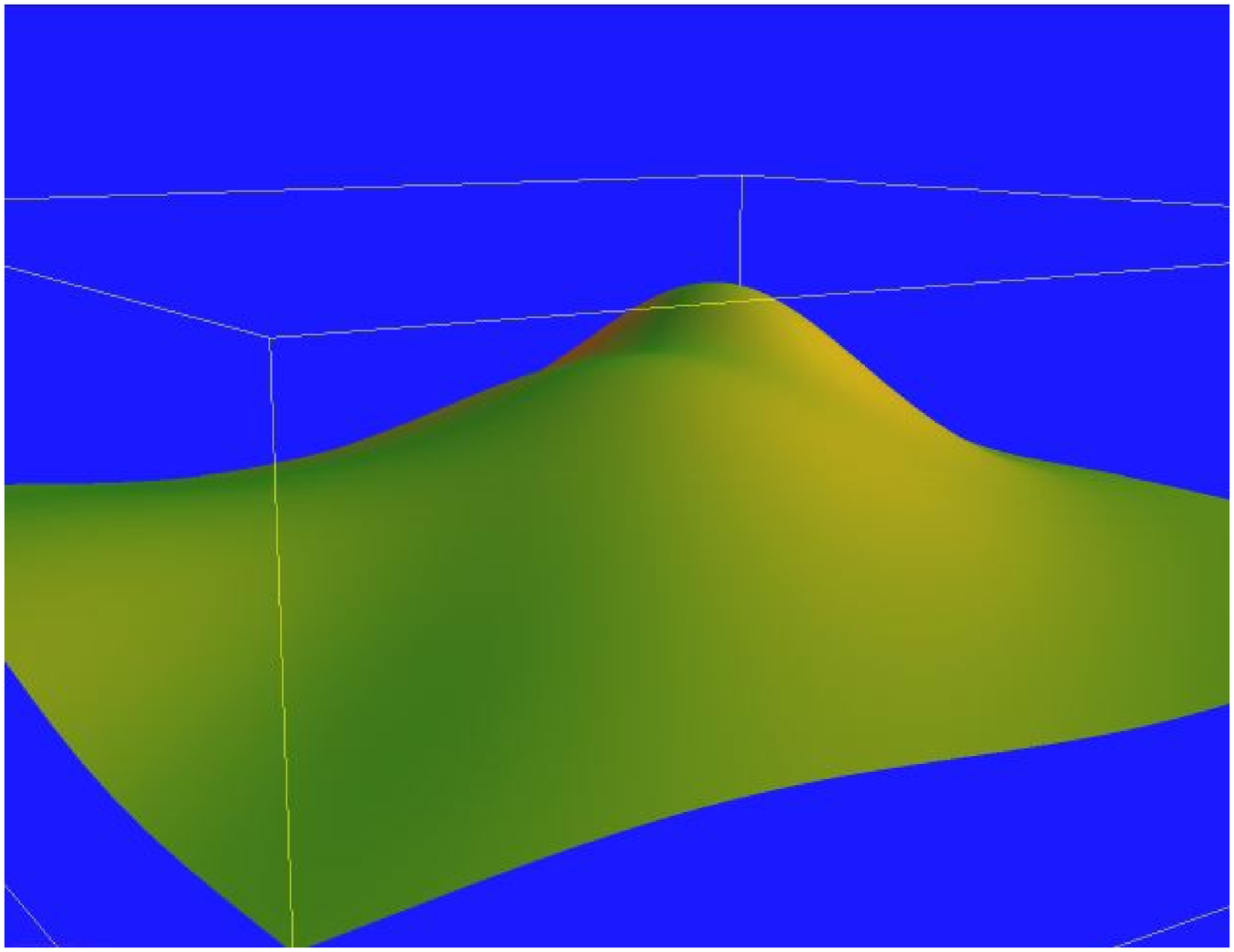}} ~\epsfysize=6.8 cm{\epsfbox{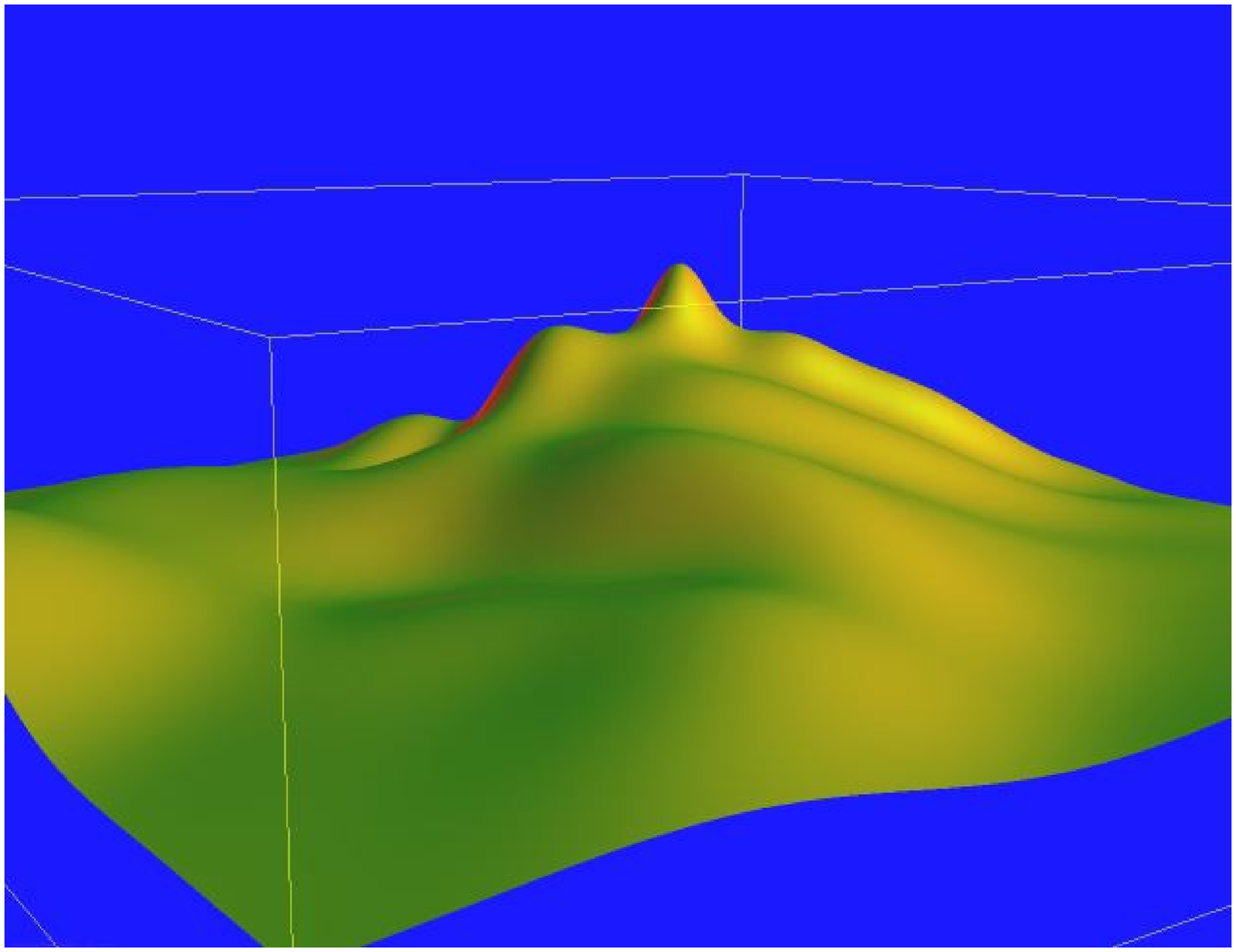}}}
\vskip 0.3cm
\centering\leavevmode{\epsfysize=6.8 cm{\epsfbox{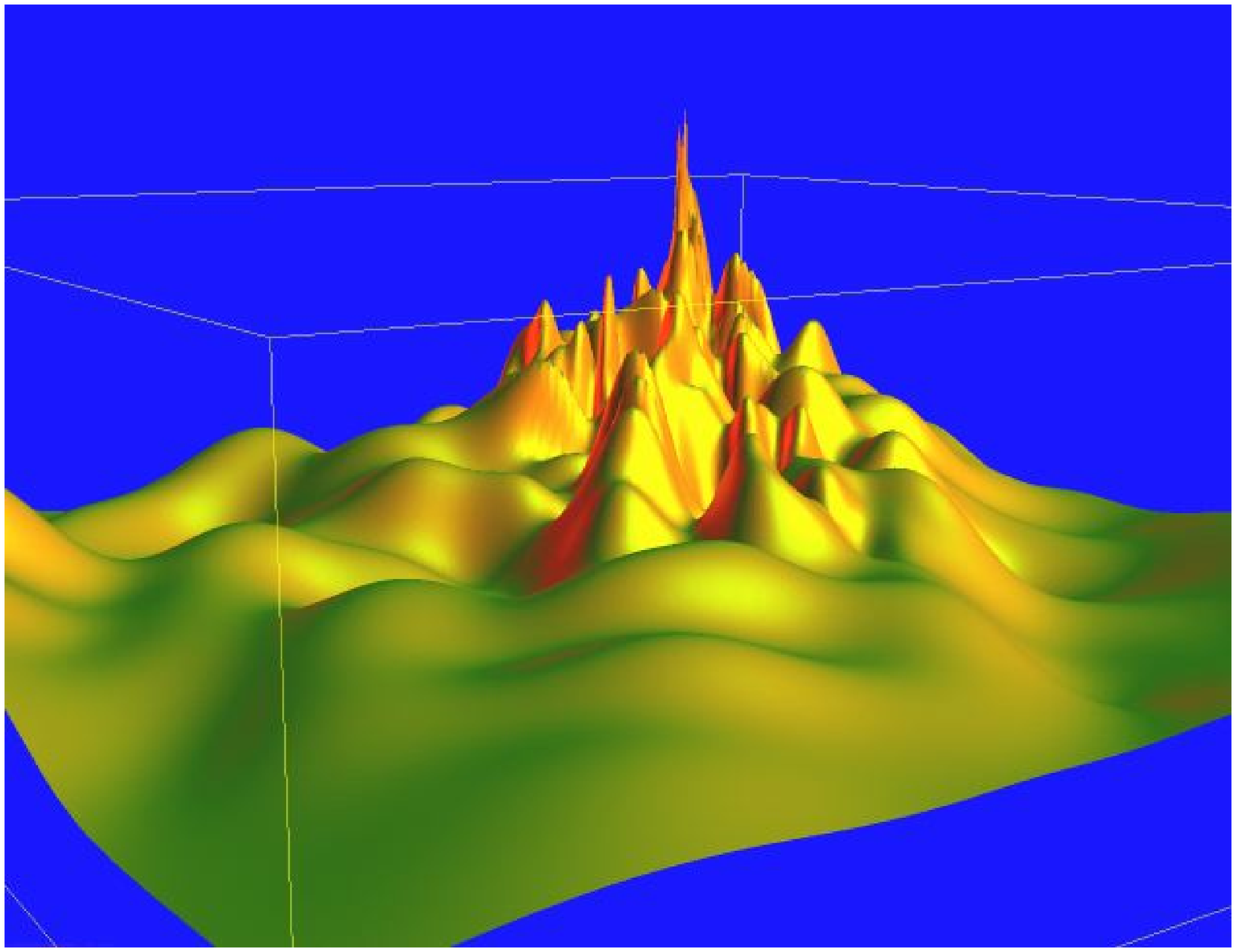}} ~\epsfysize=6.8 cm{\epsfbox{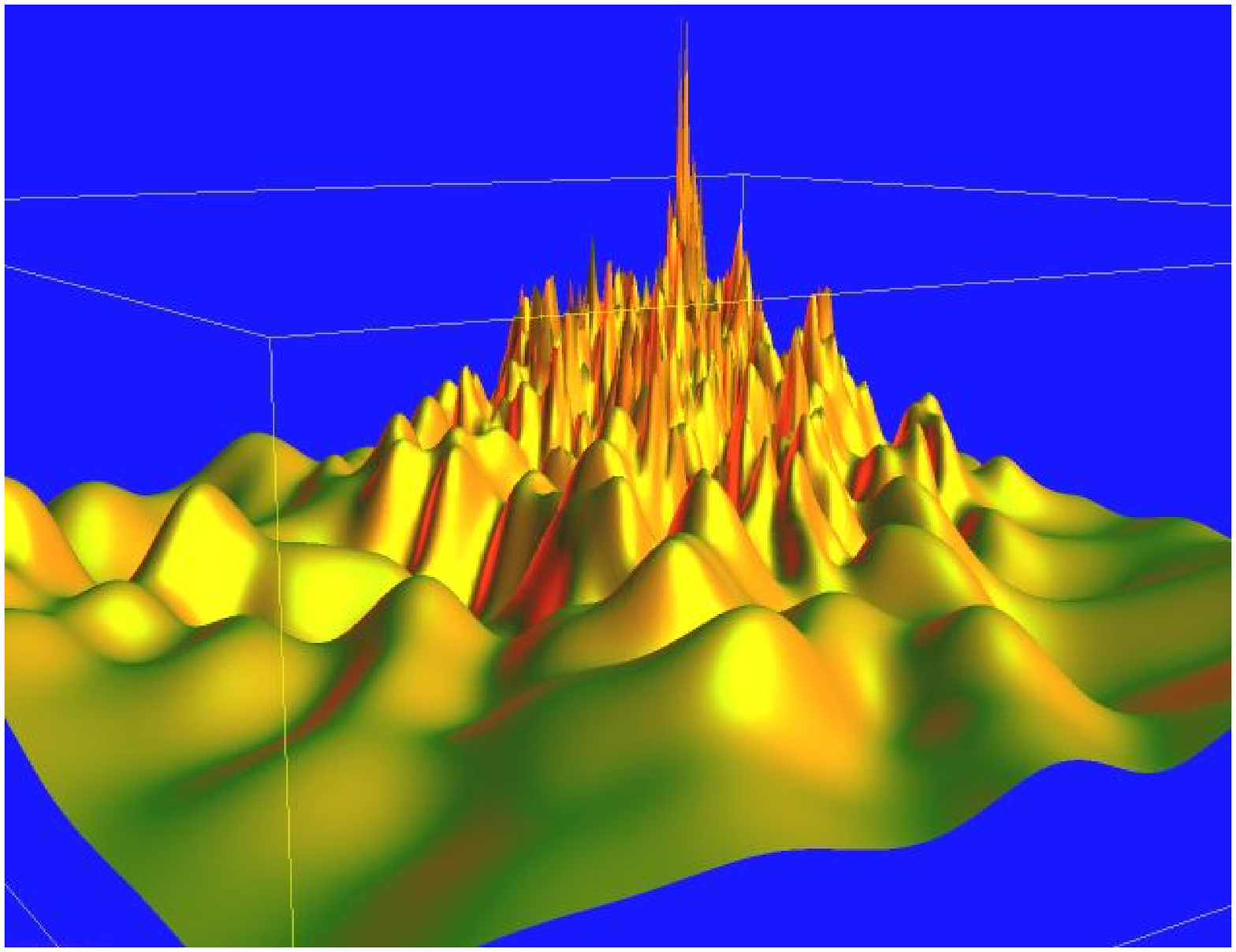}}}
\par
\ 
\caption{The evolution of the scalar field $\phi$ in a two-dimensional
inflationary Universe 
 in time $t$.}
\label{f2}
\end{figure}

 The set of Figs. {3} shows the
evolution of the scalar field in time $t$\@.  We begin our
simulations with an
almost homogeneous scalar field   $\phi \sim \phi_0$ inside a domain
of initial
size
$H^{-1}(\phi_0)$, which is represented by a grid containing
$1000\times 1000$
points, Fig.  {3}.1. Then the Universe expands, the amplitude of
the classical scalar field slowly decreases, but due to overlapping
of
waves (=wavelets) corresponding to fluctuations of the scalar field,
the
amplitude of the scalar field in some parts of the expanding domain
becomes much greater than $\phi_0$\@. This corresponds to growing
mountains on Figs. {3}.2--{3}.6. We should emphasize again that the
peaks of
these mountains are
not sharp at all; they are built from fluctuations with the physical
wavelength $\sim H^{-1}$\@. They look sharp since we are plotting
everything in comoving coordinates, in which the wavelength of
perturbations $\lambda_c$ becomes exponentially small, especially in
the parts of the Universe where $\phi$ is large and expansion is fast
(i.e. on the tops of the mountains).

In the parts of the Universe where the scalar field becomes small
there are no
high
mountains; the field continues moving towards $\phi = 0$\@. We live
in
one of such parts of the Universe. The perturbations of the scalar
field
generated at this stage of inflation are relatively small. These
perturbations
are responsible for     the  small perturbations of temperature of
the
microwave background radiation discovered by COBE.

However, if $\phi_0$ is greater than some
critical value $\phi^*$, then mountains with $\phi > \phi^*$ are
permanently produced. This corresponds to the process of the Universe
self-reproduction. The effective potential of the field $\phi$ on
tops of some
of these mountains becomes greater than the Planck density. (In these
simulations we did not introduce any boundary conditions at the
Planck
density; see however the next Section.) One may envisage  each of
such tops as
a
beginning of a new ``Big Bang''. If one wishes to reserve this name
for the
first
``Big Bang'' (if there was one), one may think about such names as a
``Small
Bang'' or ``Pretty Big Bang''. Whatever the words, the theory we have
now is
considerably different from the old Big Bang theory.

\begin{figure}[h!]
\centering\leavevmode{\epsfysize=9 cm{\epsfbox{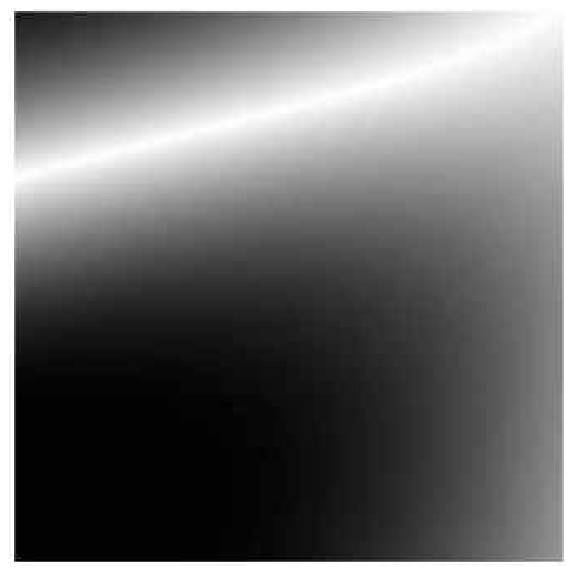}} \epsfysize=9 cm{\epsfbox{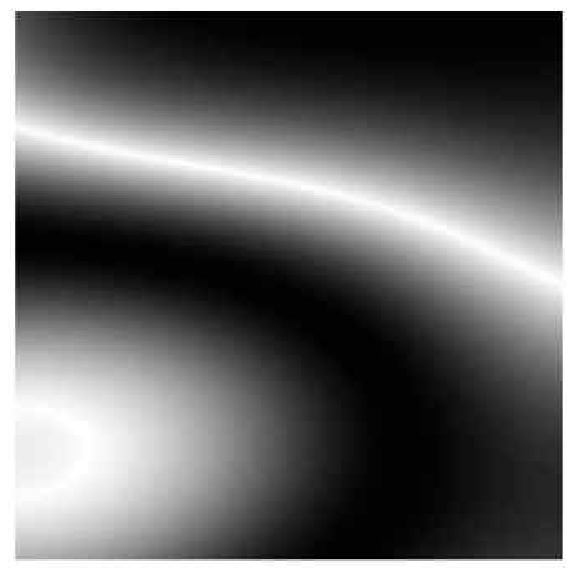}}}
 
 \vskip -2.3cm
\centering\leavevmode{\epsfysize=9 cm{\epsfbox{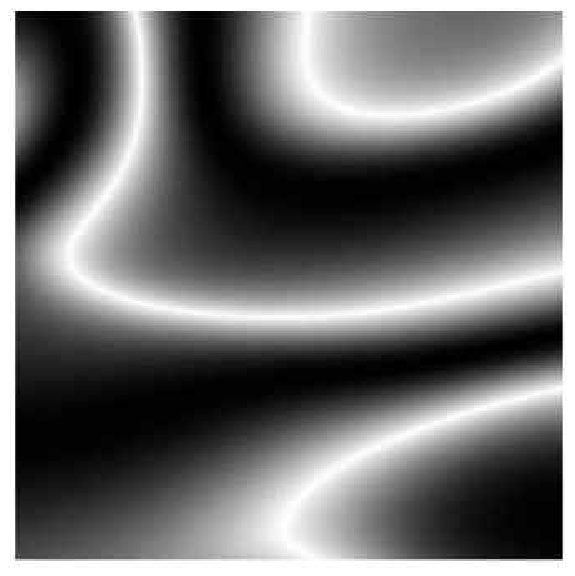}}  \epsfysize=9 cm{\epsfbox{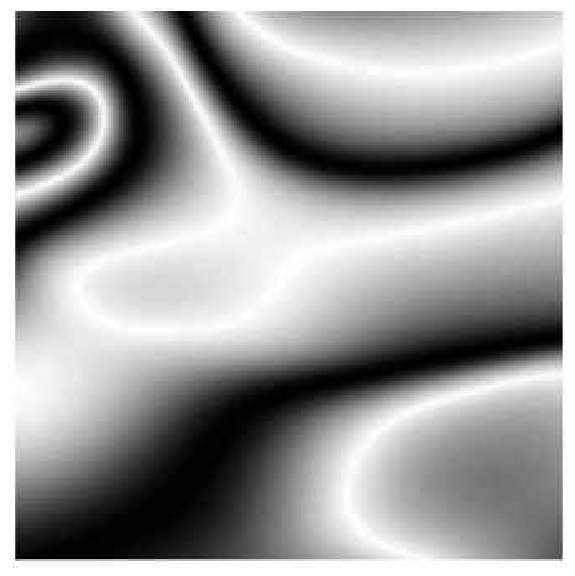}}}
\vskip -2.5cm
\caption{Generation of axion domain walls during inflation.}
\label{f2}
\end{figure}

The   series of Figures {4} corresponds to a more complicated theory
of
elementary particles, which includes another scalar field, $\Phi$\@,
with the
effective potential (\ref{c4}) with $N = 1$. In this case the shape
of the
effective potential  $V(\Phi)$ coincides with the shape of the
effective
potential in the simplest version of the axion theory. In these
figures we do
not show the evolution of the inflaton scalar field $\phi$; thus, no
mountains.
We just show  the distribution of the fluctuating scalar field $\Phi$
on the
plane $(x,y)$. White regions correspond to the minima of $V(\Phi)$ at
$\Phi/\Phi_0 = 2n\pi$, black regions correspond to the maxima at
$\Phi/\Phi_0 =
(2n+1)\pi$.

In the beginning of our calculations the color (the level of grey)
only slowly
varies in the domain of initial  size $\sim H^{-1}$, Fig. {4}.1.
Later on, in
the main part of the domain (in comoving coordinates) the field
$\Phi$ moves
towards the minima of its effective potential, but in some places the
field
diffuses, say,   from the minimum at $\Phi/\Phi_0 = 0$ to the minimum
at
$\Phi/\Phi_0 = 2\pi$. On the way it passes through the maximum at
$\Phi/\Phi_0
= \pi$, and in some places the field stays at this maximum, forming a
domain
wall between the domains with $\Phi/\Phi_0 = 0$ and with $\Phi/\Phi_0
= 2 \pi$.
 The existence of such domain walls in the axion theory was first
pointed out
by Sikivie \cite{Sik}. The inflationary mechanism of their production
described
above was discussed in \cite{LinLyth}. The   series of Figures {4}
shows the
process of formation of these  domain walls, which look as thin black
lines at
Fig. {4}. 

\begin{figure}[h!]
\centering\leavevmode{\epsfysize=11 cm{\epsfbox{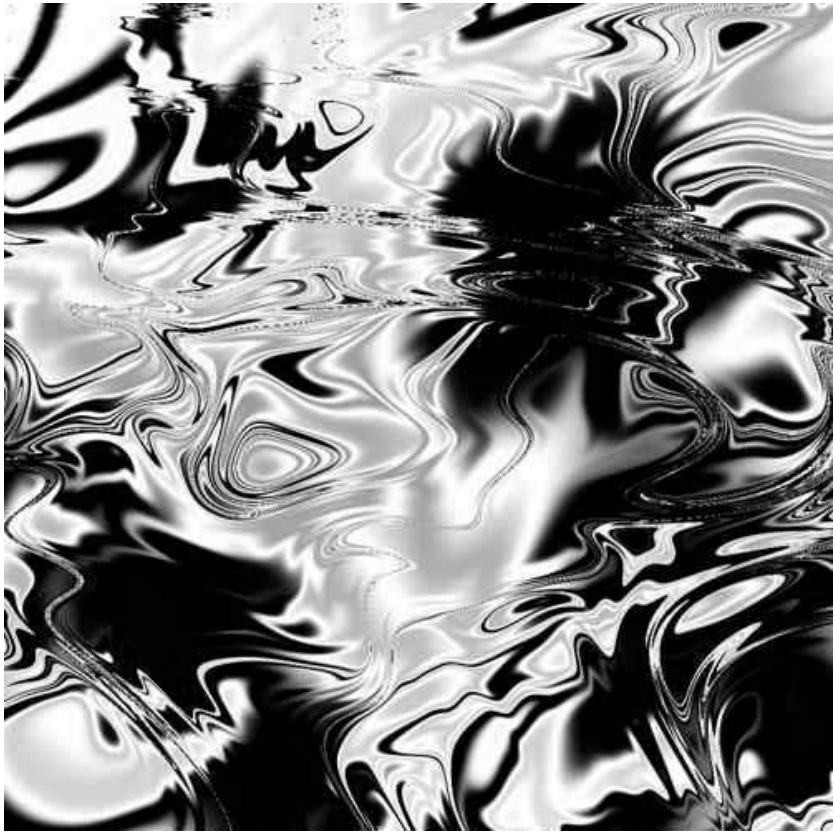}}}

\caption{ Axion domain walls in a model with a smaller value of
$\Phi_0$ (Pollock
Universe).}
\label{f2}
\end{figure}

The distribution and the properties of domain walls in the axion
theory are very
sensitive to the values of parameters of the theory. For example, by
reducing
the radius $\Phi_0$ of the effective potential  (\ref{c4}) one
obtains much
more domain walls per unit volume, see   Fig. {5}. For obvious
reasons, we
called this figure `Pollock Universe'.

Now let us consider the same model, but with $N = 2$, as we did in
our
one-dimensional calculations. In this case in the interval $0 <
\Phi/\Phi_0 <
2\pi$ we have two different minima with $V = 0$.
{}From the point of view of perturbations produced during inflation,
there is
no much difference between the cases $N=1$ and $N=2$; domain
structure will be
generated in each case. But now we assume  that some properties of
particles
interacting with the field  $\Phi$ change when the field
$\Phi/\Phi_0$ jumps
from one minimum to another. This happens, for example, in the
supersymmetric
$SU(5)$ model, where there exist many minima of the same depth
corresponding to
different types of symmetry breaking in the theory.
To distinguish between the minima at $\Phi/\Phi_0 = 2n \pi$ and the
minima at
$\Phi/\Phi_0 = (2n+1) \pi$ we will paint the first ones white and the
second
ones black, just  as we did in our one-dimensional calculations.

\begin{figure}[h!]
\centering\leavevmode{\epsfysize=11.3 cm{\epsfbox{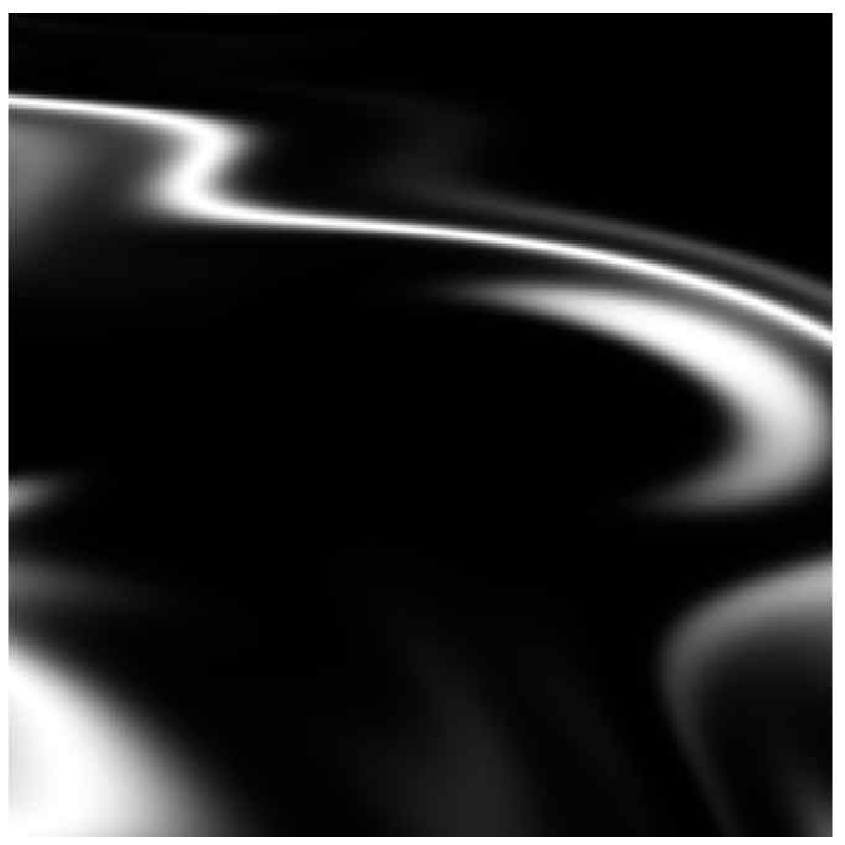}}\hskip -.6cm \epsfysize=11.3 cm{\epsfbox{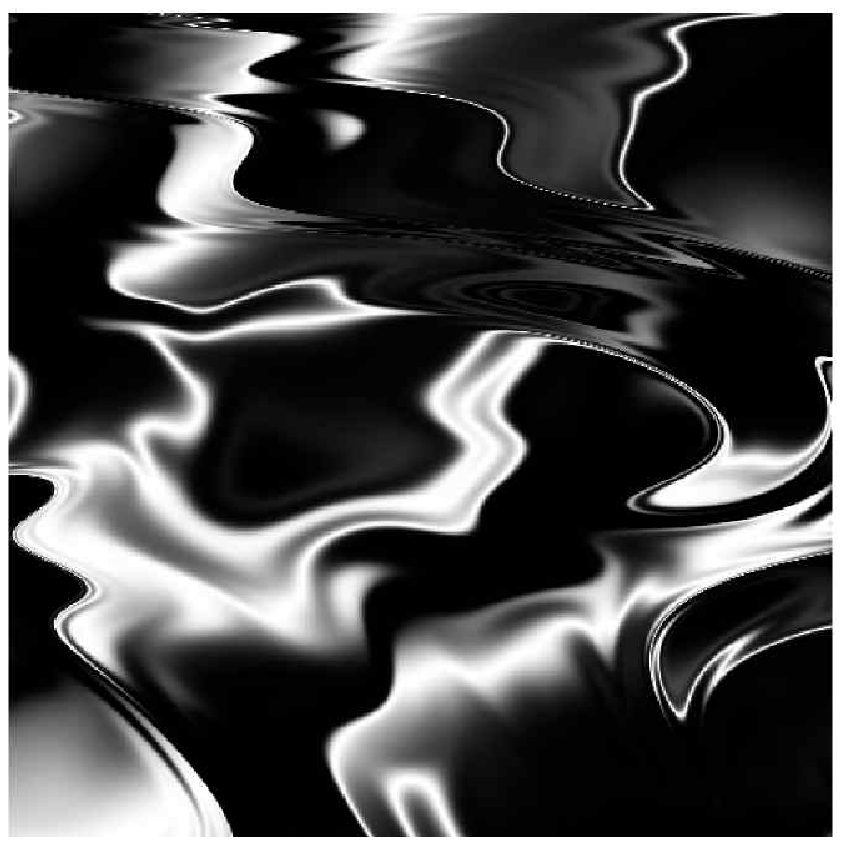}}}
\vskip -3.2cm
\centering\leavevmode{\epsfysize=11.3 cm{\epsfbox{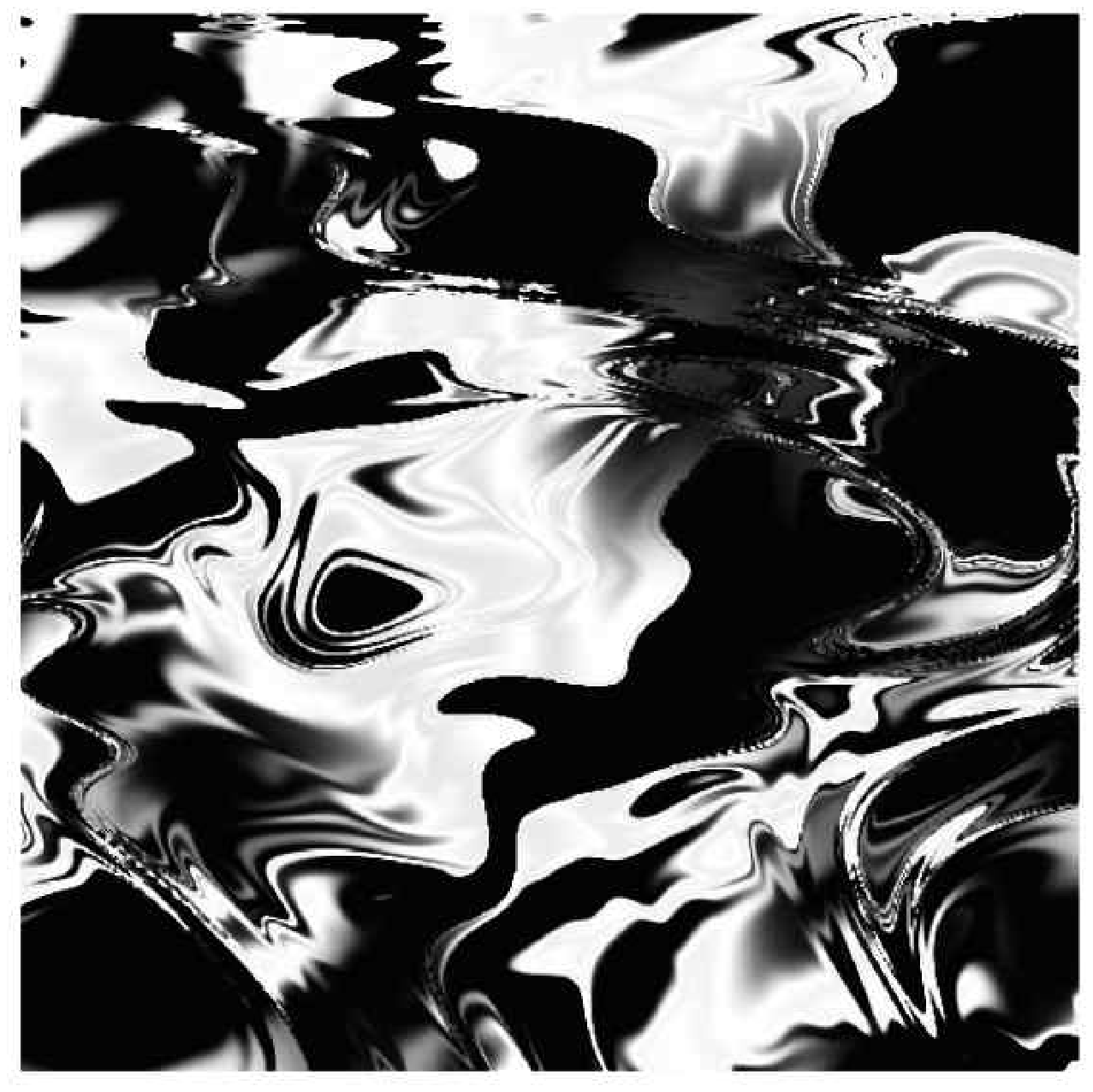}} \hskip -0.6cm \epsfysize=11.3 cm{\epsfbox{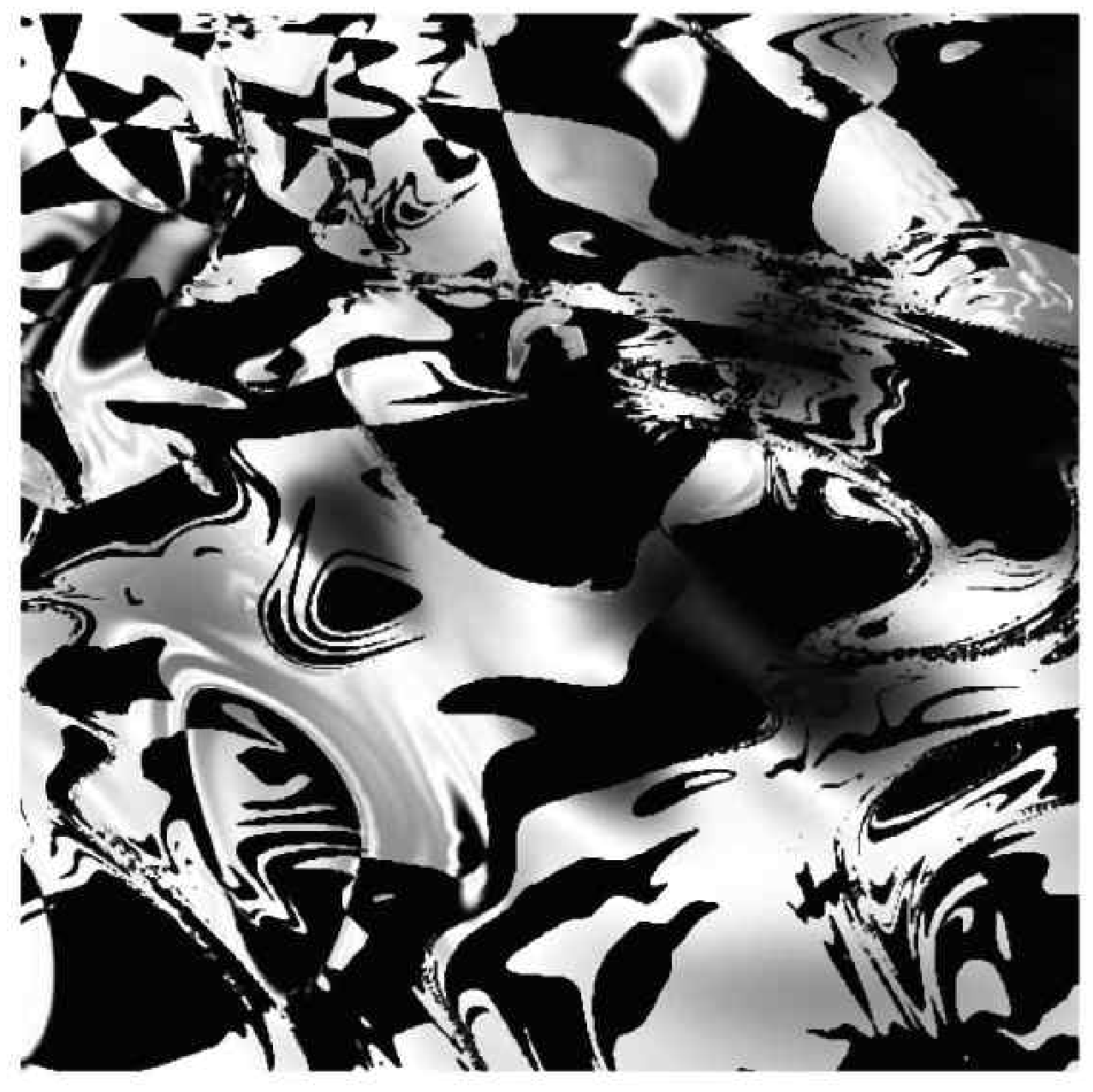}}}
\vskip -2.5cm
\caption{Generation of domain structure of the Universe during
inflation in the
axion model with $N= 2$ (Kandinsky Universe).}
\label{f2}
\end{figure}

The results of the calculations are represented by a series of
Figures {6}.
Originally the whole initial domain contains the scalar field in one
of the
minima of its effective potentials.  Then this field begins jumping
from one
minimum to another, and the Universe becomes divided into many
exponentially
large black and white domains.   Figures {6} show  that in the course
of time
the domain which originally was quite homogeneous becomes looking
like a huge
fractal.  We called this fractal `Kandinsky Universe'.

\begin{figure}[h!]

\centering\leavevmode{\epsfysize=8.5 cm{\epsfbox{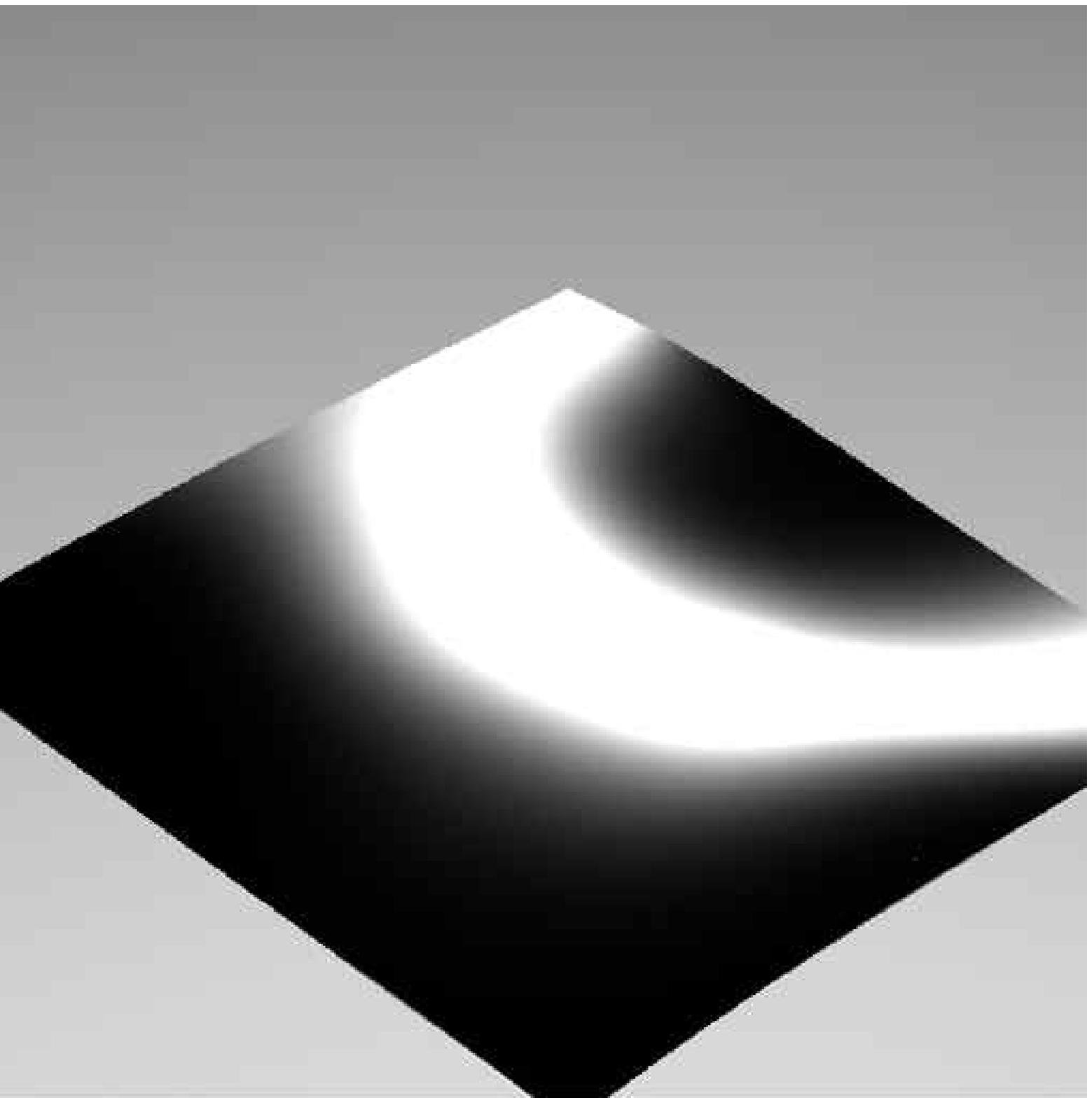}} \epsfysize=8.5 cm{\epsfbox{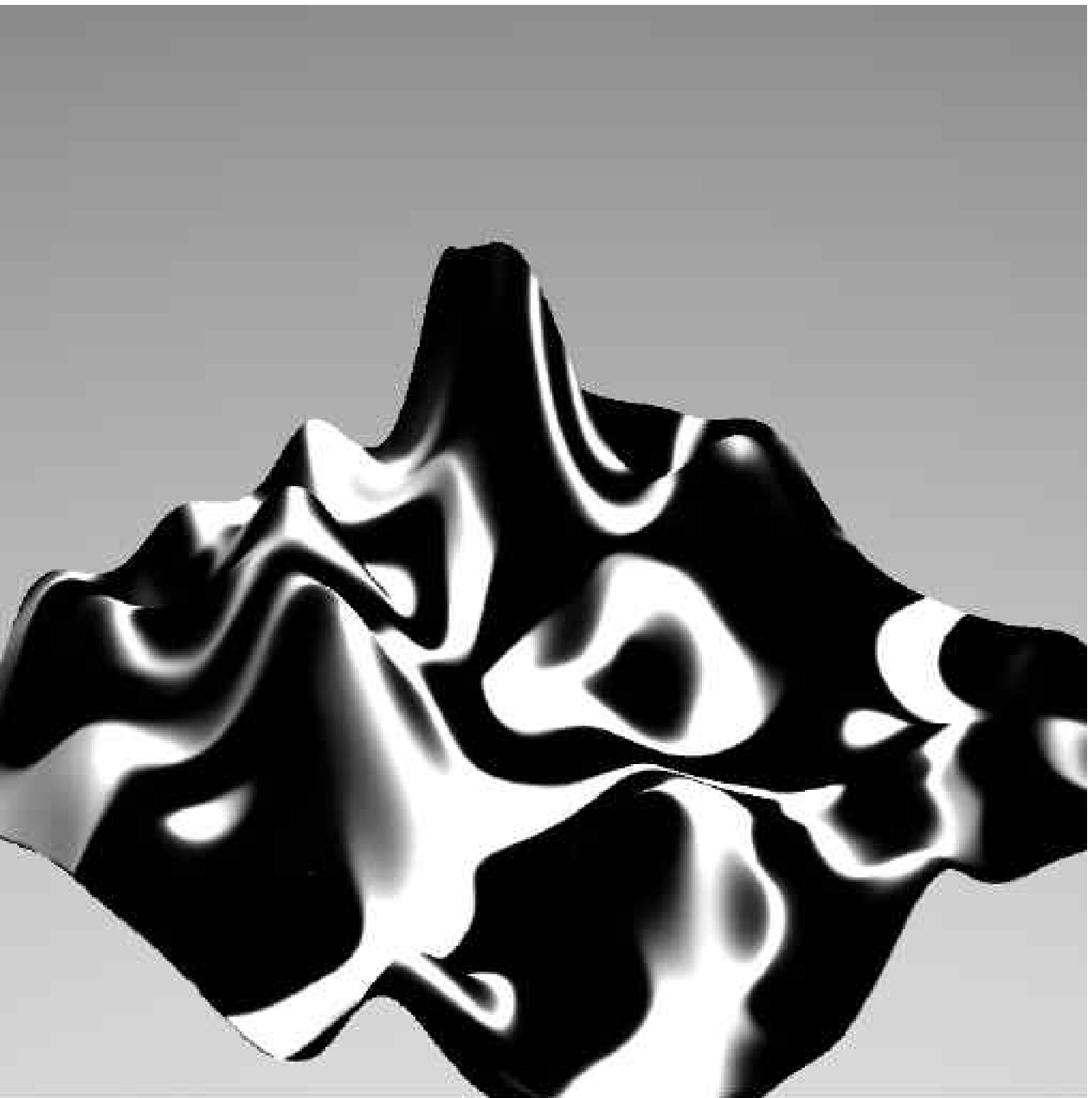}}} 
\vskip 0.1cm

\centering\leavevmode{\epsfysize=8.5 cm{\epsfbox{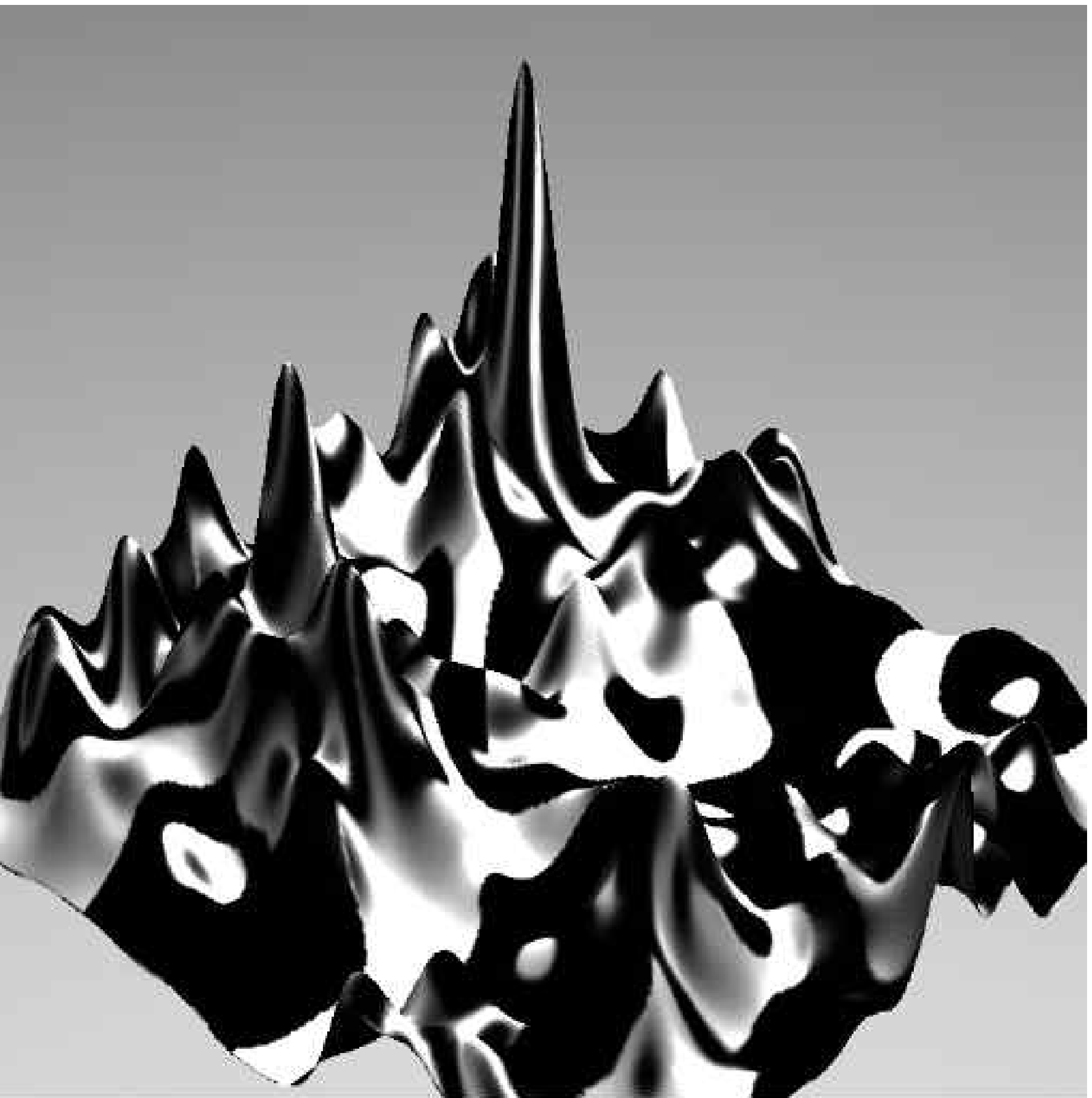}}  \epsfysize=8.5 cm{\epsfbox{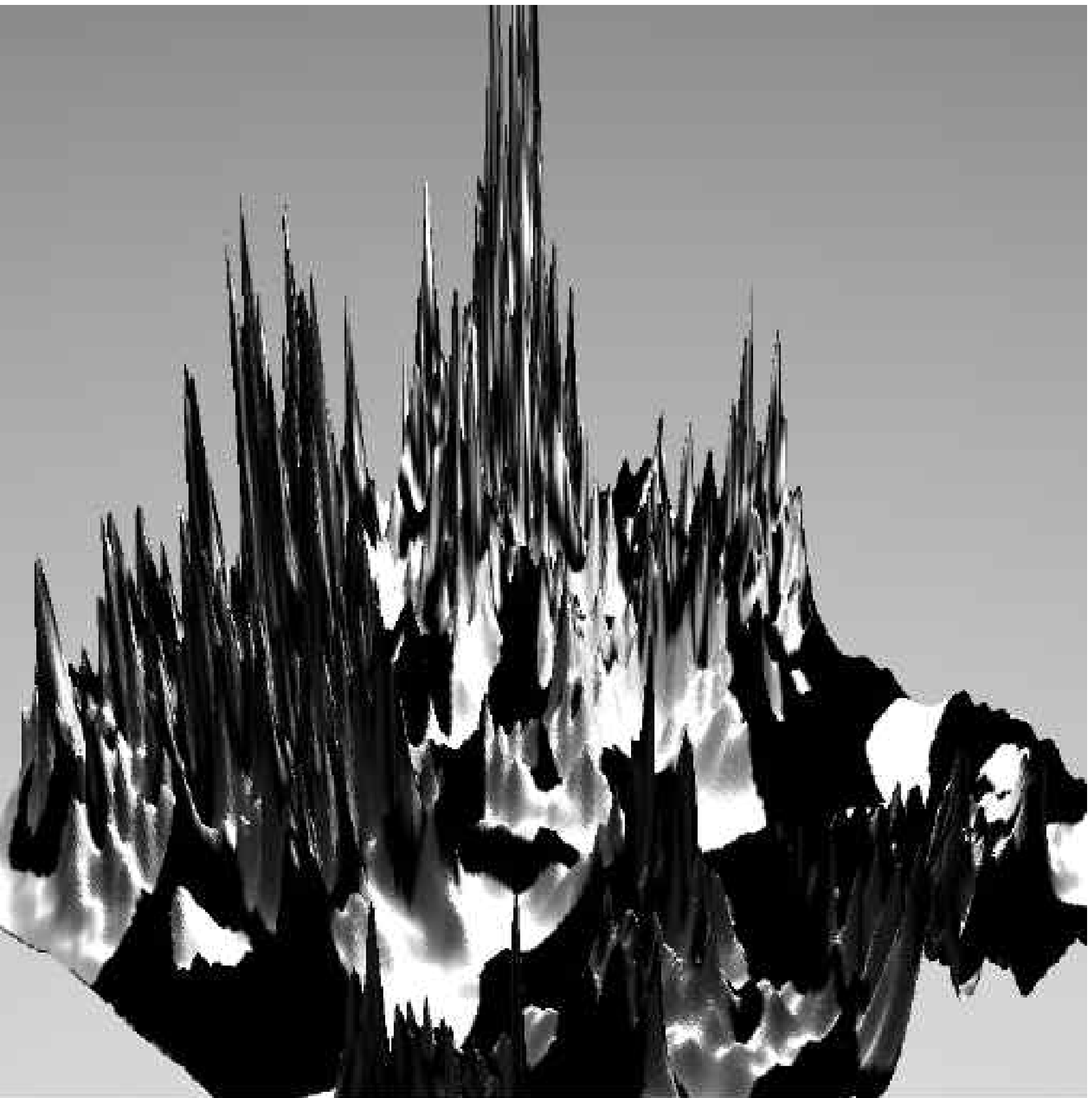}}}

\caption{The distribution of the scalar fields  $\phi$  and $\Phi$ in a
two-dimensional Universe after several steps  in time $\tau$.
}
\label{f2}
\end{figure}

In   Figs. {7}, which we obtained by studying the of stochastic
processes in the
time $\tau$, we  plot simultaneously the distribution of the inflaton
field
$\phi$ and of the field $\Phi$. By looking at these figures   one can
find
that the most chaotic and rapidly changing distribution of black and
white
domains corresponds to the regions near the  peaks of the
distribution of the
field $\phi$.  In these regions the color of the domains changes very
rapidly,
which means that the laws of the
low-energy elementary particle physics (which are related to the type
of
spontaneous symmetry breaking) are not fixed there yet. However, in
the valleys, where the inflaton field $\phi$ is small, the color
becomes fixed.
We live in one of such domains of a given
color. Other domains are exponentially far away from us. And there
are domain
walls separating domains with different colors. Laws of low energy
physics
change as one jumps through the domain walls; one should think
twice before doing so.

An interesting feature of  all these computer simulations  is the
following. If
one
takes a magnifying glass and looks at some part of the picture of a
physical
size of the order of $H^{-1}$, one will see a rather homogeneous
distribution
of the
scalar fields $\phi$ and $\Phi$\@. Then one may start computer
simulations
again. And again one will get mountains of the same type, and a
Universe
consisting of  domains of different colors. This is the first (and
the most fundamental) kind of stationarity which we certainly have
in our Universe. Not only the Universe looks like a fractal at any
given moment
of time, it  actually is   a {\it growing fractal}, which reproduces
itself
over and over again.

As we already mentioned, the fractal structure of the inflationary
Universe
in the theories where inflation occurs near a local maximum of
effective
potential $V(\phi)$  is relatively simple; one can  easily calculate
the
fractal
dimension of the inflationary Universe  \cite{ArVil}. The
corresponding
structure in the chaotic inflation  scenario with $V(\phi) \sim
\phi^n$ or
with $V(\phi) \sim e^{\alpha\phi}$ is much more complicated,
but it is also much  more interesting.  From the point of view of
computer
simulations,  the fractal structure of the Universe in this
scenario  has more ``colors'' in it, since quantum fluctuations near
the Planck
density allow more profound changes of the vacuum state.

\section{Towards the Theory of a Stationary Universe
\label{Stationary}}

\subsection{\label{1/N} Are there Any Stationary Solutions for $P_p$?}

Now let us return to the problem of stationarity of the distribution
$P_p(\phi,t)$\@. Eq. (\ref{E40}) suggests that there are no
stationary
distributions not only  for $P_c$, but for $P_p$ as well. This
conclusion, however, may fail when  the  field $\phi$  approaches
$\phi_p$ and
density approaches the Planck density
$\rho_p \sim M^4_p = 1$\@. One of the reasons is that at $\rho \ga1$
the
probability  distribution $P_c$ becomes strongly non-Gaussian,
dispersion of fluctuations of the  field $\phi$ becomes very large
and
our derivation of eq. (\ref{E40}) does not work.     To avoid this
problem, one may try to find exact solutions of eq. (\ref{E372}).

An attempt to do this was made  in  very important papers by Nambu
and
Sasaki \cite{Nambu}.  It was claimed  that in   theories  $V(\phi)
\sim
\phi^n$ with
$n < 4$ the distribution $P_p$ is always nonstationary, but in
theories
with   steeper effective potentials ($V(\phi) \sim \phi^n$ with $n
\geq
4$, or $V(\phi) \sim e^{c\phi}$) stationary distributions for $P_p$
do
exist. However, the stationary distributions which were found in
\cite{Nambu} have their maxima at densities $\rho \gg \rho_p = 1$.
These results are  unreliable for several different reasons:

\begin{enumerate}
\item Eq. (\ref{E37}) may have a slightly different form, which
corresponds to the difference between possible definitions of
stochastic term in the  equation (\ref{m1}). This difference is not
important at densities much
smaller than $1$, but at $\rho \ga 1$ it may become significant.
(Note,
however, that this is not an unsolvable problem. One may just
investigate modified equations as well. Two other problems are more
fundamental.)

\item Diffusion equations were derived in the semiclassical
approximation
which breaks down near the Planck energy density.

\item Interpretation of the processes described by these equations is
based on the notion of classical fields in a classical space-time,
which is not applicable at densities greater than $1$ because of
large
fluctuations of metric at such densities. In particular, our
interpretation of $P_c$ and $P_p$ as of probabilities to find
classical
field $\phi$ at a given point (or in a given volume) at a given time
does not
make much
sense
at $\rho > 1$.
\end{enumerate}

Thus, the results of ref. \cite{Nambu} do not help us to establish
the
existence of stationary solutions. However, it is obvious that there
exists a
class of theories where $P_p$ is stationary even though $V(\phi)$
grows
indefinitely  at large $\phi$.  For example, one can
consider a theory with the effective potential $\sim
\lambda\phi^n\exp\Bigl({\phi\over C \phi^*}\Bigr)^2$, with $C \gg 1$.
(Similar effective potentials often appear in supergravity.) In such
a theory we have both inflation and self-reproduction of the
Universe.
However, at $\phi > C\phi^*$ inflation is impossible since the
potential is too
steep. Therefore the distribution $P_p$ will be unable to move to
$\phi > C \phi^*$, and will become stationary.

There is another, more general reason  to expect the existence of
stationary
solutions. As we will argue now, inflation tends to kill itself as
the energy
density approaches the Planck density.

In our previous investigation we assumed that the vacuum energy
density is
given by $V(\phi)$ and the energy-momentum tensor is given by
$V(\phi)\, g_{\mu\nu}$\@. However, quantum fluctuations of the scalar
field give the contribution to the average value of the energy
momentum
tensor, which does not depend on mass (for $m^2 \ll H^2$) and is
given
by \cite{b99}
\begin{equation}\label{x1}
 <T_{\mu\nu}> \, = \, {3\, H^4\over 32\, \pi^2}\ g_{\mu\nu} \, =\,
 {2 \over 3}\, V^2 \ g_{\mu\nu} \ .
\end{equation}
One of the sources of this contribution is obvious. Quantum
fluctuations of the scalar field $\phi$ freeze out with the amplitude
${H\over 2\pi}$ and the wavelength $\sim H^{-1}$\@. Thus, they lead
to
the gradient energy density $(\partial_\mu\delta\phi)^2\sim H^4$.

Note, that eq. (\ref{x1}) does  not give the total contribution of
quantum fluctuations to the energy density. When the field $\phi$
is
outside the minimum of the effective potential, additional terms
appear, in particular, the term $\sim V^\prime(\phi) \delta \phi$
(\ref{E23c}) which is responsible for galaxy formation.   At $\phi <
\phi^*$, when the density perturbations responsible for galaxy
formation were produced,  the vacuum energy renormalization
(\ref{x1})
is subdominant. On the other hand, at $\phi > \phi^*$ the
contribution corresponding to
eq.\ (\ref{x1}) becomes greater than  $V^\prime(\phi) \delta \phi
\sim
V^\prime(\phi) H/2\pi$\@. For example, in the theory $V =
m^2\phi^2/2$
\begin{equation}\label{x2}
{2\, V^2\over 3 }\, = \,
{m^4\phi^4\over 6} \, \ga \,
 {V^\prime(\phi) H\over2\pi} = 2m^3\phi^2\sqrt{\pi\over 3} \
\end{equation}
for $\phi \ga \phi^* \sim 1/\sqrt m$.

An interesting property of eq. (\ref{x1}) is that the average value
of the energy-momentum tensor of quantum fluctuations does not look
like an
energy-momentum tensor corresponding to the gradients  of a
sinusoidal wave. It
looks rather like a renormalization of the vacuum energy\--momentum
tensor (it
is proportional to $g_{\mu\nu}$)\@. This result appears after
averaging over
all short\--wavelength fluctuations and over all possible outcomes of
the
process
of generation of long\--wavelength perturbations. Something similar
happens when
one calculates the energy-momentum tensor in ordinary flat
space-time: the
contribution of each wave is not proportional to $g_{\mu\nu}$, but
the
energy-momentum tensor becomes proportional to $g_{\mu\nu}$ after
integration
over all such contributions with  the invariant measure $\sim \delta
(k^2-m^2)$.  However, if we are not integrating over all
long\--wavelength
perturbations, but treat them as a classical inhomogeneous scalar
field, then
at each particular time interval $\sim H^{-1}$ in each particular
$h$-region
the energy momentum tensor of these perturbations is mainly due to
the gradient
energy density $(\partial_\mu\delta\phi)^2$, and it is not
proportional to
$g_{\mu\nu}$.

This does not lead to any interesting effects at $V \ll 1$ ($\phi \ll
\phi_p$)  since in this case $V^2 \ll V$.
However, at the density comparable with the Planck density  the
situation
becomes much more complicated. At $V > 1$ the gradient energy density
$\sim
V^2$ becomes  greater than the potential energy density $V(\phi)$\@.
A
typical
wavelength of perturbations giving the main contribution to the
gradient
energy is given by the size of the horizon, $l \sim H^{-1}$\@. This
means that the
inflationary Universe at the Planck density becomes divided into many
domains
of the size of the horizon, density contrast between   these
domains
being of the order of one. These domains evolve as separate
mini-Universes with
the energy density dominated not by the potential energy density but
by the
 energy density of gradients of the field $\phi$\@. Such domains
drop out from the process of exponential expansion. Some of them may
 reenter this process later, but many of them  collapse into black
holes  within the typical time
$H^{-1}$  and then
evaporate. Indeed, the standard criterion for the formation of
primordial black
holes is exactly the requirement that the local increase of density
$\delta\rho$ is comparable to $\rho$ on the scale of the horizon.
This
criterion is satisfied for perturbations of density produced at
$V(\phi)\ga 1$.

Of course, one may argue that all our considerations do not make
sense at
densities greater than the Planck density. When the energy  density
in
any $h$-region approaches the Planck density, it may no longer be
described in
terms of classical space-time and should be just thrown away from our
consideration. In particular, its volume should not be considered as
contributing to the total volume of the Universe. Thus, such domains
should
be neglected in our definition of $P_p$\@. In this case the
distribution
$P_p$ for the field $\phi$ will stop growing and will
approach a stationary regime when this distribution will be shifted
towards $\phi \sim \phi_p$.

We are making an even stronger statement. Even if one makes an
attempt to
consider
the domains with $V > 1$ as a part of classical space-time,  many
such domains drop out from the process of inflation. This
means
that the total volume of
inflationary $h$-regions cannot grow as fast as
$e^{3H(\phi)t}$ for $\phi > \phi_p$.

In the models describing many different scalar fields inflation may
become
self-destructing even at an energy density somewhat smaller than the
Planck
density. Let us consider, for example,  an inflationary model
describing $N$
different scalar fields, with $N \gg 1$\@. The Hubble constant in
such a theory
is determined by the sum of all effective potentials, $V = V_1 +
V_2+...$\@.
The Planck density can be defined as a density at which all higher
order
gravitational corrections are
equally important. The one-loop correction to the energy-momentum
tensor
(\ref{x1}) in our model is directly proportional to the number of
scalar
fields $N$, but the second-loop corrections do not contain such an
enhancement. Thus, one may argue that the Planck density in our model
remains the same, $\rho_p \sim 1$\@.   On the other hand, the energy
density of
gradients of the scalar field fluctuations now is proportional to
$N$:
\begin{equation}\label{xx1}
 (\partial_\mu\delta\phi)^2 \, \sim \, N\,   V^2  \ .
\end{equation}
This means that the gradient energy density  becomes greater than
the
potential
energy density $V$ at $V \sim {1\over N} \ll 1$\@.
In the model with large $N$ this does not automatically imply  black
hole
formation, since  density perturbations will be suppressed by
$1/\sqrt N$.
However, if the energy density of any domain of a size $l \ga H^{-1}$
becomes
dominated by the gradient energy, such a domain instead of inflation
enters the
regime of a slow power-law expansion. At this stage, previously
generated
fluctuations of the scalar fields either oscillate or at least
considerably
decrease. This breaks down the standard scenario of the new scalar
field
fluctuations freezing out on the top of the previously frozen
fluctuations.

In fact, some predecessors of self-destruction of inflation show up
already
at sub-Planckian densities even at small $N$\@. When we discussed the
derivation of the diffusion equation for $P_c$, we
assumed that {\it all}\, perturbations of the wavelength $H^{-1}$
give rise to
inflationary domains. However, this is not the case for very large
fluctuations of the
field $\phi$ with  the amplitude of a fluctuation  $\delta\phi > 1$.
Such fluctuations
should be present if, as we assumed, the term $\xi(t)$ in the
Langevin equation
(\ref{m1}) corresponds to the white noise. The gradient energy of the
perturbations of the wavelength $H^{-1}$ with the amplitude
$\delta\phi > 1$ is
given by ${1\over 2}\, (\partial_\mu\delta\phi)^2 > H^2 \sim
V(\phi)$\@. Domains where
such jumps occur may drop out of the process of exponential
expansion. Suppression of
inflation in the domains of a size $H^{-1}$ produced by large jumps
effectively
implies that the effective noise in the Langevin equation is not
entirely white.\footnote{A more adequate
statement would be that the noise remains
white, but when
it occasionally becomes ``too loud'', it  destroys Brownian
particles.}

The probability of such a large jump within the time $H^{-1}$ is
suppressed by
$\exp\left(-{2\pi^2 \delta\phi^2\over H^2}\right)$\@. This indicates
that at $H \ll
1$ (i.e. at sub-Planckian densities) the cut-off of noise
corresponding to  fluctuations
with the amplitude $\delta\phi \ga 1$ is not very important, since
the probability of
such fluctuations is exponentially  suppressed. However, the
closer we are to the
Planck density, the stronger is the deviation of $\xi(t)$ from the
white
noise, the less efficient is the
stochastic process producing new inflationary domains.

What are the possible consequences of this effect? First of all, we
know
that if the initial value of the field $\phi$ is in the interval
$\phi^* <
\phi < \phi_p$, then the distribution $P_p(\phi,t)$ moves towards
greater and
greater $\phi$\@. The only reason why it happens, despite the fact
that
the
distribution $P_c(\phi,t)$ moves towards small $\phi$, is an
additional
exponential growth of the number  of inflationary $h$-regions due to
expansion of the Universe. If, as we argued, the
number of inflationary $h$-regions with  $\phi \sim \phi_p$ grows at
a much
slower rate (or does not grow at all, which is the case if we just do
not
consider domains with $\phi > \phi_p$ as belonging to our classical
space-time), then the distribution $P_p(\phi,t)$ stops moving towards
large
$\phi$ as soon as the maximum of this distribution approaches
$\phi_p$\@. Thus,
the distribution $P_p(\phi,t)$ may approach some stationary regime,
being
concentrated at sub-Planckian densities. Now let us study this
possibility at a more quantitative level.

\subsection{\label{Branching}\sloppy Self-Reproduction   of Inflationary
Domains as a  Branching   Diffusion Process}

When describing the process of self-reproduction of inflationary
domains one
should keep in mind that this is not an ordinary diffusion process,
but
process where diffusion of the scalar field in each particular
$h$-region is
accompanied with their branching into many independent $h$-regions.
As was argued in  \cite{MezhMolch}, a good candidate for the
mathematical model
describing such behavior is the   theory of   branching diffusion
processes
\cite
{Watanabe}.

 There are two main sets of
questions which may be asked concerning such processes. First of all,
one may
be interested in the probability $P_p(\phi, t|\phi_0)$ to find a
given field
$\phi$ at a
given time $t$ under the condition that initial value of the field
was equal to
some $\phi_0 = \phi(t=0) $\@. In what follows we will denote $\phi_0$
as $\chi$\@.

On the other hand, one may wish to know, what is the probability
$P_p(\phi,t|\chi)$  that the given final value of the field $\phi$
appeared as
a process of diffusion and
branching of
a domain containing some field $\chi$\@. Or, more generally, what are
the typical
properties of branching Brownian trajectories which end up at a
hypersurface
of a given $\phi$?

In the end of the Section \ref{Stochastic} we have shown that the
probability to find a
given field $\phi$  {\it in a given volume} moves towards Planckian
densities with the growth of time $t$, if $\phi_0 > \phi^*$\@. We
have
shown also
that a typical trajectory producing domains of a given density for a
long  time
fluctuates at very large $\phi$, close to  the Planck density. Thus,
we obtained
qualitative answers for the two questions mentioned above. To obtain
a more
quantitative description of the process of self-reproduction of the
Universe, we
will first develop  more general methods of investigation using the
simplest characteristics of the branching diffusion processes which
involve only those parameters of the model \cite{MezhMolch} that can
be
unambiguously identified at this stage of our studies.

Many properties of a branching diffusion process
 can be reproduced with the help of the following function
\begin{equation} \label{eq3}
 u(\chi, t, {\cal D})\,  = \  \left\langle\left. {\mu(t, {\cal
D})}  \right| \, \phi_0 =
\chi \right\rangle  \ .
\end{equation}
Here $\mu(t, {\cal D})$ is the number of ``particles'' with the
coordinates $x$  in
the interval \, ${\cal D}$ at the moment \, $t$\@.   In our case the
analog of a particle with a coordinate $x$ is the $h$-region with a
given field
$\phi$\@. The interval of all possible values of $\phi$ corresponding
to inflation
in classical space-time is ${\cal D}_{max}=[\phi_{e}, \phi_{p}]$.
However, one may
choose to consider as ${\cal D}$ in eq. (\ref{eq3}) any smaller
segment
of the
interval $[\phi_{e}, \phi_{p}]$\@.

By $\left\langle \left. {\mu(t, {\cal D})}  \right| \,
\phi_0 =
\chi \right\rangle$    we denoted the mathematical
expectation value (or, in other words, the mean value) of the number
of
``particles'' in the interval  ${\cal D}$ at the moment $t$
under the condition that initially there was one ``particle'' with
$\phi_0 = \chi$.

The last condition implies that
\begin{equation} \label{eq13}
u(\chi, 0, {\cal D}) = \left\{ \begin{array}{ll}
                 1, & \mbox{if $ \chi \in {\cal D}$} \ , \\
                 0, & \mbox{otherwise} \ .
                 \end{array}    \right.
\end{equation}

There is some subtlety in definition of the  `number of
particles' $\mu(t, {\cal D})$ in the stochastic approach to
inflation. Depending on the way one formulates the problem,
this may be either the {\it number}\,  ${\cal N}(t, {\cal D})$ of
$h$-regions or their total  {\it
volume}\,  ${\cal V}(t, {\cal D})$ divided by some normalization
factor $H^{-3}(\phi_0)$.\footnote{We denote volume by  ${\cal V}$ to
distinguish it from the effective potential $V$.} The
difference between these two quantities stems from the fact that the
total volume
of each $h$-region is proportional to  $H^{-3}(\phi)$, where $\phi$
is the local
value of the scalar field in a given $h$-region. This difference
typically is not
very important: it is proportional to the third power of
${H(\phi)\over
H(\phi_0)}$, whereas most of the distributions we will obtain depend
on
$H(\phi)$ exponentially. Nevertheless, whenever appropriate, we will
make a distinction between these two understandings of $\mu(t, {\cal
D})$\@. In what follows, we will understand by $\mu(t, {\cal D})$ the
volume ${\cal V}(t, {\cal D})$ of
the $h$-regions with the scalar field $\phi \in {\cal D}$\@.  (For an
equation
describing the number of $h$-regions ${\cal N}$ see
also \cite{Mijic}).

Now we will try to find an equation describing the time evolution of
$u(\chi, t, {\cal D})$. The space-time structure of the branching
diffusion process may be represented by a growing tree of world lines
of diffusing particles. The section of the tree at some level
represents
the spatial distribution of the particles at the corresponding moment
of
time, while the increasing number of ``branches'' corresponds to the
increasing number of particles due to the processes of ``splitting''
of
one ``parent'' particle into several ``daughter'' ones. Here we
produce
a derivation of the reduced form of the equation describing the
growth
of this tree \cite{MezhMolch} in a way which does not involve any
new model parameters in
addition to those already presenting in the chaotic inflation.

Consider the whole tree of the branching diffusion
process from $t=0$ to $t + \Delta t$ and divide it into two parts ---
a part from $t=0$ to $t = \Delta t$ and a part from $t = \Delta t$ to
$t + \Delta t$\@.  After the time $\Delta t$, the original domain
filled by the scalar field $\chi =
\phi_0 = \phi(t = 0)$ becomes filled by the slightly inhomogeneous
field
$\phi(\Delta t)$, and its volume grows $\exp({3H(\chi)\Delta t)}$
times. To first
order in $\Delta t$, the new domain can be represented as a
combination of two
domains: the domain of original volume ${\cal V}$ containing the
field
$\phi(\Delta t)$, plus the new domain of the volume ${\cal V} \cdot
3H(\chi)\Delta t$\@ with an unchanged field $\phi = \chi$. Very soon,
after the
characteristic time $\sim H^{-1}$,
the interior of the second domain will loose any contact with the
interior of the first
one (``no-hair'' theorem for de Sitter space), and its subsequent
evolution would
proceed totally independently. This means that when we  take averages
calculating $u( \chi, t+ \Delta t, {\cal D})$, they  split into two
parts (in the first order in $\Delta t$):
\begin{equation} \label{eq4}
u(\chi, t+ \Delta t, {\cal D}) =\left\langle \left. {u(\phi(\Delta
t),  t,
{\cal
D})} \right| \,
\phi_0 =
\chi \right\rangle  + 3H(\chi) \Delta t \cdot u(\chi, t, {\cal D}) \
{}.
\end{equation}
This yields
\begin{equation} \label{eq4'}
\frac{\partial }{\partial t} u(\chi, t, {\cal D}) = 3H(\chi)  \cdot
u(\chi, t, {\cal D}) +  \lim_{\Delta t \rightarrow 0}\frac{\langle
\left.  u(\phi(\Delta t), t, {\cal D}) - u(\chi, t,
{\cal D})
\right| \, \phi_0 = \chi \rangle}{\Delta t} \ .
\end{equation}

According to the theory of stochastic  processes \cite{Watanabe}, the
last term in this equation can be represented in the following way:
\begin{equation}      \label{eq5}
\lim_{\Delta t \rightarrow 0}
\frac{\langle \left.  u(\phi(\Delta t), t, {\cal D}) - u(\chi, t,
{\cal D})
\right| \, \phi_0 = \chi \rangle}{\Delta t} = \hat{A}
{\mbox{$u(\chi,t,z,{\cal
D})$}}\ ,
\end{equation}
where the operator $\hat{A}$ (the generating operator of diffusion)
can be constructed in a standard way if the corresponding Langevin
equation (\ref{m1}) is known:
\begin{equation} \label{eq6}
 \hat{A} f(\chi)=\frac{H^{3/2}(\chi)}{8\pi^2} \frac{\partial
}{\partial\chi}
 \left( H^{3/2}(\chi)\, \frac{\partial  f(\chi)}{\partial\chi}
\right)
 - \frac{V'(\chi)}{3H(\chi)} \frac{\partial  f(\chi)}{\partial\chi} \
{}.
\end{equation}
Thus, the  function $u(\chi, t, {\cal D})$ satisfies the following
equation
\begin{equation} \label{eq7}
\frac{\partial }{\partial t}\, u(\chi, t, {\cal D}) = \hat{A} \,
u(\chi, t, {\cal D}) +
3H(\chi)  \cdot  u(\chi, t, {\cal D}) \ .
\end{equation}

As we have discussed, in our approach this function gives the
expectation value of the volume
${\cal V}$ of all $h$-regions with the scalar field $\phi \in {\cal
D}$\@. Now we will consider the interval ${\cal D}(\phi) = [\phi_{e},
\phi]$ and define the probability
distribution
\begin{equation} \label{eq11}
P_p(\phi,t|\chi) = \frac{\partial \, u(\chi, t, {\cal
D}(\phi))}{\partial \phi}  \ .
\end{equation}

One can easily understand that this is the same (unnormalized)
distribution
$P_p(\phi,t)$ which we studied in the previous section, but from now
on we will
also keep track of the dependence of this function upon $\phi_0 =
\chi$\@. In
particular, according to eqs. (\ref{eq7}), (\ref{eq11}), this
function satisfies equation
\begin{equation} \label{eq11aau}
\frac{\partial }{\partial t}\, P_p(\phi,t|\chi) = \hat{A} \,
P_p(\phi,t|\chi) +
3H(\chi)  \cdot P_p(\phi,t|\chi)  \ ,
\end{equation}
or, in an expanded form,
\begin{equation} \label{eq11aa}
\frac{\partial  P_p(\phi,t|\chi)}{\partial t} =
\frac{H^{3/2}(\chi)}{8\pi^2} \frac{\partial }{\partial\chi}
 \left( H^{3/2}(\chi) \frac{\partial  P_p(\phi,t|\chi)
}{\partial\chi}\right)
 - \frac{V'(\chi)}{3H(\chi)} \frac{\partial
P_p(\phi,t|\chi)}{\partial\chi}  +
3H(\chi)  \cdot P_p(\phi,t|\chi)  \ ,
\end{equation}
This is the branching diffusion analog of the backward Kolmogorov
equation known for the ordinary diffusion. The
probability distribution  $P_p(\phi,t|\chi) $\,
satisfies also the forward Kolmogorov equation (i.e.\ Fokker-Planck
equation)
\begin{equation} \label{eq11aaa}
\frac{\partial }{\partial t}\, P_p(\phi,t|\chi) = \hat{A}^{\dag} \,
P_p(\phi,t|\chi) +
3H(\phi)  \cdot P_p(\phi,t|\chi)  \ ,
\end{equation}
where $\hat{A}^{\dag}$ is the operator which is adjoint to the
diffusion generating
operator $\hat{A}$ (\ref{eq6}) \cite{Watanabe}. From the standard
definition of
the adjoint operator,
\begin{equation}\label{1a1}
\int F(\phi)\hat{A}\, f(\phi)\, d\phi =
\int \{\hat{A}^{\dag}\,F(\phi)\} f(\phi)\, d\phi  \ ,
\end{equation}
  one can easily obtain the
following expression for $\hat{A}^{\dag}$
\begin{equation}\label{eq19}
 \hat{A}^{\dag} f(\phi)=\frac{\partial }{\partial\phi}
 \left( {H^{3/2}(\phi)\,\over 8\pi^2} \frac{\partial }{\partial\phi}
\left(
 {H^{3/2}(\phi)}f(\phi) \right)
 +  \frac{V'(\phi)}{3H(\phi)} \, f(\phi) \right) \ .
\end{equation}
Thus, the forward Kolmogorov equation is
\begin{equation} \label{eq11aaaa}
\frac{\partial P_p(\phi,t|\chi)}{\partial t}  = \frac{\partial
}{\partial\phi}
 \left( {H^{3/2}\over 8\pi^2} \frac{\partial }{\partial\phi} \left(
 {H^{3/2}}P_p(\phi,t|\chi) \right)
 +  \frac{V'}{3H} \, P_p(\phi,t|\chi) \right)
 +  3H \cdot P_p(\phi,t|\chi)  \ .
\end{equation}
This equation coincides with the equation (\ref{E372}),
which was
earlier derived in a different way.

\subsection{Boundary conditions} \label{Boundary}

We will give now a brief discussion of boundary conditions imposed on
 $P_p(\phi,t|\chi)$\@. A more
detailed discussion will be given in the forthcoming paper
\cite{MezhLin}\@.

As we will see,  in many models the distribution  $P_p(\phi,t|\chi)$
is sharply
peaked at a very large field $\phi  \sim \phi_p \gg \phi_e$, and the
maximal
value of $P_p(\phi,t|\chi)$  is  much greater than
$P_p(\phi_e,t|\chi)$. In
such models the form of boundary conditions at $\phi_e$
is almost irrelevant, and we may simply assume that
\begin{equation}\label{eq8}
P_p(\phi_e,t|\chi) = \left.\frac{\partial \, u(\chi, t, {\cal
D}(\phi))}{\partial \phi}\right|_{\phi_e} = 0 \ .
\end{equation}
A similar condition can be imposed on $P_p(\phi,t|\chi)$ at $\chi =
\phi_e$.

However, this simple trick does not have a universal validity for all
potentials $V(\phi)$ and all possible time parametrizations.
Therefore we
should find a more general way to impose boundary conditions at
$\phi_e$.

Note that at the boundary $ \phi = \phi_e$ inflation ends, there
is no diffusion back from the region $\phi < \phi_e$, and the field
$\phi$ in the domain with $\phi < \phi_e$ continues rolling down to
  even smaller  values of $\phi$\@. The best way to describe it is to
say that
the
diffusion coefficient vanishes at $\phi < \phi_e$. This means that
the form of
the diffusion equation changes discontinuously at $ \phi = \phi_e$.
However,
neither the probability distribution nor the probability current can
be
discontinuous at this point.

The simplest way to describe this situation is to consider first the
distribution $P_c$, for which we have the probability conservation
equations
(\ref{E37111a}) and (\ref{E37111b}). (For a description of this
method
see \cite{Gardiner}.)

The continuity condition  for the probability distribution $P_c$ and
the
probability current $J_c$ at $\phi_e$ can be written as follows:
\begin{equation} P_c({\phi_{e_+}}) = P_c({\phi_{e_-}}); ~~~
J_c({\phi_{e_+}})=
J_c({\phi_{e_-}})\ .
\end{equation}
 The notations $\phi_{e_+}$ and $\phi_{e_-}$ are used to show that we
may
approach $\phi_e$ either from $\phi > \phi_e$ or from   $\phi <
\phi_e$. The
last condition can be written as
\begin{equation}\label{curr}
\left.V^{3(1-\beta)/2}(\phi) \
 \frac{\partial}{\partial
\phi}\left({V^{3\beta/2}(\phi)}P_c\right)\right|_{{\phi_{e_+}}} +
{3V'(\phi)\over
8V^{1/2}(\phi)}\, P_c({\phi_{e_+}})
  =  {3V'(\phi)\over 8V^{1/2}(\phi)}\,
P_c({\phi_{e_-}}) \ .
\end{equation}
 Here we have  used our assumption that  diffusion does not
contribute to $J_c$
at $\phi < \phi_e$. For generality we restored the parameter $\beta$,
corresponding to the ambiguity of the definition of stochastic force.

Using the continuity condition $P_c({\phi_{e_+}})
= P_c({\phi_{e_-}})$, we obtain the following boundary condition for
$P_c$:
\begin{equation}
\left.\frac{\partial}{\partial
\phi}\left({V^{3\beta/2}(\phi)}P_c\right)\right|_{{\phi_{e_+}}}=0 \ ,
\end{equation}
or, equivalently,
\begin{equation}\label{beta}
{P'_c({\phi_{e_+}})\over P_c({\phi_{e_+}})} = - {3\beta\over 2}\
{V'\over V} \
{}.
\end{equation}
 The stationary solutions we are going to obtain will not be very
sensitive to
the choice of $\beta$, unless we choose it extremely large. In the
particular
case $\beta = {1\over 2}$,  which we consider throughout the paper,
the
boundary condition is
\begin{equation}
{P'_c({\phi_{e_+}})\over
P_c({\phi_{e_+}})} = - {3\over 4}\ {V'\over V} \ .
\end{equation}

 A similar approach can be used for the case of the probability
distribution
$P_p$. Here we have a slight complication, since the analog of the
probability
current in this case is not conserved due to the production of new
$h$-regions.
This is precisely the reason for the appearance of  the additional
term $3HP_p$
in the diffusion equation. However, one can derive the same boundary
conditions
in this case as well. In order to do it, one should just integrate
equation
(\ref{E371}) over $\phi$ in the infinitesimal interval from
${\phi_{e_-}}$ to
${\phi_{e_+}}$. Since the term $3HP_p$ is not singular in this
interval, we
recover eq. (\ref{curr}) for $P_p$, and finally obtain the same
boundary
condition as before,
\begin{equation}\label{bound}
\frac{\partial}{\partial\phi}P_p({\phi_{e_+}}) = - {3\over 4}\
{V'\over V} \
P_p({\phi_{e_+}})
 \ .
\end{equation}

 Finally, we should fix the boundary conditions for the backward
Kolmogorov
equation  (\ref{eq11aa}) at $\chi = \chi_e \equiv  \phi_e$. There
exists a
regular way to reconstruct these boundary conditions from the
boundary
conditions for the forward Kolmogorov equation  (\ref{bound})
\cite{Gardiner}.
One should accurately reconstruct the diffusion operator $\hat{A}$
from its
adjoint $\hat{A}^{\dag}$ (\ref{eq19}), taking into account that the
functions
on which these operators act may not disappear at the boundaries.
This will
produce additional boundary terms in the expression for $\hat{A}$.
Then one
should take such boundary conditions for the backward Kolmogorov
equation,
which would make  these additional terms vanish. As we will show in a
subsequent publication,   the resulting boundary condition has a very
interesting and general form, which does not depend either on the
parameter
$\beta$ or on a particular choice of time
parametrization:
\begin{equation}\label{TunnBound}
\left.\frac{\partial
}{\partial\chi}\left( P_p\  \exp\Bigl({3\over
8V(\chi)}\Bigr)\right)\right|_{\chi_{\phantom{}_{e_+}}} = 0 \ .
\end{equation}
This
equation implies that the probability distribution near $\chi_e$ is
always
given by the square of the tunneling wave function,
\begin{equation}\label{TunnSolution}
P_p(\chi \sim \chi_{e})  \sim
\exp\Bigl(-{3\over 8V(\chi)}\Bigr)  \ .
\end{equation}
In fact, as we will see soon, in some models to be considered this
remarkable
relation holds (in a sense to be discussed later)  in a very large
region,
almost up to the Planck boundary $\chi \sim \phi_p$, and this result
remains
valid even if we considerably modify the boundary condition
(\ref{TunnBound}).

Now we should discuss the boundary conditions near the Planck
boundary, where
$V(\phi_p) \sim 1$.
As we have shown in the  Section \ref{1/N},  the situation at this
boundary is
much more complicated and ambiguous. Description of evolution of the
scalar field at $\phi > \phi_p$ in terms of classical space-time is
impossible, and our diffusion equations do not make much sense there.
There are several possibilities to be considered.

\begin{enumerate}
\item First of all, we may exclude from our investigation all
$h$-regions which
jump into the state with $\phi > \phi_p$, since we cannot study their
evolution
 in any case. In other words, we may restrict our attention to the
part of space which {\it at all times}\, $t > 0$ remains classical.
There are
several formal ways to do so. The simplest one is to introduce a
phenomenological description of destruction of all domains which jump
to
the region $\phi > \phi_p$. This trick will effectively remove these
domains
from our consideration. This can be done by the same way we
introduced
branching. To do it, we added the term  $3H(\phi) P_p$ to the
diffusion
equation. Now we may, for example, multiply this term by a function
$F(\phi)$, which is equal to $1$ for $\phi < \phi_p$, and which
becomes
large and negative for $\phi > \phi_p$. By doing so, we do not
prohibit the
diffusion process in the region $\phi > \phi_p$, but we discard all
domains
which diffuse there.

\item  There is also another possibility which we discussed in the
previous
section: inflation may lead to a self-destruction of inflationary
domains with
 $\phi > \phi_p$\@. Phenomenological description of this regime may
be
achieved by the same trick as in the previous case, except for the
behavior of the function $F(\phi)$ at $\phi > \phi_p$ may be less
dramatic; it may just introduce a small coefficient in front of
$3H(\phi)$.
However, if this coefficient is small enough, the final effect will
be
basically the same: it makes the distribution $P_p$ stationary and
concentrated
at $\phi \la \phi_p$. This would also correspond to something like an
absorbing   boundary conditions (or, more generally, to ``elastic
screen''
type boundary conditions, which are intermediate between  the
absorbing and
reflecting ones)\@.

\item  Finally, one should take into account that in the realistic
theories
the effective potential may be steep at large $\phi$. We have already
discussed
one of such possibilities, which appears if the effective potential
is $\sim
\lambda\phi^n\exp\Bigl({\phi\over C \phi^*}\Bigr)^2$, with $C > 1$.
In such a
theory
we have both inflation and self-reproduction of the Universe.
However, at
$\phi > C\phi^*$ inflation is impossible since the potential is too
steep. Therefore the distribution $P_p$ will be unable to move to
$\phi > C \phi^*$ and will become stationary.
\end{enumerate}

There may be other reasons why inflation cannot penetrate into the
region $\rho
> 1$. It well may be so that space-time with $\rho > 1$ simply cannot
exist
\cite{Markov,Brand}, or that the stringy nature of interactions does
not allow
us to penetrate to distances smaller than the Planck scale
\cite{Gross,Vafa}.
Whatever the reasons are, to describe their phenomenological
consequences for
the distribution $P_p$ one should either add some terms,
corresponding to
destruction of inflationary domains to the diffusion equation, or to
modify the
effective potential $V$, the diffusion coefficient $\cal D$ and the
mobility coefficient $\kappa$ at $\phi >\phi_p$, or to impose some
kind of
absorbing or reflecting boundary conditions so as to prevent
penetration
of the fluctuating field  into the domain with $\phi \gg \phi_p$.

 Our investigation  has shown \cite{MezhLin} that the final results
for the
distribution $P_p$ are not very sensitive to the method one uses   to
prevent
penetration of   the  field $\phi$ into the domain with $\phi \gg
\phi_p$ and
on the type of the boundary conditions imposed (whether they are
absorbing,
reflecting, etc.)\@. They depend only on the value of the field $\phi
=
\phi_b$, where the boundary conditions are to be imposed, and this
dependence
is rather trivial. Our arguments suggest that $\phi_b \sim
\phi_p$\@. In what follows, when solving the diffusion equations we
will simply
assume  that the
functions $u(\chi, t, {\cal D}(\phi))$ and $P_p(\phi,t|\chi)$ satisfy
absorbing
boundary conditions at $\phi=\phi_p$, where $V(\phi_p)=1$. In
particular, we
will assume that
\begin{equation} \label{eq8a} P_p(\phi_p,t|\chi)  = 0 \ ,
\end{equation}
where $V(\phi_p)=1$. A similar condition should be imposed at the
Planck
boundary for the field $\chi$:
\begin{equation} \label{eq8aa}
P_p(\phi,t|\chi_p)  = 0 \ ,
\end{equation}
where $\chi_p \equiv \phi_p$. However, we will keep in mind  for
future
discussion that, strictly speaking,  $\phi_p$ in these
equations is just a phenomenological parameter, not necessarily
corresponding
to $V(\phi_p)=1$\@.  A detailed investigation of  different boundary
conditions
near the Planck boundary will be contained in \cite{MezhLin}.

\subsection{Stationary Solutions}

There are several ways to proceed further. One may try to obtain
solutions of equations (\ref{eq11aa}) and (\ref{eq11aaaa}) directly.
We  prefer a different strategy, looking for a solution of eqs.
(\ref{eq11aa}) and (\ref{eq11aaaa}) in a form of
 the following series of biorthonormal system of eigenfunctions
of the pair of adjoint linear operators
$\hat{A} + 3 H$ and $\hat{A}^{\dag} + 3 H$:
\begin{equation} \label{eq14}
P_p(\phi,t|\chi) =
\sum_{s=1}^{\infty} { e^{\lambda_s t}\, \psi_s(\chi)\, \pi_s(\phi) }
\ .
\end{equation}
Indeed, this gives us a solution of eqs. (\ref{eq11aa}) and
(\ref{eq11aaaa}) if
\begin{equation} \label{eq15}
 \frac{1}{2} \frac{H^{3/2}(\chi)}{2\pi} \frac{d
}{d\chi}
 \left( \frac{H^{3/2}(\chi)}{2\pi} \frac{d }{d\chi}
\psi_s(\chi) \right)
 - \frac{V'(\chi)}{3H(\chi)} \frac{d }{d\chi} \psi_s(\chi)
 + 3H(\chi)  \cdot \psi_s(\chi) =
\lambda_s \, \psi_s(\chi)  \ .
\end{equation}
and\begin{equation} \label{eq17}
\frac{1}{2}  \frac{d }{d\phi}
 \left( \frac{H^{3/2}(\phi)}{2\pi} \frac{d }{d\phi}
\left(
 \frac{H^{3/2}(\phi)}{2\pi} \pi_j(\phi) \right) \right)
 + \frac{d }{d\phi} \left( \frac{V'(\phi)}{3H(\phi)} \,
\pi_j(\phi) \right)
 + 3H(\phi)  \cdot \pi_j(\phi) =
\lambda_j \, \pi_j(\phi)  \ .
\end{equation}
The orthonormality condition reads
\begin{equation} \label{eq20}
\int_{\phi_e}^{\phi_p} { \psi_s(\chi) \, \pi_j(\chi) \, d\chi }
= \delta_{sj}\ .
\end{equation}

In our case (with regular boundary conditions) one can easily show
that the
spectrum of $\lambda_j$ is discrete and bounded from above. Therefore
the
asymptotic solution for $P_p(\phi,t|\chi)$ (in the limit $t
\rightarrow \infty$) is given by
\begin{equation} \label{eq22}
P_p(\phi,t|\chi) = e^{\lambda_1 t}\, \psi_1(\chi) \,
\pi_1(\phi)\, \cdot \left(1 + O\left( e^{-\left(\lambda_1 - \lambda_2
\right) t} \right) \right) \ .
\end{equation}
Here $\psi_1(\chi)$ is the only positive eigenfunction of eq.
(\ref{eq15}),
$\lambda_1$ is the corresponding (real) eigenvalue, and $\pi_1(\phi)$
(invariant density of branching diffusion) is the eigenfunction
of the conjugate operator (\ref{eq17}) with the
same eigenvalue $\lambda_1$\@. Note, that $\lambda_1$ is
the largest eigenvalue, $\mbox{Re} \left( \lambda_1 - \lambda_2
\right) > 0 $\@. This is the reason why the asymptotic equation
(\ref{eq22}) is valid at large $t$\@. We found \cite{MezhLin} that
in realistic theories of inflation a typical time of
relaxing to the asymptotic
regime, $\Delta t \sim (\lambda_1 - \lambda_2)^{-1}$, is extremely
small. It
is only about a few thousands Planck times, i.e. about
$10^{-40}~sec$\@.

Now we see that the average volume of the Universe
filled by  the inflaton field $\phi$ in the interval ${\cal
D}=[\phi_{e}, \phi_{p}]$,
grows exponentially. And the exponent $e^{\lambda_1 t}$
does not depend on $\phi$ and $\chi$.
This means, that the normalized distribution
\begin{equation} \label{eq22aa}
\tilde{P}_p(\phi,t|\chi) = e^{-\lambda_1 t}
\,P_p(\phi,t|\chi)
\end{equation}
rapidly converges to the time-independent  distribution
\begin{equation} \label{eq22ax}
 \tilde{P}_p(\phi,t \rightarrow \infty|\chi) =  \psi_1(\chi) \,
\pi_1(\phi) \ .
\end{equation}
It is this stationary distribution that we were looking for. The
remaining problem is to find the functions $\psi_1(\chi)$ and
$\pi_1(\phi)$,
and to check that all assumptions about the boundary conditions which
 we made on the way to eq. (\ref{eq22}) are actually satisfied. The
boundary
conditions on $P_p$ in terms of $\pi_1$ and $\phi_1$ can be written
as
follows:
\begin{equation}\label{TunnBoundPsi} \psi'_1({\chi_{e}})  =  {3\over
8}\ {V'\over V^2} \cdot \psi_1({\chi_{e}}) \ ,~~~~~ \psi_1(\chi_p) =
0\ ,
\end{equation}
and
\begin{equation}\label{TunnBoundPi}
 \pi'_1({\phi_{e}})  = -\, {3\over 4}\ {V'\over V} \cdot
\pi_1({\phi_{e}})\
,~~~~~ \pi_1(\phi_p) = 0\  .
\end{equation}

 The general program of finding solutions for $\psi_1$ and $\pi_1$
for a wide
class of inflationary models will be pursued in \cite{MezhLin}. In
what follows
we will present   solutions of these equations for the  theories $V =
{\lambda\over 4} \phi^4$ and $V = V_0\,e^{\alpha\phi}$.

 Equations (\ref{eq15}), (\ref{eq17}) for $\psi_1(\chi)$ and
$\pi_1(\phi)$ in
the theory ${\lambda\over 4} \phi^4$ look as follows:
\begin{equation}\label{r}
 \psi_1'' - \psi_1'\Bigl({6\over\lambda\chi^5} - {3\over\chi}\Bigr) +
\psi_1
\Bigl({36\pi \over\lambda\chi^4}  -{\lambda_1\over \pi \chi^6}
\Bigl({6\pi\over\lambda}\Bigr)^{3/2}\Bigr) = 0\ ,
\end{equation}
\begin{equation}\label{r1}
\pi_1'' + \pi_1'\Bigl({6\over\lambda\phi^5} +
{9\over\phi}\Bigr) + \pi_1 \Bigl({6\over\lambda\phi^6} +
{15\over\phi^2} +
{36\pi\over\lambda\phi^4} -{\lambda_1\over \pi \phi^6}
\Bigl({6\pi\over\lambda}\Bigr)^{3/2}\Bigr) = 0\ .
\end{equation}
The corresponding boundary conditions  (\ref{TunnBoundPsi}) and
(\ref{TunnBoundPi}) for this theory are
\begin{equation}\label{TunnBoundPsi1}
\psi'_1({\chi_{e}})  =  {6\over \lambda\chi_e^5}\,
\psi_1({\chi_{e}}) \ ,~~~~~
\psi_1(\chi_p) = 0\ ,
\end{equation}
and
\begin{equation}\label{TunnBoundPi1}
\pi'_1({\phi_{e}})  = -\,
{3\over \phi_e}\, \pi_1({\phi_{e}})\ ,~~~~~ \pi_1(\phi_p) = 0\  ,
\end{equation}
where we take for definiteness $\phi_e = \chi_e =0.3$ and
$\phi_p = \chi_p = \Bigl({4\over \lambda}\Bigr)^{1/4}$. (Note the
difference
between the coupling constant $\lambda$ and the eigenvalue
$\lambda_1$.)

The analytic solution of this equation  shows that in the limit of
small
$\lambda$ the eigenvalue $\lambda_1 = 2\sqrt{6\pi} \approx 8.681$
\cite{MezhLin}, if one identifies the upper boundary $\phi_p$ with
the value of
$\phi$ at which $V(\phi) = 1$. However, this limit is approached very
slowly.
We  solved this equation numerically with the boundary conditions
(\ref{TunnBoundPsi1}), (\ref{TunnBoundPi1}).  We have found the
eigenvalues
$\lambda_1$
corresponding to different values of the coupling constant $\lambda$:
\begin{center}
\begin{tabular}{|c|c|c|c|c|c|c|c|}
\hline \hline
 $\lambda$ & $1$ & $10^{-1}$ & $10^{-2}$ & $10^{-3}$ & $10^{-4}$ &
$10^{-5}$ & $10^{-6}$ \\
\hline
 $\lambda_1$ & 2.813 & 4.418 & 5.543 & 6.405 & 7.057 & 7.538 & 7.885
\\ \hline \hline
\end{tabular}
\end{center}

As a test of self-consistency of our approach, one may solve one of
the
equations first, find the eigenvalue $\lambda_1$, and then check that
the same
eigenvalue can be obtained from the second equation with the boundary
conditions specified above. We performed this check, and confirmed
our results.

One can find also the second eigenvalue $\lambda_2$. For example, for
$\lambda = 10^{-4}$ one gets  $\lambda_2=6.789$. This means that for
$\lambda =
10^{-4}$ the time of relaxation  to the stationary distribution is
$\Delta t
\sim (\lambda_1-\lambda_2)^{-1} \sim 4 M_p^{-1} \sim 10^{-42}$
seconds --- a
 very short time indeed. One should note, however, that the complete
time for
establishing the stationary distribution depends on initial
conditions, and in
some cases it may be much longer than $(\lambda_1-\lambda_2)^{-1}$
\cite{MezhLin}.

Note that the parameter $\lambda_1$ shows the rate of exponential
expansion of
the volume filled by a given field $\phi$. {\it This rate does not
depend on
the field $\phi$}, and has the same order of magnitude as the rate
of
expansion at the Planck density. Indeed, $\lambda_1$ should be
compared to
$3H(\phi)= 2\sqrt{6\pi V(\phi)}$, which is  equal to $2\sqrt{6\pi}$
at the
Planck density. As we already mentioned,  in the limit
$\lambda
\to 0$ the eigenvalue $\lambda_1$ also becomes equal to
$2\sqrt{6\pi} \approx
8.681$. The meaning of this result is very simple: in the limit
$\lambda \to 0$
our solution becomes completely concentrated near the Planck
boundary, and
$\lambda_1$ becomes equal to $3H(\phi_p)$.

At  first glance, independence of the rate of expansion of volume
$e^{\lambda_1 t}$ on the value of the field $\phi$ may seem
counterintuitive.
The meaning of this result is that the domain filled with the field
$\phi$
gives the largest contribution to the growing volume of the Universe
if it
first diffuses towards the Planckian densities, spends there as long
time as
possible {\it expanding with  nearly Planckian rate}, and then
diffuses back
to its original value $\phi$.

But what  about the field $\phi$ which is already at the Planck
boundary? Why
do the corresponding domains not  grow exactly with the Planckian
Hubble
constant $H(\phi_p) = 2\sqrt{6\pi}/3$\,? It happens partially due to
diffusion
and slow rolling of the field towards smaller $\phi$. However, the
leading
effect  is the destructive diffusion towards the space-time foam with
$\phi >
\phi_p$. One may  visualize this process  by painting white all
domains with
$V(\phi) < 1$, and by painting black domains filled by space-time
foam with
$V(\phi) > 1$.  Then each time $H^{-1}(\phi_p)$ the volume of white
domains
with $\phi \sim \phi_p$ grows approximately $e^3$ times, but some
`black holes'
appear in these domains, and, as a result, the total volume of white
domains
increases only $e^{3\lambda_1/2\sqrt{6\pi}}$ times. This suggests (by
analogy
with \cite{ArVil})   calling the factor $d_{fr} =
3\lambda_1/2\sqrt{6\pi}$
`the fractal dimension of classical space-time', or  `the fractal
dimension of
the inflationary Universe'. (Note that
$d_{fr} < 3$ for $\lambda \not = 0$; for example, $d_{fr} = 2.6$ for
$\lambda =10^{-5}$.)
However, one should keep in mind that the fractal structure of the
inflationary
Universe in the chaotic inflation scenario in general is more
complicated than
in the new or old inflation and cannot be completely specified just
by one
fractal dimension \cite{MezhLin}.

The numerical solution for $\pi_1(\phi)$   is shown in Fig. {8}.
It has some interesting properties. As we already mentioned, it is
concentrated heavily at the highest allowed values of the inflaton
field. This
concentration becomes more and more pronounced with a decrease of the
coupling
constant $\lambda$. At small values of the field $\phi$ this function
rapidly
vanishes. This result is in a  contrast with the behavior of the
square of the
Hartle-Hawking wave function, which is  extremely sharply peaked at
small
$\phi$.  Note that the shape of the solution, as
well as the eigenvalues $\lambda_1$, prove to be practically
independent on
the boundary conditions at $\phi =\phi_e$.

\begin{figure}[h!]
\centering\leavevmode{\epsfysize=5.5 cm{\epsfbox{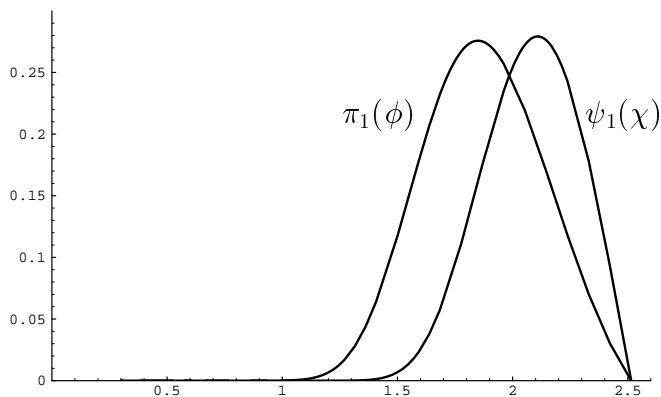}}}
\par
\ 
\caption{ Functions $\pi_1(\phi)$ and $\psi_1(\chi)$, representing the
stationary solution $\tilde P_p(\phi, t|\chi)$ for the theory
${\lambda\over
4}\phi^4$ for $\lambda = 0.1$.}
\label{f2}
\end{figure}

The function $\psi_1(\chi)$ looks  similar to  $\pi_1(\phi)$, see
Fig.
{8}, but  its functional dependence on $\chi$ is, in fact,  quite
different,
revealing an important similarity to the square of the tunneling wave
function
$\sim \exp\Bigl(-3/8V(\chi)\Bigr)$.
This is not unexpected: the boundary conditions (\ref{TunnBound}),
(\ref{TunnSolution}) suggest precisely that kind of behavior.
However, this
`easy explanation' is, to some extent, misleading. The solution for
$\psi_1(\chi)$ is not very sensitive to the boundary conditions at
$\chi =
\chi_e$, and its relation to the tunneling wave function has  deeper
reasons,
to be explained in \cite{MezhLin}.

 To verify (and clarify) this statement by numerical methods, we will
consider
the function $\Psi_1(\chi)$ related to the function $\psi_1$ as
follows:
$\psi_1(\chi) = \Psi_1(\chi) \,
\exp\Bigl(-3/8V(\chi)\Bigr)=\Psi_1(\chi)
\,\exp\Bigl(-{3\over 2\lambda\chi^{4}}\Bigr)$. According to
(\ref{r}),
(\ref{TunnBoundPsi1}), this function obeys the equations
\begin{equation}\label{rexp} \Psi_1'' +
\Psi_1'\Bigl({6\over\lambda\chi^5} +
{3\over\chi}\Bigr) + \Psi_1 \Bigl({36\pi \over\lambda\chi^4}-
{12\over
\lambda\chi^6}  -{\lambda_1\over \pi \chi^6}
\Bigl({6\pi\over\lambda}\Bigr)^{3/2}\Bigr) = 0\ ,
\end{equation}
\begin{equation}\label{TunnBoundPsiExp} \Psi'_1({\chi_{e}})  =  0 \
,~~~~~
\Psi_1(\chi_p) = 0\ .
\end{equation}

\begin{figure}[h!]
\centering\leavevmode{\epsfysize=5.5 cm{\epsfbox{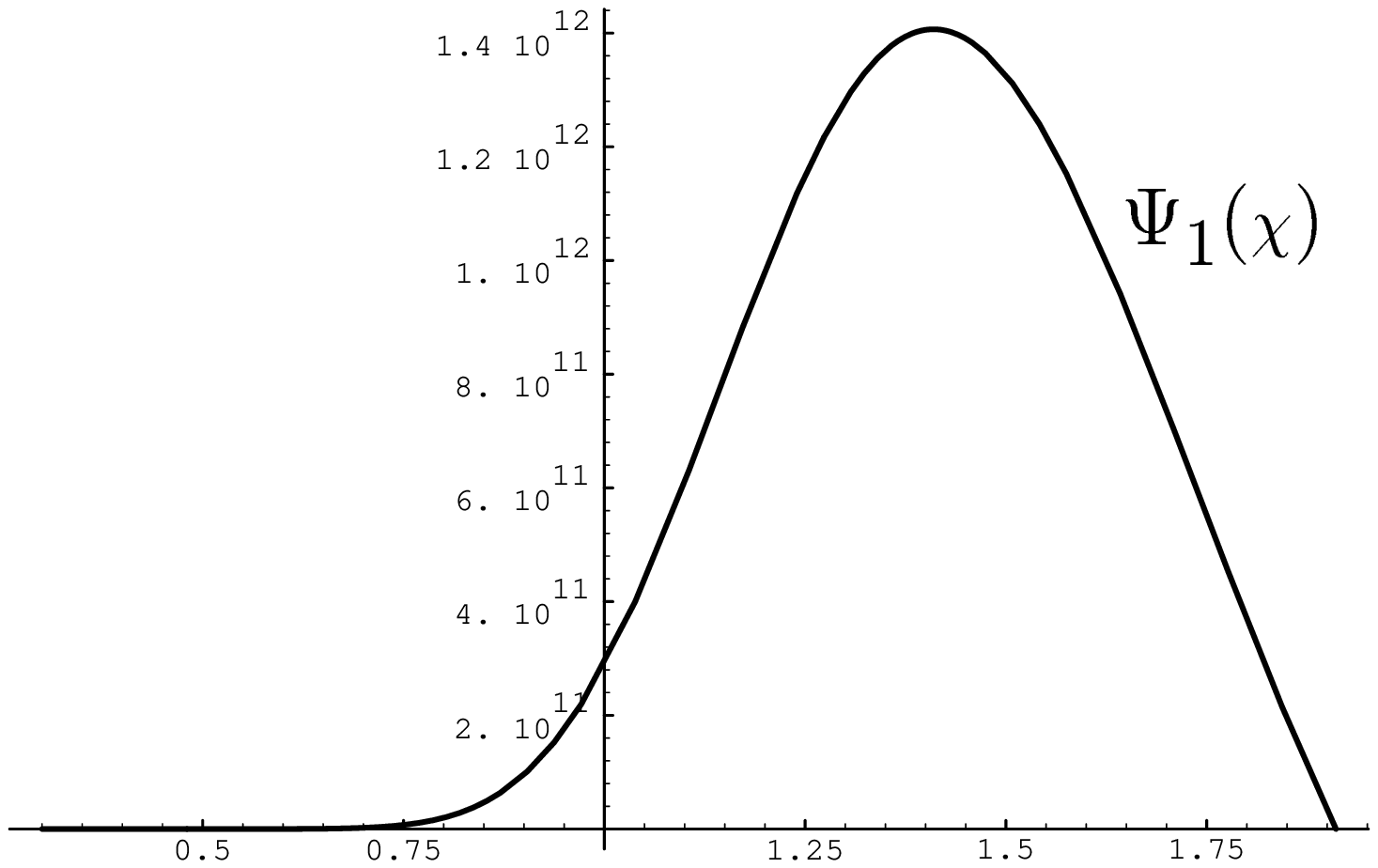}}}
\par
\ 
\caption{Behavior of the function  $\Psi_1(\chi) = \psi_1(\chi) \,
\exp\Bigl(3/8V(\chi)\Bigr)$ for the theory ${\lambda\over 4}\phi^4$
for $\lambda
= 0.3$; time $t$.}
\label{f2}
\end{figure}

 The solution is shown in Fig. {9}. For definiteness, we normalized
$\Psi_1({\chi_{e}}) = 1$. One can easily see that, e.g., for $\lambda
= 0.3$,
the function $\Psi_1$ grows $\sim 10^{12}$ times before it reaches
its maximum.
However, in the same interval, the function
$\exp\Bigl(-3/8V(\chi)\Bigr) =
\exp\Bigl(-{3\over 2\lambda\chi^{4}}\Bigr)$ grows $ \sim  10^{1072}$
times.
This means that the leading contribution to $\psi_1(\chi)$ in the
whole
interval from $\chi_e$ to $\chi_p$ is given not by $\Psi_1$, but by
the square
of the tunneling wave function. This amazing fact is in a complete
agreement
with a tentative interpretation of the tunneling wave function: Its
square
gives the probability that a given domain of an inflationary Universe
originally (at $t = 0$) was (created) in a state  with a given field
$\phi(t=
0) = \chi$.

 However, one should not be too excited about it.  The most dramatic
growth of
$\psi_1$ occurs at relatively small values of $\chi$, close to
$\chi_e$. In
this region the square of the tunneling wave function correctly
describes
$\psi_1$. Similarly, the change of $\psi_1$ in the interval from
$\chi_e$ to
$\chi_p$ in the leading approximation is given by the square of the
tunneling
wave function.This means that the square of the tunneling wave
function
reasonably well describes the exponential suppression of probability
that the
inflationary Universe was originated in a state with a very small
field  $\chi
\sim \chi_e$. However, at  $\chi-\chi_e \gg \chi_e$ the tunneling
wave function
changes
slower than $\Psi_1(\chi)$. In particular, the tunneling wave
function
becomes almost constant near the maximum of $\psi_1$. Therefore if
one wishes
to find the behavior of $\psi_1$ near its maximum (which is one of
the most
interesting problems), the tunneling wave function becomes
irrelevant, and
instead of it one should find  a complete solution for
$\psi_1(\chi)$. This was
one of the main purposes of our investigation.

 Now let us find the distribution $P_p(\phi,t|\chi)$ for the theory
with the
exponential potential $V(\phi) = V_0 \ e^{\alpha\phi}$. The
corresponding
equations in this case  are
\begin{equation}\label{r2}
 \psi_1'' + \psi_1'\left({3\alpha\over 4} - {3\alpha\over
8V_0}e^{-\alpha\chi}\right) + \psi_1 \left({9\pi \over V_0}
e^{-\alpha\chi}
-{\lambda_1\over \pi } \Bigl({3\pi\over
2V_0}\Bigr)^{3/2}e^{-3\alpha\chi/2}\right) = 0\ ,
\end{equation}
\begin{equation}\label{r3}
\pi_1'' + \pi_1'\left({9\alpha\over 4} +
{3\alpha\over 8V_0}e^{-\alpha\phi}\right) + \pi_1 \left({9\alpha^2
\over 8} +
{3\alpha^2 \over 16V_0} e^{-\alpha\phi} + {9\pi \over V_0}
e^{-\alpha\phi}
-{\lambda_1\over \pi } \Bigl({3\pi\over
2V_0}\Bigr)^{3/2}e^{-3\alpha\phi/2}\right) = 0\ .
\end{equation}
The boundary conditions (\ref{TunnBoundPsi}), (\ref{TunnBoundPi}) in
this case
look as follows:
\begin{equation}\label{TunnBoundPsiExpo} \psi'_1({\chi_{e}})
=  {3 \alpha e^{-\alpha\chi_e}\over 8 V_0} \, \psi_1({\chi_{e}})\
,~~~~~
\psi_1(\chi_p) = 0\ ,
\end{equation}
and
\begin{equation}\label{TunnBoundPiExp}
\pi'_1({\phi_{e}})  = -\, {3\alpha
\over 4}\, \pi_1({\phi_{e}})\ ,~~~~~ \pi_1(\phi_p) = 0\  .
\end{equation}
Without any loss of generality, one may take $\chi_e = \phi_e = 0$ in
these
equations, see next paragraph. The solutions of these equations look
qualitatively similar to the solutions to equations for the theory
${\lambda\phi^4\over 4}$, see Fig. {10}. However, here one should
issue
a warning, which is not very important with this time
parametrization, but will
be more important in the next Section.

\begin{figure}[h!]
\centering\leavevmode{\epsfysize=5.5 cm{\epsfbox{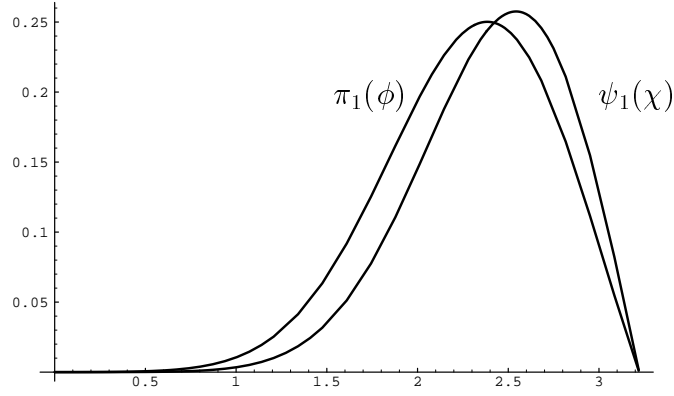}}}
\par
\ 
\caption{ Functions $\pi_1(\phi)$ and $\psi_1(\chi)$, representing
the
stationary solution $\tilde P_p(\phi, t|\chi)$ for the theory $V_0\,
e^{\alpha\phi}$ with $V_0 = 0.2$, $\alpha = 0.5$.}
\label{f2}
\end{figure}

 The boundary conditions at the end of inflation in the theory
${\lambda\phi^4\over 4}$ were natural in the sense that inflation by
itself
ended at $\phi = \phi_e$. In the theory  $V_0 \ e^{\alpha\phi}$
inflation never
ends by itself; one must change the shape of the effective potential
`by hand'
at some place, say, at $\phi = 0$. Without such a cut-off one may get
all kinds
of unphysical stationary solutions. The
cut-off may be effectively performed, e.g., by considering models
with the
potential   $4V_0\,  \sinh^2{\alpha  \phi \over 2}$. This potential
at large
positive  $\phi$ looks  like $V_0 \ e^{\alpha\phi}$, but at small
$\phi$ it looks
like $\frac{1}{2} m^2\phi^2$ (where $m^2 = 2 \alpha^2 V_0$), which
implies that
inflation ends at $\phi \la 0.2$.

 Alternatively, one may introduce an abrupt cut-off at $\phi =
\phi_e$. This
can be accomplished by a discontinuous increase of the derivative
$V'(\phi)$ at
$\phi < \phi_e$. However, such a change would lead to an extra term
in the
boundary conditions (\ref{curr}):
\begin{equation}\label{curr2} \left. V^{3/4}(\phi) \
\frac{\partial}{\partial
\phi}\left({V^{3/4}(\phi)}P_p\right)\right|_{{\phi_{e_+}}}
  =  {3\over 8V^{1/2}(\phi)}\,\Bigl(V'({\phi_{e_-}})-
V'({\phi_{e_+}})\Bigr)\,  P_p({\phi_{e_+}})\ .
\end{equation}
 Since the solutions we found extremely rapidly decrease at small
$\phi$, this
modification was not important for us. However, with different time
parametrizations, the corresponding solutions may take greater values
at
$\phi_e$ or $\chi_e$. In such case one should either take into
account improved
boundary conditions, or  consider
 potentials of the type of  $4V_0\,  \sinh^2{\alpha  \phi \over 2}$
instead of the exponential potential $V_0 \ e^{\alpha\phi}$. We
will return to this question in the next Section.

\section{Stationary solutions with other time parametrizations
\label{OtherTimes}}

In this section we will briefly describe our results concerning the
stationary regime in the $\tau$-parametri\-za\-tion of time, where
$\tau  =
\ln{{a\left(x, t \right) \over a(x,0)}} =  \int_0^t{H(\phi(x,t_1))\
dt_1}$, see
Section \ref{tau}. In this case equations analogous to  (\ref{eq15})
and
(\ref{eq17}) look as follows (compare with eq. (\ref{eq37aau})):
\begin{equation} \label{eq15a}
{1\over 3\pi}\ \left(\sqrt {V(\chi)}\
\frac{d}{d \chi} \Bigl({\sqrt {V(\chi)}}\  {\frac{d
\psi_s(\chi) }{d \chi}}\Bigr) - {3V'(\chi)\over 8
 V(\chi)}\, \frac{d
\psi_s(\chi)}{d \chi}\right)
  = -\bigl(3-\lambda_s\bigr) \, \psi_s(\chi)  \ ,
\end{equation}
and
\begin{equation} \label{eq17a}
{1\over 3\pi}\ \frac{d}{d \phi} \left(\sqrt {V(\phi)}\
\frac{d }{d \phi}\Bigl({\sqrt {V(\phi)}}\ \pi_j(\phi)\Bigr)+
{3V'(\chi)\over 8
 V(\phi)}\, {\pi_j(\phi)}\right)   = -\bigl(3-\lambda_j\bigr) \,
\pi_j(\phi)  \ .
\end{equation}

As before, the
asymptotic solution for $P_p(\phi,\tau|\chi)$ (in the limit $\tau
\rightarrow \infty$) is given by
\begin{equation} \label{eq22x}
P_p(\phi,\tau|\chi) = e^{\lambda_1 \tau}\, \psi_1(\chi) \,
\pi_1(\phi)\, \cdot \left(1 + O\left( e^{-\left(\lambda_1 - \lambda_2
\right) \tau} \right) \right) \ .
\end{equation}
Here $\psi_1(\chi)$ is the only positive eigenfunction of eq.
(\ref{eq15a}),
$\lambda_1$ is the corresponding  eigenvalue, and $\pi_1(\phi)$
 is the eigenfunction
of the conjugate operator (\ref{eq17a}) with the
same eigenvalue $\lambda_1$\@. The stationary normalized solution we
are
looking for is given by
\begin{equation} \label{eq22ab}
  \tilde{P}_p(\phi,\tau \rightarrow \infty|\chi) =  \psi_1(\chi) \,
\pi_1(\phi) \ ,
\end{equation}
 where $ \tilde{P}_p(\phi,\tau |\chi) =  P_p(\phi,\tau|\chi)\,
e^{-\lambda_1 \tau} $.

 Interestingly enough, the boundary conditions for $\psi_1(\chi)$ in
the new
time parametrization remain unchanged. However, due to the change of
the
expression for the probability current in the $\tau$-parametrization,
the
boundary condition on $\pi_1(\phi_e)$ now looks slightly different:
\begin{equation}
\left.\frac{d}{d \phi}\left({V^{1/2}(\phi)}\,
\pi_1(\phi)\right)\right|_{{\phi_{e}}}=0 \ ,
\end{equation}
or, equivalently,
\begin{equation}\label{beta3}
\pi'_1({\phi_{e}}) = -\, {1\over 2}\ {V'\over V}
\, \pi_1({\phi_{e}})\ . \end{equation}

Let us show how the equations for $\psi_1(\chi)$ and $\pi_1(\phi)$
look in the
theory $V(\phi) = {\lambda\over 4} \phi^4$. To simplify these
equations,  we
will make a change of variables, $\xi = \lambda^{1/4} \chi$, $\varphi
=
\lambda^{1/4} \phi$ and use the notation $\alpha_1 =
12\pi\,\lambda^{-1/2}\  (3-\lambda_1)$. Then the
corresponding equations acquire the following form:
 \begin{equation}
\label{eq15aa}
\psi_1'' - {6\over \xi^5}\, \Bigl( 1 -{\xi^4\over 3} \Bigr) \psi_1' +
{\alpha_1\over \xi^{4}} \psi_1 = 0 \ , \end{equation}
\begin{equation} \label{eq17aa}
\pi_1'' + {6\over \varphi^5} \, \Bigl( 1 +\varphi^4 \Bigr)\pi_1' -
{6\over
\varphi^6} \, \Bigl( 1 -\varphi^4 - {\alpha_1\over 6}\varphi^2 \Bigr)
\pi_1 =
0 \ . \end{equation}
The boundary conditions are
\begin{equation}\label{TunnBoundPsiTau}
  \psi'_1({\xi_{e}})  =  {6 \over \xi_e^5}\, \psi_1({\xi_{e}}) \
,~~~~~ \psi_1(\xi_p) = 0\ ,
\end{equation}
and
\begin{equation}\label{TunnBoundPiTau}
 \pi'_1({\varphi_{e}})  = -\, {2\over \varphi_e}\, \pi_1({\phi_{e}})\
,~~~~~
\pi_1(\phi_p) = 0\  .
\end{equation}
 The Planck boundary in the new variables corresponds to $\xi_p =
\varphi_p =
\sqrt 2$. Inflation in the theory ${\lambda\over 4} \phi^4$ ends at
$\phi_e
\sim 0.3 $, which corresponds to $\xi_e = \varphi_e = 0.3
\lambda^{1/4}$.

\begin{figure}[h!]
\centering\leavevmode{\epsfysize=5.5 cm{\epsfbox{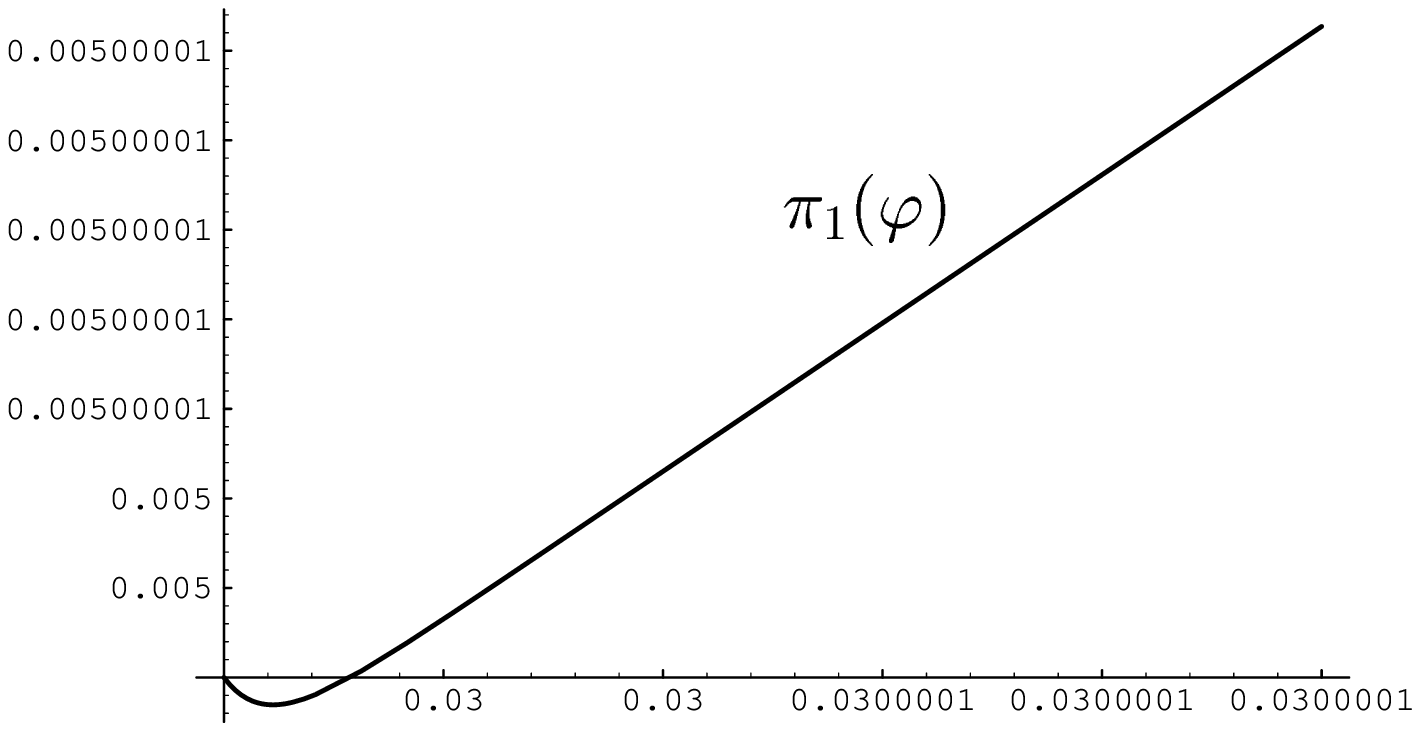}}}
\par
\ 
\caption{ Behavior of the function $\pi_1(\varphi)$ near $\varphi_e$
in the
theory  ${\lambda\over 4}\phi^4$; time $\tau$. Here $\varphi_e =
\lambda^{1/4}
\phi_e$, $\lambda = 10^{-4}$.}
\label{f2}
\end{figure}

 We begin with the equation for $\pi_1$. As in the previous Section,
its
solutions prove to be rather stable with respect to the boundary
conditions at
$\varphi_e$. This is illustrated by Fig. {11}. In the beginning
the curve goes down, in accordance with the boundary conditions
(\ref{TunnBoundPiTau}). However, almost immediately it turns up and
approaches
the asymptotic regime
\begin{equation} \label{eq17aaaa}
\pi_1 = C\    \varphi \
\exp\Bigl(-{\alpha_1\varphi^2\over 12}\Bigr) = c\    \phi \
\exp\Bigl(-{\pi\,
(3-\lambda_1)\phi^2}\Bigr) \ ,
\end{equation}
where $C, c$ are some normalization constants. The change of the
slope of this
curve at  $\varphi_e$ by a factor of two would just slightly modify
the value
of $\varphi$ where the curve reaches its asymptote (\ref{eq17aaaa}).

\begin{figure}[h!]
\centering\leavevmode{\epsfysize=5.5 cm{\epsfbox{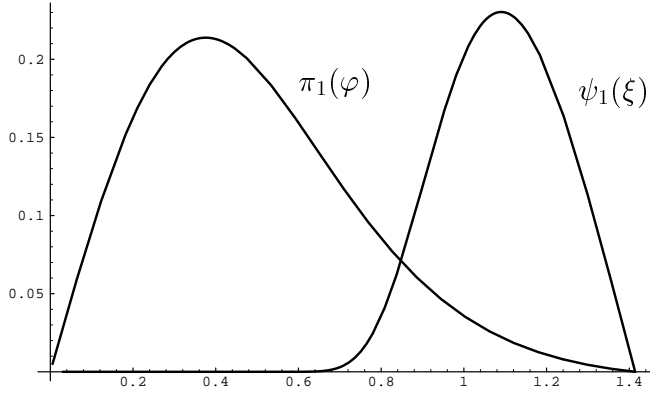}}}
\par
\ 
\caption{Functions $\pi_1(\varphi)$ and $\psi_1(\xi)$, representing
the
stationary solution $\tilde P_p(\phi, \tau|\chi)$ for the theory
${\lambda\over
4}\phi^4$ in terms of variables $\varphi  = \lambda^{1/4} \phi $ and
$\xi  =
\lambda^{1/4} \chi $.
}
\label{f2}
\end{figure}

The result of the numerical solution of eq.  (\ref{eq17aa}) is
shown in Fig. {12}. Interestingly enough, the asymptotic solution
 (\ref{eq17aaaa}) gives an excellent approximation (with an accuracy
of few
percent) to the exact  solution for $\pi_1(\varphi)$  in the whole
interval
from $\varphi_e$  to $\varphi_p$. We have found that the parameter
$\alpha_1$
corresponding to the eigenvalue $\lambda_1$ is given by 41.95.  This
leads to
the following expression both for the eigenvalue $\lambda_1$ and for
the
`fractal dimension' $d_{fr}$ (which in this case refers both to the
Planck
boundary at $\phi_p$ and to the end of inflation at $\phi_e$):
\begin{equation}
\label{eq15u}
  \lambda_1 = d_{fr} = 3 - {41.95  \over 12\pi}\, \sqrt \lambda
\approx 3 - 1.1\,
\sqrt \lambda \ .
\end{equation}

 The solution of eq. (\ref{eq15aa}) for $\psi_1(\xi)$ is also shown
in Fig.
{12}. Note that  due to our redefinition of variables $\chi \to \xi$,
$\phi \to
\varphi$ all solutions here remain the same for {\it all} values of
$\lambda$. To study this solution in a more detailed way, it is very
instructive to represent it in the form
  $\psi_1(\xi) = \Psi_1(\xi) \,
\exp\Bigl(-3/8V(\xi)\Bigr)=\Psi_1(\xi)
\,\exp\Bigl(-{3\over 2 \xi^{4}}\Bigr)$, as we did in the previous
Section. According to  (\ref{eq15aa}), (\ref{TunnBoundPsiTau}), this
function
obeys equations
\begin{equation}\label{rexp1}
\Psi_1'' +
\Psi_1'\Bigl({6\over\xi^5} + {2\over\xi}\Bigr) - \Psi_1 \Bigl(
{18\over \xi^6}
-{\alpha_1\over  \xi^4}\Bigl) = 0\ ,
\end{equation}
 \begin{equation}\label{TunnBoundPsiExpTau}
\Psi'_1({\xi_{e}})  =  0 \ ,~~~~~
\Psi_1(\xi_p) = 0\ .
\end{equation}

\begin{figure}[h!]
\centering\leavevmode{\epsfysize=5.5 cm{\epsfbox{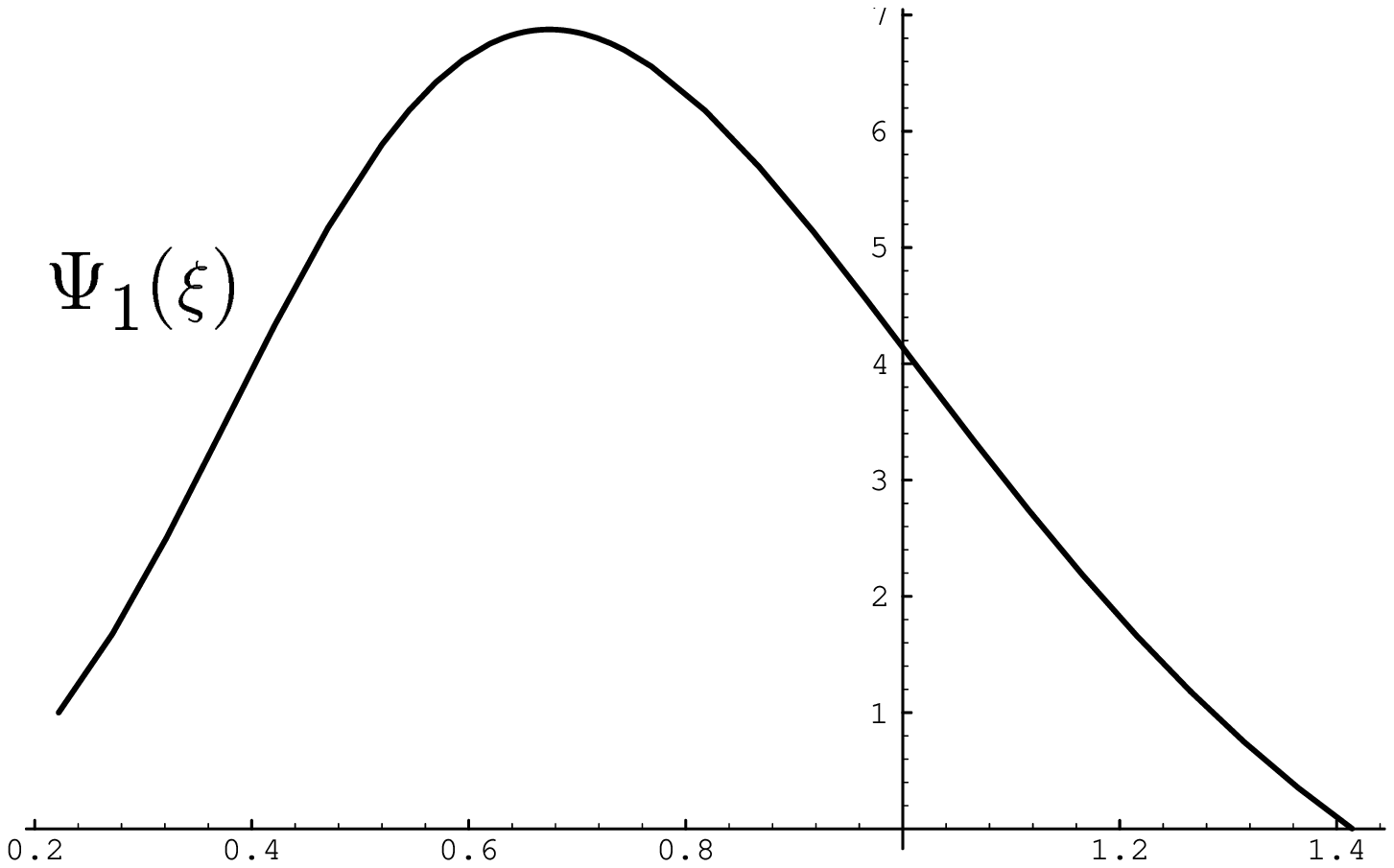}}}
\par
\ 
\caption{  Behavior of the function  $\Psi_1(\xi) = \psi_1(\xi) \,
\exp\Bigl(3/8V(\xi)\Bigr)$ for the theory ${\lambda\over 4}\phi^4$
for $\lambda
= 0.3$; time $\tau$.
}
\label{f2}
\end{figure}

The solution is shown in Fig.  {13}. As before, we normalized
$\Psi_1({\xi_{e}}) = 1$. One can easily see that, e.g., for $\lambda
= 0.3$,
the function $\Psi_1({\xi})$ grows only $7$ times before it reaches
its
maximum. In the same interval, the function
$\exp\Bigl(-3/8V(\xi)\Bigr) =
\exp\Bigl(-{3\over 2\, \xi^{4}}\Bigr)$ grows $ \sim  10^{1072}$
times.
For $\lambda = 10^{-4}$, the function $\Psi_1$ grows only $2300$
times, whereas
the function $\exp\Bigl(-3/8V(\xi)\Bigr) $ grows $ \sim
10^{3216996}$ times.
This difference is much more dramatic than in the standard
$t$-parametrization
of time. It reflects the fact that in almost the whole interval from
$\xi_e$ to
$\xi \sim 0.65$, where the function $\psi_1({\xi})$ reaches its
maximum, it is
correctly described by the square of the tunneling wave function. In
terms of
the original variable $\chi$, this implies that in about one-third of
the whole
interval from $\chi_e = 0.3$ to $\chi_p = ({4/\lambda})^{1/4}$ the
function
$\psi_1({\chi})$ is given by the square of the tunneling wave
function:
\begin{equation}
\psi_1(\chi) \sim \exp\Bigl(-{3\over 8V(\chi)}\Bigr)=
\exp\Bigl(-{3\over 2\lambda\,\chi^{4}}\Bigr) ~.
\end{equation}
Thus, the square of
the tunneling wave function does play an extremely important role in
quantum
cosmology. A more accurate fit to the function $\psi_1(\chi)$, which
can be
used in the whole interval from $\chi_e$ to $\chi_p$,  is given by
$\exp\Bigl(-{3\over 8V(\chi)}\Bigr)\, \Bigl({1\over V(\chi)+0.4} -
{1\over1.4}\Bigr)$.

The resulting  stationary
distribution is
\begin{eqnarray}
\tilde{P}_p(\phi,\tau \rightarrow \infty|\chi) &\sim &
\exp\Bigl(-{3\over
8V(\chi)}\Bigr)\,
\Bigl({1\over V(\chi)+0.4} - {1\over1.4}\Bigr)\, \cdot \,   \phi
\,\exp\Bigl(-{\pi\, (3-\lambda_1)\phi^2}\Bigr) \nonumber \\
 &=& \exp\Bigl(-{3\over 2 \lambda\, \chi^4}\Bigr)\, \Bigl({4\over
\lambda
\chi^4+1.6} - {1\over1.4}\Bigr)\, \cdot \,   \phi \, \exp\Bigl(- 3.5
\sqrt\lambda\phi^2\Bigr)\ .
\end{eqnarray}
This expression is valid  in the whole
interval from
$\phi_e$ to  $\phi_p$ and it correctly describes asymptotic behavior
of
$\tilde{P}_p(\phi,\tau|\chi)$ both at  $\chi \sim \chi_e$ and at
$\chi \sim
\chi_p$.

 A similar investigation can be carried out for the theory $V(\phi) =
V_0\
e^{\alpha\phi}$. The corresponding  equations are
\begin{equation}\label{r4}
 \psi_1'' + \psi_1'\left({\alpha\over 2} - {3\alpha\over
8V_0}e^{-\alpha\chi}\right) + \psi_1\ {3\pi (3-\lambda_1)\over V_0}
e^{-
\alpha\chi} = 0\ ,
\end{equation}
\begin{equation}\label{r5}
 \pi_1'' + \pi_1'\left({3\alpha\over 2} + {3\alpha\over
8V_0}e^{-\alpha\phi}\right) + \pi_1 \left({\alpha^2 \over 2} + {3\pi
(3-\lambda_1)\over V_0} e^{- \alpha\phi} \right) = 0\ .
\end{equation}
The (naive) boundary conditions are
\begin{equation}\label{BoundPsiExpTau}
 \psi'_1({\chi_{e}})  =  {3 \alpha e^{-\alpha\chi_e}\over 8 V_0} \,
\psi_1({\chi_{e}})
\ ,~~~~~ \psi_1(\chi_p) = 0\ ,
\end{equation}
and
\begin{equation}\label{BoundPiExpTau}
\pi'_1({\phi_{e}})  = -\, {\alpha \over 2}\, \pi_1({\phi_{e}})\ ,
{}~~~~~ \pi_1(\phi_p) = 0 ~.
\end{equation}
 The solution for $\psi_1(\chi)$ in the whole interval from $\chi_e$
to
$\chi_p$ looks as follows:
\begin{equation}\label{xyz1}
\psi_1 =  \exp\Bigl(-{3\over
8V(\chi)}\Bigr)\,
 \Bigl({1\over V(\chi)} - {1}\Bigr) = \exp\Bigl(-{3\,
e^{-\alpha\chi}\over
8V_0}\Bigr)\, \Bigl(\frac{e^{-\alpha\chi}}{V_0} - 1 \Bigr)  ~.
\end{equation}
The  solution for $\pi_1(\phi)$
in the whole  range from $\phi_e$ to $\phi_p$ with an accuracy of few
percent
is given by the following simple function:
\begin{equation}\label{xyz}
\pi_1({\phi}) =\Bigl({1\over V(\phi)}
- {1}\Bigr)\, V^{-1/2}(\phi)  =V^{-3/2}_0\, \Bigl(e^{-\alpha\phi}
- V_0\, \Bigr) \,  e^{-\alpha\phi/2} \ .
\end{equation}

\begin{figure}[h!]
\centering\leavevmode{\epsfysize=5.5 cm{\epsfbox{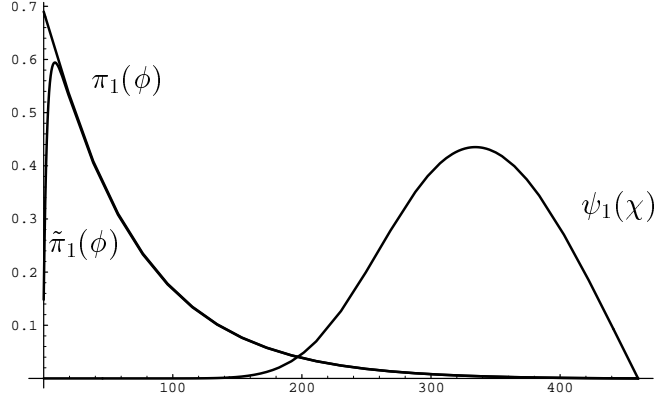}}}
\par
\ 
\caption{Functions $\pi_1(\phi)$ and $\psi_1(\chi)$, representing the
stationary solution $\tilde P_p(\phi, \tau|\chi)$ for the theory
$V_0\,
e^{\alpha\phi}$ with $V_0 = 0.1$, $\alpha = 0.1$. The function
$\tilde\pi_1(\phi)$ corresponds to the solution with improved
boundary
conditions.}
\label{f2}
\end{figure}

 The solutions are shown in Fig. {14}.  Note that now the solution
for
$\pi_1(\phi)$ is concentrated near the end of inflation, at $\phi
\sim \phi_e$.
The reason why we obtained such a solution is related to the
`prescribed'
nature of the end of inflation in the theory with the exponential
potential:
Motion of the field is slow and diffusion is large until the very end
of
inflation at $\phi = \phi_e = 0$. If the speed of motion of the field
$\phi$
near the end of inflation had been vanishingly small,   the
distribution
$\pi_1(\phi)$ would be equal to the  square of  the Hartle-Hawking
wave
function. For the exponential potential, however, this is not the
case, see eq.
(\ref{xyz}).

 As we mentioned at the end of the previous Section, when one
describes
inflation in the theory with the exponential potential, some care
should be
taken of boundary conditions. One should either modify the boundary
conditions
for the case of the exponential potential, or modify the effective
potential
itself. For example, one may study  the theory with the effective
potential
$4V_0\,  \sinh^2{\alpha  \phi \over 2}$. However, this potential  is
considerably different from the exponential potential in a very large
region $0
< \phi \la \alpha^{-1}$. This  leads to a profound modification of
the
solutions, shifting the maximum of  $\pi_1(\phi)$ towards $\phi >
\alpha^{-1}
\gg 1$. On the other hand, there is another way to modify the theory,
by adding
to it one more scalar field, which triggers a phase transition with
an
instantaneous end of inflation at $\phi = 0$. Such a theory is
described in
\cite{Axions}.  In this case the potential of the field $\phi$
remains
exponential at $\phi >0$. In order to obtain a phenomenological
description of
the end of inflation in this model one may assume that the derivative
$V'$
abruptly increases when the field $\phi$ becomes negative. In this
case the
boundary condition for the function $\pi_1$ becomes  (compare with
(\ref{curr2}))
\begin{equation}\label{curr3}
V^{1/2}(\phi) \
\left.\frac{\partial}{\partial
\phi}\left({V^{1/2}(\phi)}\pi_1\right)\right|_{{\phi_{e_+}}}
  =  {3\over 8V(\phi)}\,  \pi_1({\phi_{e_-}})\Bigl(V'({\phi_{e_-}})-
V'({\phi_{e_+}})\Bigr)\ ,
\end{equation}
 where in our case $\phi_e = 0$.  The r.h.s. of this equation gives a
positive
contribution to the value of $\pi_1'(\phi_e)$ in eq.
(\ref{BoundPiExpTau}). The
value of this contribution depends on the jump of $V'$ at the point
$\phi_e$.
Numerical solution of the corresponding equation for different values
of this
jump   shows that the solution  remains unchanged
in the main part of the interval from $\phi_e$ to $\phi_p$, except
for a small
vicinity of the point $\phi_e$, see Fig. {14}. This again confirms
that our
solutions are
rather robust with respect to the change of the boundary conditions
at the end
of inflation.

A complete stationary probability distribution is given by
\begin{eqnarray}
\tilde{P}_p(\phi,\tau \rightarrow \infty|\chi) &\sim &
\exp\Bigl(-{3\over
8V(\chi)}\Bigr)\,
\Bigl({1\over V(\chi)} - {1}\Bigr)\, \cdot \,  \Bigl({1\over V(\phi)}
- {1}\Bigr)\, V^{-1/2}(\phi)\nonumber \\
 &=&  V^{-5/2}_0\, \exp\Bigl(-{3\, e^{-\alpha\chi}\over 8V_0}\Bigr)\,
\Bigl(e^{-\alpha\chi}  - V_0\, \Bigr)  \cdot \,
\Bigl(e^{-\alpha\phi}
- V_0\, \Bigr)\,  e^{-\alpha\phi/2}\ .
\end{eqnarray}
This expression gives a  rather good approximation for
$\tilde{P}_p(\phi,\tau \rightarrow \infty|\chi)$ for all $\phi$ and
$\chi$,
except for a small vicinity of the point $\phi_e$ mentioned above.

Note, that the distributions which we obtained do depend on the
choice of the time parame\-tri\-zation. This is not unexpected; each
event may look differently being described in different coordinate
systems. However, it would be very desirable to find an invariant
description of our results. Fortunately, the most important
qualitative results we discussed, the existence of the
self-reproduction of the Universe and the existence of a stationary
probability distribution  $P_p$, do not depend on our choice of $t$-
or $\tau$-parametrization of time.

\section{Comments and Interpretation \label{Jumps}}

 The results obtained in this paper may seem strange and sometimes
even
counterintuitive. Therefore we are going to discuss their
interpretation and an
alternative derivation in a subsequent publication \cite{MezhLin}.
However, it
is necessary to make some comments right now.

\subsection{Do we need Planck density to have stationarity?
\label{Doweneed}}

There are two possible kinds of the stationarity. First of all,
self-reproduction of inflationary domains implies that even in a very
distant
future there will be many inflationary domains in the Universe
containing all
possible values of scalar fields compatible with inflation. According
to the
no-hair theorem for de Sitter space, each such domain of a radius
greater than
$H^{-1}$ ($h$-region) will expand practically independently of the
processes in
the nearby domains. This means that the Universe will repeatedly
reproduce
inflationary domains, which will have statistically  same properties
as the
similar domains produced billions of years ago.

This kind of stationarity is the most fundamental. Its existence is
related
only to the existence of the regime of self-reproduction.  In all
models we
considered in this paper self-reproduction of inflationary domains
occurs at
$\phi > \phi^*$, where $V(\phi^*) \ll 1$ \cite{b19}.
Self-reproduction occurs
also in the models where  inflation is possible when the field $\phi$
is near a
local  maximum of its effective potential, at $V(\phi) \ll 1$
\cite{ArVil,NewStationary}; such models originally were used in the
new
inflationary Universe scenario. Thus, the very existence of the
process of
self-reproduction of inflationary domains, and, consequently, the
existence of
the first, most fundamental stationarity, does not depend on unknown
processes
at $V(\phi) > 1$. We will call this stationarity  {\it local}, or
{\it
microstationarity}, to distinguish it from the {\it global}
stationarity, or
{\it macrostationarity}, which refers to   stationarity of
probability
distributions over the whole Universe.

This second kind of stationarity may or may not exist. Its existence
does not
follow from any general considerations. This is similar to the
situation in
quantum statistics, where the description in terms of  a
macrocanonical
ensemble sometimes is impossible despite the existence of a good
description in
terms of a microcanonical ensemble.  On the other hand, in the
situations where
the Universe is globally stationary, the description of its evolution
can be
considerably simplified.

As we have seen, in all realistic inflationary models the probability
distribution in comoving coordinates $P_c$ is not stationary.
Fortunately, the
probability distribution $P_p$ is stationary in many interesting
cases.

First of all, it is stationary in the models where the inflaton field
is
fluctuating near a local maximum of its effective potential
\cite{ArVil,NewStationary}. This is already extremely important.
However, for
the reason to be discussed in this Section, it would be most
interesting to
obtain stationary solutions in the theories where inflation is
possible very
close to $V(\phi) \sim 1$.

In this paper (see also \cite{MezhLinSmall}) we have found stationary
solutions
for $P_p$ in a class of models of chaotic inflation with  the
effective
potentials ${\lambda\over 4}\phi^4$ and $V_0\,e^{\alpha\phi}$; a
generalization
for other theories will be considered in a separate publication
\cite{MezhLin}.
The main assumption which we made to obtain these solutions is that
the
self-reproduction of inflationary domains is impossible (or at least
is
strongly hampered) at the density higher than the Planck density. We
gave three
different reasons why this assumption may be reasonable:

\begin{enumerate}
\item  Diffusion equations for $P_p$ and our interpretation of their
solutions
it terms of the distribution of a classical field in a classical
space do  not
work for $V(\phi) \ga 1$. Therefore it seems that at the present
level of our
understanding of the Planckian physics the only thing one can do is
to study
only  inflationary domains   with $V(\phi) \la 1$ and discard those
domains
which jump to  $V(\phi) \ga 1$.

\item  Large  energy density concentrated in the spatial gradients of
the
scalar field fluctuations hampers the process of self-reproduction of
inflationary domains with
 $V(\phi) \ga 1$.

\item  The only natural mass scale near the Planck density is the
Planck mass.
Therefore even if the effective potential is not very steep at small
energy
density, nothing can protect it from becoming very curved due to
quantum
gravity effects near the Planck density. In such a case inflation (or
the
process of self-reproduction of inflationary domains) will cease to
exist at
$V(\phi) \ga 1$.
\end{enumerate}

These arguments suggest that inflation (or at least self-reproduction
of
inflationary domains)  cannot exist at $V(\phi) \ga 1$. Therefore one
should
impose some boundary conditions which do not allow $P_p$ to penetrate
deeply
into the regions with $V(\phi) > 1$. This immediately leads to the
existence of
stationary solutions for $P_p$ concentrated at $V(\phi) \la 1$. The
exact form
of these solutions does depend on the processes near the Planck
boundary.
However, as we will show in a separate publication \cite{MezhLin},
in many
cases this dependence is rather trivial and does not change the
qualitative
behavior of the stationary distribution $P_p$.  For example,  one can
show that
for each $\phi_p$ there exists such $\tilde\phi_p \sim \phi_p$ that
imposing
absorbing boundary conditions at $\phi_p$ gives the same solution for
$\pi_1(\phi)$  as imposing reflecting boundary conditions at
$\tilde\phi_p$.
This means that  the choice between absorbing and reflecting boundary
conditions is equivalent to the corresponding redefinition of the
position of
the Planck boundary. Thus, even though a complete understanding of
physical
processes near the Planck boundary is important for obtaining exact
stationary
solutions for $P_p$, we do not expect that the qualitative features
of these
solutions in the region $\phi \ll \phi_p$ will be dramatically
different from
those which we obtained in the present paper.

One should keep in mind that   in some theories the effective
potential may
become very steep at  $\phi > \phi_b$ where $V(\phi_b)\ll1$, i.e.
without any
relation to quantum gravity effects and the Planck boundary.
If this happens at $\phi_b > \phi^*$, then we will have a stationary
distribution which is concentrated at
$\phi \sim \phi_b$, and all our results will remain
qualitatively correct after substituting $\phi_b$ instead of
$\phi_p$. In
particular, there will be a fractal geometry with a dimension $d_{fr}
=
\lambda_1$ (for $\tau$-parametrization of time), where $\lambda_1$
will be
slightly less than $3$ (and, correspondingly, $d_{fr}
=\frac{\lambda_1}{H(\phi_b)} < 3$ for $t$-parametrization).

On the other hand, there are some theories where the effective
potential at
large $\phi$ approaches some constant value $V_0 \ll 1$. In such
models nothing
will prevent the distribution $P_p$ from moving towards indefinitely
large
$\phi$ without ever reaching the Planck boundary, and  the global
stationarity
may be absent. Another   class of models which may have runaway
solutions are
the models including interaction terms $\xi\phi^2R$.   In such models
 the
distribution $P_p$ is  concentrated near the Planck boundary, but the
Planck
boundary is   a line rather than a point, and there is a room for a
runaway
diffusion along this boundary \cite{LindeBellido}.

 \subsection{\sloppy Is it possible to have stationarity and
self-reproduction of the
Universe at $\phi < \phi^*$?}

In the first part of the paper we have shown that self-reproduction
of the
Universe occurs only if we have an inflationary domain with the field
$\phi >
\phi^*$. The critical field $\phi^*$ is typically  much greater than
$\phi_e$.
For example,  $\phi^* \sim 1/\sqrt{m} \gg \phi_e$ in the theory
${m^2\over 2}
\phi^2$. However, we obtained a stationary probability distribution
for all
$\phi$ in the interval from $\phi_e$ to $\phi_p$. Does this mean that
now we
are taking our words back and saying that the self-reproduction of
inflationary domains may occur at $\phi \ll \phi^*$ as well?

 The answer to this question consists of two parts.  First of all,
for the
existence of  stationary distribution $\tilde{P}_p(\phi,t|\chi)$ at
some field
$\phi = \phi_0$ one does not need  inflation and self-reproduction of
the Universe at $\phi_0$. The distribution will remain stationary
even for
$\phi \ll \phi_e$, i.e. after the end of inflation. The stationarity
of
distribution at small $\phi$ is an automatic consequence of  the
stationarity
of
distribution at large $\phi$ and of the  diffusion and rolling of the
inflaton field from large
$\phi$ towards small $\phi$. This is the reason why we obtained the
same speed
of growth of volume $e^{\lambda_1t}$ for {\it all} $\phi$. This
implies also
that the total volume of
{\it all} parts of the Universe with $\rho \sim 10^{-29}$g$\cdot
$cm$^{-3}$
should increase with the same speed. Indeed, according to our
results, the
total volume of the domains with $\phi = \phi_e$ at all times
increases as
$e^{\lambda_1t}$. But $10^{10}$ years later, the density of matter
inside all
such domains will become  $\sim 10^{-29}$g$\cdot $cm$^{-3}$. It is
obvious that
since the total volume of the domains with $\phi = \phi_e$
permanently grows
as  $e^{\lambda_1t}$, the total volume of the all domains with
$\rho \sim
10^{-29}$g$\cdot$cm$^{-3}$  will grow with exactly the same speed
(with the
time delay of $10^{10}$ years). Consequently, the relative fraction
of volume
of the parts of the Universe with any given properties (i.e. the
total volume
of
the parts with the given properties, divided by $e^{\lambda_1t}$) is
time-independent, even if these parts are post-inflationary.

 Independently of this question, it is interesting to address the
issue of
self-reproduction of the parts of the Universe with $\phi < \phi^*$.

 The critical field $\phi^*$ plays an important role when we study
the
possibility that the process of self-reproduction of inflationary
domains
begins from {\it one} domain of a size $O(H^{-1})$. Indeed, if we
have one
$h$-region with the field $\phi > \phi^*$, this is already enough to
ensure the
permanent self-reproduction of inflationary domains. However,  soon
after the beginning of this process there will be exponentially many
domains
with all values of the field $\phi$, including the field $\phi \sim
\phi_e$.
Typical quantum jumps of the field $\phi$ in such domains within the
time
$\Delta t \sim H^{-1}$ have the amplitude $\sim {H\over 2\pi}$, which
is much
smaller than the average classical decrease $\Delta \phi$ of the
scalar field
during this time
\begin{equation}\label{j1}
\Delta\phi = - {V'\over 3 H^2} =-
{V'\over 8\pi V} \ .
\end{equation}
The end of inflation in a typical scenario is defined by the
condition that the
decrease of $V(\phi)$ within the Hubble time becomes comparable with
$V(\phi)$:
\, $|\Delta V(\phi)| \sim V' |\Delta\phi| \sim V(\phi)$. This gives
$V' \sim
\sqrt{8\pi} V$, and $|\Delta\phi| \sim 1/\sqrt{8\pi}$ at the end of
inflation.
In all realistic models of inflation this decrease is much greater
than the
typical amplitude of fluctuations at the end of inflation, ${H\over
2\pi} \ll
1$.

 However, if we already have many domains with $\phi \sim \phi_e$,
then in some
of these domains large jumps with an amplitude $\delta\phi \ga
|\Delta\phi|
\sim
1/\sqrt{8\pi}$ may occur. The probability of such jumps is
exponentially
suppressed, but once they occur and bring the field towards $\phi >
\phi^*$, an
infinite process of self-reproduction of inflationary domains begins
again.

 Let us first check whether such jumps are possible at all. There may
exist
quantum fluctuations  with any amplitude. However, if the amplitude
is too large, then the gradient energy density of these fluctuations
becomes
much greater than the potential energy density $V(\phi)$, and the
standard
approach we used in this paper should be considerably modified
\cite{Creation}.  The boundary at which our `white noise' becomes
`not so
white' is determined by the condition ${1\over
2}(\partial_i\delta\phi)^2\sim
{1\over 2}(H \delta\phi)^2 \la V(\phi)$, which gives $\delta\phi \la
\sqrt{6}/\sqrt{8\pi}$.

 This means that the standard description of the fluctuations with
$\delta\phi
\sim \Delta\phi$ is valid for $\phi >\phi_e$, since  $|\Delta\phi| <
1/\sqrt{8\pi}$ at $\phi >\phi_e$. Therefore the portion of the
original volume
where the field $\phi$ within the typical  time $\Delta t = H^{-1}$
experiences a jump up by $\delta\phi = C |\Delta\phi| \sim
C/\sqrt{8\pi} $ is
given by the Gaussian distribution
\begin{equation}\label{j2}
P(\phi \to \phi+\Delta\phi+\delta\phi) \sim
\exp{\Bigl(-{(\delta\phi)^2\over 2
<\phi^2>}\Bigr)} = \exp{\Bigl(-{2\pi^2(\delta\phi)^2\over
H^2(\phi)}\Bigr)} =
\exp{\Bigl(-{3 C^2\over 32 V(\phi)}\Bigr)} \ .
\end{equation}
Here $C$ is some
constant, which should be somewhat greater than $1$ if we wish that
the field
$\phi$ increases despite its decrease due to   rolling down by
$|\Delta\phi| \sim
1/\sqrt{8\pi}$:~~ $\delta\phi > |\Delta\phi|$. If we want the field
$\phi$ to
experience a subsequent jump by the same value $\delta\phi =
C/\sqrt{8\pi}$, we
should multiply our result by $\exp{\Bigl(-{3 C^2\over 32
V(\phi+\Delta\phi+\delta\phi)}\Bigr)}$.  However, each of the
subsequent jumps
will be much more probable than the first one, since the degree of
suppression
is exponentially sensitive to the increase of
$V(\phi+\Delta\phi+\delta\phi)$.
Moreover, now the field should not jump so high to compete with the
classical
rolling,
since the value of $\Delta\phi$ decreases at large $\phi$. Therefore,
to a
reasonable approximation, the probability of the first jump
(\ref{j2}) gives us
the whole result. (This estimate is particularly good for small
initial $\phi
\sim \phi_e$, which is  comparable to  $1/\sqrt{8\pi}$.)  Thus, the
portion of
the original volume occupied by some field  $\phi (t = 0) \equiv
\chi$, which
experiences a series   of jumps    to $\phi^*$ (or to any other field
$\phi >
\chi$), is given by
\begin{equation}\label{j3}
P(\chi \to \phi) \sim
\exp{\Bigl(-{3 C^2\over 32 V(\chi)}\Bigr)} \ .
\end{equation}
With an accuracy
of the factor $C= 2$, this is just the square of the
tunneling wave function!

 This result implies that if we have an inflationary domain (or a
collection of
inflationary domains) of a total volume $\sim H^{-3}(\phi)
\exp{\Bigl(+{3
C^2\over 32 V(\chi)}\Bigr)}$, some parts of this domain will enter
eternal
process of the self-reproduction of the Universe even if the field
inside this
domain initially was smaller than $\phi^*$.

\subsection{What  about the Big Bang?}

The main result of our work is that under certain conditions the
properties of
our   Universe  can be described by a time-independent probability
distribution, which we have found for theories with polynomial and
exponential
effective potentials. A lot of work still has to be done to verify
this
conclusion. However, once this result is taken
seriously, one should consider its interpretation and  rather unusual
implications.

When  making cosmological observations, we study our part of the
Universe and
find that in this part inflation ended  about $t_e \sim 10^{10}$
years ago. The
standard assumption of the first models of inflation was that the
total
duration of the inflationary stage was  $\Delta t \sim 10^{-35}$
seconds. Thus
one could come to an obvious conclusion that our part of the Universe
was
created in the Big Bang, at the time  $t_e + \Delta t \sim 10^{10}$
years ago.
However, in our scenario the answer is somewhat different.

Let us consider an inflationary domain which gave rise to the
self-reproduction of new inflationary domains. As we argued in
Section 5.2, one
can visualize self-reproduction of inflationary domains  as a
branching diffusion
process. During this process, the first
inflationary
domain of  initial radius $\sim H^{-1}(\phi)$ within the time
$H^{-1}(\phi)$
splits into $e^{3} \sim 20$ independent inflationary domains of
similar size.
Each of them contains a slightly different field $\phi$,  modified
both by
classical motion down to the minimum of $V(\phi)$ and by
long-wavelength
quantum fluctuations of amplitude $\sim H/2\pi$. After the next time
step
$H^{-1}(\phi)$, which will be slightly different for each of these
domains,
they split again, and so on. The whole process now looks like a
branching
tree
growing from the first (root) domain. The radius of each branch is
given by
$H^{-1}$;  the total volume of all domains at any given time $t$
corresponds to
the `cross-section' of all branches of the tree at that time, and is
proportional to the number of branches. This volume  rapidly grows,
but when
calculating it, one should   take into account  that those branches,
in which
the field becomes  greater than $\phi_p$,  die and fall down from the
tree, and
each branch in which the field becomes  smaller than  $\phi_e$, ends
on an
`apple' (a part of the Universe where inflation ended and life became
possible).

One of our results is that even after we discard at each given moment
the dead
branches and the branches with  apples at their ends, the total
volume of
live (inflationary) domains will continue growing exponentially, as
$e^{\lambda_1
t}$. What is even more interesting, we have found that very soon the
portion of
branches with given properties (with given values of scalar fields,
etc.)
 becomes time-independent. Thus, by observing any finite part of a
tree  at any
given time $t$ one cannot tell how old   the tree is.

To give the most dramatic representation of our conclusions, let us
see
where
most of the apples  grow. This can be done simply by integrating
$e^{\lambda_1
t}$ from $t=0$ to $t = T$ and taking the limit as $T\to \infty$. The
result
obviously diverges at large $T$ as
 $\lambda_1^{-1}\, e^{\lambda_1 T}$, which
means that most  apples grow at an indefinitely large distance from
the root.
In other words, if we ask what is the total duration of inflation
which
produced a typical apple, the answer is that it is indefinitely
long.

This conclusion may seem very strange. Indeed, if one takes a typical
point in
the root domain, one can show that inflation at this point ends
within a finite
time $\Delta t \sim 10^{-35}$ seconds. This is a correct (though
model-dependent) result which can be confirmed by stochastic methods,
using the
distribution $P_c(\phi,\Delta t|\chi)$ \cite{b19}.  How could  it
happen that
the duration of inflation was any longer than $10^{-35}$ seconds?

The answer is related to the choice between $P_c$ and $P_p$, or
between roots
and fruits. Typical points in the root domain drop out from the
process of
inflation within $10^{-35}$ seconds. The number of those points which
drop out
from inflation at a much later stage is exponentially suppressed,
but they
produce the main part of the total volume of the Universe.
Note that the length of each particular branch continued back in time
may well
be finite \cite{Vilenkin}. However, there is no overall upper limit
to the
length
of
branches, and, as we have seen, the longest branches produce almost
all
parts of
the Universe with properties similar to the properties of the part
where we
live now. Since by local observations we can tell nothing about our
distance in
time from the root domain, our probabilistic arguments suggest that
the root
domain is, perhaps, indefinitely far away from us. Moreover, nothing
in our
part of the Universe depends on the distance from the root domain,
and,
consequently, on the distance from the Big Bang.

Thus,  inflation  solves many problems of the Big Bang theory and
ensures that
this theory provides an excellent description of the local structure
of the
Universe. However,  after making all kinds of improvements of  this
theory, we
are now winding up with a model of a stationary Universe, in
which the
notion of the Big Bang    loses its dominant position, being removed
to the indefinite past.

  But from inflation
it also follows that on a much larger scale the Universe is
extremely {\it in}homogeneous. In some parts of the Universe
the energy density $\rho$ is now of the order of one (in Planck
units),
which is 123
orders of magnitude higher than the $\sim 10^{-29}
g\cdot cm^{-3}$ we can see nearby. In such a scenario there is
no reason to assume that the Universe was initially
homogeneous and that all of its causally disconnected parts
started their expansions simultaneously.

There is one subtle  point in this discussion. According to our
picture, the
main part of the volume of the  Universe is being produced by
inflation not
very
far from the  Planck boundary. Indeed, imagine that we imposed our
absorbing
or reflecting boundary conditions not at $V(\phi_p) = 1$, but at
$V(\bar\phi_p) = 1/4$, thus excluding from our consideration those
branches
where the energy density may become greater than $V(\bar\phi_p) =
1/4$. This
would reduce the eigenvalue $\lambda_1$ approximately by a factor of
$2$:
$\bar\lambda_1\sim {1\over 2} \lambda_1$. Correspondingly, the total
volume
of the Universe would grow not as $e^{\lambda_1 t}$, but at a much
slower rate
$e^{\bar\lambda_1 t}$. This means that the main contribution to the
total
volume of the Universe is given by the branches which from time to
time come
very close to the state of the maximal possible energy density
compatible with
the self-reproduction of inflationary domains. In Section 4 we called
such
states  `Small Bangs'; they correspond to the peaks of the mountains
on Figs.
3.6  and 7.5.

Therefore even though the original Big Bang singularity may be
removed to the
indefinite past, the typical time lapse from the last  Small Bang  to
the end
of inflation may be as small as $10^{-35}$ seconds.
{}From the point of view of a present observer, there is no much
difference
between the last Small Bang and the original Big Bang; thus one may
still use
the old Big Bang theory for a phenomenological description of the
observational data. However, if one wishes to understand the global
structure
of the Universe, its beginning and its fate,
one should use a more complete theory including quantum cosmology.

The difference between the Big Bang and the Small Bangs is especially
clear in
the theories where there exists a large difference between the
maximal possible
value of $V(\phi)$ compatible with inflation and the Planck density.
This is
the case  in the theories with the potentials which have a local
maximum at
some $\phi$, or which approach a plateau with $V(\phi ) = V_0 \ll 1 $
at large
$\phi$, or which become very steep at $\phi > \phi_b$, where $\phi^*
\ll \phi_b \ll \phi_p$. In such
theories inflationary quantum fluctuations never bring the Universe
back to the
Planck density. However, as we have argued, this difference may exist
even in
such theories as  $\phi^n$ and $e^{\alpha\phi}$.
The Big Bang  is believed to be the single event of creation of the
very first domains of classical space-time;  the description of these
domains in terms of classical space-time was impossible before the
Big Bang. Meanwhile,  each branch may come through the state which we
call
 `Small Bang' many times; the more times each branch approaches the
state of
maximal density compatible with inflation, the greater contribution
to the
total volume of the Universe it gives.

\subsection{Initial conditions for inflation, different versions of
inflationary
theory and observational data}

 Even though the properties of the Universe around us do not depend
on the
time from the Big Bang to the present epoch, one may still try to
examine the
problem of initial conditions near the cosmological singularity.
However, as we
argued above, a more relevant problem would be to find  typical
trajectories
(branches) which could produce a part of the Universe of our type.
Some aspects
of this
problem can be studied with the help of the   probability
distribution
${P}_p(\phi,t|\chi)$, or with the help of  the function
$\psi_1(\chi)$.\\

Let us consider for example the classical potential of the
Coleman-Weinberg
type used in the new inflationary theory
\begin{equation}\label{u}
V(\phi) = \lambda\left(\phi^4\ln{\phi\over\sigma} - {\phi^4\over 4} +
{\sigma^4\over 4}\right) ~.
\end{equation}
Here $\phi \geq 0$;\, the field  $\sigma \ll 1$ corresponds to the
minimum of
the effective potential (\ref{u}). In the new inflationary Universe
scenario it
was assumed that the high temperature effects put the field $\phi$
onto the top
of the effective potential at $\phi = 0$, and then inflation begins
as the
temperature drops down \cite{b16}. However, it was soon realized that
this
scenario usually does not work, and one should consider chaotic
inflation
instead. According to this scenario, one can have two different
possibilities
to
obtain inflation in the theory (\ref{u}): One may consider
inflationary domains
where the
Universe from the very beginning was in a state close to $\phi = 0$,
or domains
with $\phi \ga 1$ \cite{b17,Chaot2}. Whereas both possibilities in
principle
can be realized, the second one appears to be much more probable from
the
point of view of initial conditions. As we argued in  Section 2,
initial
conditions which are necessary for inflation naturally appear if
inflation may
begin at $V(\phi)\sim 1$, i.e.  close to the Planck density. On the
other hand,
the probability
of inflation seems to be exponentially suppressed if it may begin
only at
$V(\phi) \ll 1$, which typically is the  case if it begins at $\phi =
0$. This
conclusion follows from many different arguments, including the
qualitative
discussion of possible initial conditions near the singularity
\cite{b17,ExpChaot}, calculation of the tunneling wave function of
the Universe
\cite{b56}--\cite{b59}, and even from the computer simulation of
different
regimes of the Universe expansion \cite{Piran}. Did we learn anything
new
about it after the present work?

First of all, our results give some additional indication of
importance of the
tunneling wave function, since its square appears in the expression
for
$\psi_1(\chi)$.  This may be considered as a  new
confirmation of our conjecture that the most natural realization of
chaotic
inflation scenario in the theory (\ref{u}) occurs if inflation begins
near the
Planck density, at $\lambda\phi^4\ln{\phi\over\sigma} \sim 1$, i.e.
at $\phi
\gg 1$.

However, now  we can say even more. If we examine the solutions of
the
diffusion equation for $P_p$ in this model we would find that these
solutions
consist of two independent branches. The field $\phi$ in the domains
with
$\phi < \sigma$ never diffuses to $\phi > \sigma$, and {\it vice
versa}.
Correspondingly, the distribution $P_p(\phi,t|\chi)$, which gives the
total
volume occupied by the field $\phi$, will consist of two branches:
\begin{equation}\label{u1} P^+_p(\phi,t|\chi) = e^{\lambda_1^+
t}\pi_1^+(\phi)\psi_1^+(\chi)~~~~ \mbox{for}~~~~  \phi, \chi > \sigma
{}~,
\end{equation}
 and
\begin{equation}\label{u2}
P^-_p(\phi,t|\chi) =  e^{\lambda_1^-
t}\pi_1^-(\phi)\psi_1^-(\chi)~~~~
\mbox{for}~~~~  \phi, \chi < \sigma ~.
\end{equation}
The eigenvalue $\lambda_1^-$ is approximately equal to  $3H(\phi =
0)\ll 1$
\cite{ArVil,NewStationary}, whereas from our results it follows that
$\lambda_1^+\sim 1$. This means that if the Universe originally
contained
both domains with $\phi  > \sigma$ and with $\phi< \sigma$, then
within a very
short time of the order of the Planck time $t \sim 1$, the main part
of the
volume of the Universe becomes totally dominated by domains with
$\phi >
\sigma$.

Using the analogy between the Universe and the growing tree, let us
imagine
that the branches with   $\phi  > \sigma$ are green, and  the
branches with
$\phi< \sigma$ are red. Our results imply that if the Universe
originally
contained at least one green
branch, then very soon it becomes all green, independently of the
original
number of red branches (which in fact should be smaller than the
original
number of green branches if our previous arguments concerning initial
conditions are correct).

This conclusion may be somewhat modified if one takes into account a
possibility of large jumps of the field $\phi$ over the
non-inflationary
region
near $\sigma$. Note that the width of this region is very large,
$\Delta\phi
\sim 1$, and since this region is not inflationary, our methods do
not allow
any
description of such jumps. If such jumps are possible, for example,
due to the
process described in \cite{Creation}, or due to Euclidean tunneling,
then green
branches will produce red off-springs, and the total volume of the
red domains
will grow with the same speed as the total volume of the green
domains.
However, even if this process is possible, the total volume of
domains $\phi<
\sigma$ will remain exponentially small as compared with the volume
of
domains with $\phi  > \sigma$ because of  the exponentially small
probability
of such large jumps \cite{Creation}.

Suppose now that the effective potential $V(\phi)$ (\ref{u}) at large
  $\phi$
grows not as $\lambda\phi^4\ln{\phi\over\sigma}$ but, for example, as
$e^{\alpha\phi}$ with $\alpha \ga \sqrt{16\pi}$, or in any other way
which
precludes inflation (or self-reproduction of inflationary domains) at
 $\phi>
\sigma$. Then, in the absence of any competition, the red branches
(beginning
at $\phi \ll \sigma$) will win. Of course, this will happen only if
initial
conditions are good enough for inflation to begin at $\phi = 0$. As
we just
argued, this probability  is exponentially small. However, we are
speaking
about the conditional probability that our  part of the Universe was
formed by
inflationary expansion, as compared with the probability that it was
formed  by
some non-inflationary process.  This suggests that in order to
compare these
probabilities one should multiply the probability of having given
initial
conditions by the subsequent increase  of the volume of any part of
the
Universe
with these initial conditions, and then normalize the total
probability by
dividing it by the total volume of the Universe. This was our method
of
obtaining the normalized probability distribution
$\tilde P_p$. If this  is correct, then the self-reproducing
inflationary
branches always win. Indeed, whatever is the probability of proper
initial
conditions for inflation beginning at  $V(\phi) \ll 1$, the main
contribution
to
the total volume of the Universe will be given by self-reproducing
inflationary
domains, since this contribution even now continues growing
exponentially as
$e^{\lambda_1t}$, and only this contribution will survive  after the
division
by
$e^{\lambda_1t}$.

Thus, it may be possible to have a consistent cosmological theory
even if
inflation occurs only at $V(\phi) \ll 1$. This is the case, e.g., for
the
`natural
inflation' \cite{Natural} and for the hyperextended inflation
\cite{Hext}.
However, as we  have seen, realization of inflation in these models
requires
somewhat trickier reasoning than in the theories where inflation is
possible at $V(\phi) \sim 1$. What if the Universe was created  in a
non-inflationary state, because of  the exponential suppression $\sim
e^{-3/8V(\phi)}$ of the probability to produce an inflationary
Universe with small $V(\phi)$? Does
it make sense to multiply the volume of an unborn inflationary
Universe by
$e^{\lambda_1t}$? Even though we believe that  this is a wrong
objection and
that the indefinitely large growth of volume produced by inflation
does solve
the problem of initial conditions for every theory where the
self-reproduction
of the Universe is possible \cite{Mijic,MyBook}, it would be
desirable to have
an
alternative realization of  chaotic inflation scenario, where initial
conditions for inflation would appear in a natural way even if
inflation driven
by the field $\phi$ may occur   only at $V(\phi) \ll 1$.

Fortunately, this task can be easily fulfilled
\cite{Axions,TwoFields}. The
simplest way to do it is
to introduce an additional    scalar field $\Phi$, which may   have
no
interaction with the field $\phi$, but which can drive inflation at
$V(\Phi)
\sim 1$. In such a theory inflation may begin  at $V(\Phi) \sim 1$ in
a quite
natural way.
This may be  a `bad' inflation, producing large perturbations
${\delta\rho\over\rho} \gg 10^{-5}$. However, in the process of
indefinitely
long inflation driven by the field $\Phi$, fluctuations of the scalar
field
$\phi$ are produced, which in some domains of the Universe put the
field
$\phi$ on the top of its effective potential.  In those domains where
the heavy
field $\Phi$ rolls down to the minimum of $V(\Phi)$, inflation
continues due
to the light field $\phi$, which determines the properties of the
observable
part of our Universe. Models of this type were called `hybrid
inflation' in
ref. \cite{Axions}.

Let us consider, for example, the `natural inflation' model, which
has an
inflaton potential of the type \begin{equation}\label{u3}
V(\Phi) = V_0~\left (1 - \cos {\Phi\over \Phi_0 }\right )~.
\end{equation}
This is exactly the model which we considered in our computer
simulation of
 production of the axionic domain walls during inflation driven by
the field
$\phi$, see Section 3.4. Black
lines in Fig. {5} correspond to the parts of the Universe where $\Phi
= \pi
(2n + 1)$, i.e. where quantum fluctuations have driven  the field
$\Phi$ onto
the top of its effective potential. In the comoving coordinates these
black
lines
look very thin, but one should remember that they are always much
thicker
than $H^{-1}$. In those parts of the Universe where the field $\phi$
eventually
rolls down to the minimum of its effective potential, inflation still
continues in
the black regions with $\Phi = \pi (2n + 1)$. The Universe in these
regions
enters the process of eternal self-reproduction. After a while, the
parts of
the
Universe produced by these black regions enter a stationary regime,
in which
nothing depends on the first stages of inflation driven by the field
$\phi$.
This exactly corresponds to the scenario we discussed in the previous
paragraph, up to an obvious redefinition $\phi \leftrightarrow \Phi$.

Of course, there will be many other parts of the Universe where at
the end of
inflationary stage driven by the field $\phi$, the field $\Phi$ will
not stay on
the top of the effective potential (\ref{u3}).  The volume of such
parts at the
end of the first stage of inflation  will be greater than the volume
of the parts
with $|\Phi - \pi (2n + 1)| \la H$  by a factor $\sim \Phi_0/H \gg
1$. This is
reflected by the relatively large area painted  white in Fig.  {5}.
However,
the regions with $|\Phi - \pi (2n + 1)| \la H$ (painted black in Fig.
 {5}) enter
the
process of   self-reproduction, and very soon the main part of the
volume of
the Universe  (at a given value of the field $\phi$) will be
dominated by
domains produced by inflation of these   black regions.

Thus, even though the `natural inflation' model does not look natural
at all
from the point of view of initial conditions, it can be  incorporated
into a
chaotic inflation scenario of a more general type where the problem
of initial
conditions can be easily solved. According to this scenario, we live
in the
remnants of the domain walls produced at the first stage of
inflation.

Our main conclusion is that whenever one may have two different
inflationary
branches, the main contribution to the total volume of the Universe
will be
given by the branch on which inflation may occur at a greater value
of
$V(\phi)$. The initial conditions for inflation on this branch are
also more
natural. However, it is possible to construct consistent inflationary
models in
which our part of the Universe is produced by a branch where
inflation may
occur only at very small $V(\phi)$.
\vskip 0.5cm

The main motivation for our investigation was a desire to get a
complete and
internally consistent picture of the global structure of the
Universe. However,
sometimes the knowledge of the global structure of the Universe may
tell us
something nontrivial about its local structure.  For example, let us
consider
the ratio of adiabatic perturbations to   gravitational waves, or the
ratio
between scalar and tensor perturbations of metric. Is it possible to
find a
model-independent relation between these perturbations
\cite{Davis,Kolb}?

In some models   this relation  may depend  not only on the choice of
the
potential, but also  on the  initial conditions for inflation. In
particular,
the amplitude and spectrum of
adiabatic perturbations produced at the last stages of inflation in
the theory
(\ref{u}) do  not depend
strongly on whether inflation begins at $\phi \ll \sigma$ or at $\phi
\gg
\sigma$. In both cases ${\delta\rho\over \rho} \sim \sqrt \lambda ~
\log^{3/2}
l$~ \cite{MyBook}. However, the amplitude of gravitational waves
produced
during the
rolling from $\phi > \sigma$ will be proportional to $H(\phi \sim
M_p) =
O(\sqrt\lambda)$, whereas the amplitude of gravitational  waves
produced
during the rolling from $\phi < \sigma$ is proportional to $H(\phi =
0) \sim
\sqrt\lambda\phi^2_0 \ll \sqrt \lambda$. Without saying anything
about
initial conditions and about the process of self-reproduction of the
Universe in
this theory, one cannot have a definite prediction of the amplitude
of
gravitational waves and of the ratio of tensor perturbations of
metric to the
scalar ones in this model. However, as we have argued, the main part
of the
volume of the Universe is produced by inflation along the branches
with $\phi >
\sigma$. This enables us to make a definite prediction  of the
amplitude of
tensor perturbations in this model: this amplitude should be
proportional to
$\sqrt\lambda$ rather than to $\sqrt\lambda\phi^2_0$.

This example shows that the investigation of the global structure of
inflationary Universe, which at the first glance deals with the
scales which we
will never observe, may lead to  testable predictions.  In what
follows we will
discuss a much more speculative possibility, which, however, may
deserve future
investigation.

Until now, most of the studies of properties of the observable part
of the
Universe were to some extent related to the distribution $P_c$. It
was assumed
that one should take a typical sample of inflationary Universe, study
its
evolution and predict its properties. However, our investigation
shows that
the field distributions which seem typical  from the point of view of
 $P_c$,
may happen to be very unusual from the point of view  of    $P_p$,
and {\it
vice versa}. What if we live in a part of the Universe which is very
unusual
from the `normal' point of view, reflected in the distribution $P_c$?

We have discussed already the most dramatic difference between $P_c$
and $P_p$.
A  trajectory which is typical from the point of view of $P_c$
corresponds to
the duration of inflation $\sim 10^{-35}$ seconds, whereas  from the
point of
view of $P_p$ a typical trajectory spends an indefinitely long time
in the
inflationary regime  near the Planck boundary. Analogously, the
standard
investigation of  density perturbations suggests that the typical
deviation of
the density of the Universe from the critical density  cannot exceed
${\delta\rho\over \rho}\sim 10^{-5}$ on the scale of the horizon. The
question
we would like to address is whether there may exist some reason for
us to live
in a not very typical domain, where the deviation of density from the
critical
density might be greater than $10^{-5}$. The reason why this might
happen can
be explained as follows. If  one tries to find out those trajectories
which
give the dominant contribution to the total volume of the Universe
within a
given time interval $t$, one may conclude that if the time $t$ is
sufficiently
large, then the main contribution will be given by the trajectories
which first
rush up towards the highest possible values of $V(\phi)$, spend there
as long
time as possible, and at the very last moment rush down towards small
$\phi$
{\it with the speed even somewhat greater than the speed of the
classical
rolling.} Indeed, even though the probability of such a regime is
exponentially
suppressed, those lucky trajectories, which happen to follow these
rules of the
game and saved some time in the process of rolling down, may become
generously
rewarded by the additional exponentially large growth of volume
during the
extra time they spend near the highest possible values of $V(\phi)$.

If this is the case, the main contribution to the volume of the
Universe at a
given time $t$ in a state with a given density $\rho$ will be given
by the rare
regions where the field $\phi$ at the last stages of inflation jumped
down with
the speed exceeding its speed of the nearby domains. Thus, the energy
density
inside these regions should be somewhat smaller than the energy
density in
their neighborhood. However, not every jump down may compensate the
delay in
the homogeneous classical rolling. The proper jump should result in a
nearly
homogeneous decrease of density on a scale comparable to the scale of
the
horizon.  A region where such a jump took place
  will look now like a part  of an open Universe. Such regions will
be
spherically symmetric, since the probability of formation of
asymmetric regions
by quantum jumps will be suppressed by a large  exponential factor,
just as the
probability of formation of asymmetric bubbles during the first order
phase
transitions\footnote{For a discussion of   a related possibility to
explain
homogeneity of the observable part of the Universe see also
\cite{Creation,Sasaki}.}.

Needless to say, this scenario is extremely speculative.  Moreover, a
preliminary investigation  of this effect indicates that in the
simple theories
with $V(\phi) \sim \phi^n$ the corresponding effect is insignificant.
The only
theories where this  effect  might  be considerable are those where
the Hubble
constant at the last stages of inflation is very small. However,  the
possibility that we may live in  a   locally open part of
inflationary
Universe, i.e. in a part with $\Omega < 1$,   should not be
overlooked.  In
order to study it in a proper way, one needs to know not only the
probability
distribution $P_p$, but also the correlation functions for the values
of the
field $\phi$ at different points, taking into account different rate
of
expansion  of the Universe along different inflationary trajectories.
  We hope
to return to a discussion of this possibility in a separate
publication.

\section{Brief Summary \label{Summary}}

The history of the development of cosmology seems to follow a very
nontrivial
path. At the beginning of the century many people tried to find a
stationary
solution of the Einstein  equations, with the hope that General
Relativity
would
resolve the inability of   Newton's   theory to provide us with a
stationary
cosmological model. Einstein even introduced the cosmological
constant into his
theory for this purpose. The non-stationary character of the Big Bang
theory
advocated by Gamov on the basis of Friedmann  cosmological models
seemed
very unpleasant to many scientists in the  fifties. Then, the
discovery of the
cosmic microwave background turned the situation upside down.
Physicists
began to treat with contempt any attempts to find stationarity
(remember the
`steady-state' model)\@.  After several decades of the reign of the
Big Bang
theory, the inflationary scenario appeared, which solved many of the
intrinsic
problems of the Big Bang cosmology and apparently removed the last
doubts
concerning its validity.

However, it was realized
soon  afterwards, that inflation is even more dynamic than the old
Big Bang
theory. In inflationary cosmology, in addition to the ordinary
classical
evolution of the Universe governed by the Einstein equation, quantum
mechanical evolution proves to be  extremely important, being
responsible for
the large-scale structure formation and even for the global structure
of the
Universe. This quantum mechanical evolution can be approximately
described by stochastic methods, and some of the solutions of the
corresponding stochastic equations prove to be stationary!
Surprisingly
enough, after the dramatic development of the Big Bang  theory during
the
last ten years, we are coming now to a new formulation of the
stationary
cosmology, on a new level of understanding and without losing a
single
achievement of our predecessors. The observable part of the Universe
can be
very well described by   the homogeneous isotropic Big Bang model.
However,
on  extremely large scales (far beyond the visible horizon) the
Universe is
very inhomogeneous. On even larger  scales this inhomogeneity
produces a
kind of fractal structure, repeating itself on larger and larger time
and
length scales. The statistical properties of this structure are what
we
have found to be stationary.

Of course, we are far from giving final answers to fundamental
questions of cosmology. In fact,  we are just beginning to learn how
to ask
proper questions in the context of the new cosmological paradigm.
We need
to make sure that our results have correct quantum mechanical
interpretation.
It would be important  to find an invariant way of formulating our
results, which
would not depend on a particular choice of   time parametrization.
There are
many other problems related to our approach which are to be solved.
We know, however, that at the first stage of any investigation in
quantum field
theory or in quantum statistics it makes a lot of  sense to find a
vacuum state,
or a state of thermal equilibrium, or any  stationary state which may
play the
role of a ground state of the system.  We hope very much  that the
existence
of the stationary regime of the evolution of the Universe described
in this
paper  may help us to find a proper framework for the future
investigation
of quantum cosmology.

\vspace{0.2cm}

It is a pleasure to thank  J. Bond, S. Hawking, M. Miji\'c, D. Page,
A. Starobinsky and A. Vilenkin  for valuable discussions.  This work was
supported in part  by NSF grant PHY-8612280.

%\vfill

%\newpage

\end{document}